\documentclass[useAMS,usenatbib]{mn2e}
\usepackage{graphicx,amsmath,ctable,amssymb}
\usepackage{mathrsfs}

\newcommand{\Msun}{\mathrm{M}_{\odot}}

\newcommand{\Mpc}{\mathrm{Mpc}}

\newcommand{\da}{\delta}
\newcommand{\dd}{\mathrm{d}}
\newcommand{\pd}{\partial}

\newcommand{\fNL}{f_{\mathrm{NL}}}
\newcommand{\MM}{\mathcal{M}}
\newcommand{\fcoll}{f_{\mathrm{coll}}}
\newcommand{\avgfcoll}{f_{\mathrm{coll,0}}}
\newcommand{\fcollgm}{f_{\mathrm{coll}}^{\mathrm{G}}}
\newcommand{\avgfcollgm}{f_{\mathrm{coll,0}}^{\mathrm{G}}}

\newcommand{\Afd}{\mathcal{A}}
\newcommand{\Bfd}{\mathcal{B}}
\newcommand{\Cfd}{\mathcal{C}}

\newcommand{\dcoll}{\delta_{\mathrm{fcoll}}}
\newcommand{\bcoll}{b_{\mathrm{fcoll}}}
\newcommand{\mmin}{M_{\mathrm{min}}}
\newcommand{\Tvir}{T_{\mathrm{vir}}}
\newcommand{\mumol}{\mu_{\mathrm{mol}}}
\newcommand{\Smin}{S_{\mathrm{min}}}
\newcommand{\damin}{\delta_{\mathrm{min}}}

\newcommand{\rhoH}{\rho_{\mathrm{H}}}
\newcommand{\rhoHI}{\rho_{\mathrm{HI}}}
\newcommand{\rhoHII}{\rho_{\mathrm{HII}}}
\newcommand{\drhoHII}{\delta_{\rho_\mathrm{HII}}}
\newcommand{\bHII}{b_{ \rho \mathrm{HII}}}
\newcommand{\nHII}{n_{\mathrm{HII}}}

\newcommand{\nH}{n_{\mathrm{H}}}
\newcommand{\bnH}{\bar{n}_{\mathrm{H}}}
\newcommand{\DHII}{\Delta_{\mathrm{HII}}}
\newcommand{\ngam}{n_{\gamma}}
\newcommand{\nHI}{n_{\mathrm{HI}}}
\newcommand{\alphaB}{\alpha_{\mathrm{B}}}
\newcommand{\clumpone}{C^{(1)}_{\gamma \mathrm{H}}}
\newcommand{\clumpHII}{C_{\mathrm{HII}}}
\newcommand{\clumpGH}{C_{\gamma \mathrm{H}}}
\newcommand{\clumptwo}{C^{(2)}_{\gamma \mathrm{H}}}
\newcommand{\Tdelta}{\tilde{\delta}}
\newcommand{\TDs}{\tilde{\Delta}_s}
\newcommand{\TDHII}{\tilde{\Delta}_{\mathrm{HII}}}
\newcommand{\TDgam}{\tilde{\Delta}_{\gamma}}
\newcommand{\TalphaB}{\hat{\alpha}_{\mathrm{B}}}
\newcommand{\betaSI}{\beta_{\mathrm{S.I.}}}

\title[The scale-dependent signature of PNG in the large-scale structure of cosmic reionization]{The scale-dependent signature of primordial non-Gaussianity in the large-scale structure of cosmic reionization}

\author[D'Aloisio, Zhang, Shapiro, and Mao]{Anson
  D'Aloisio$^{1}$\thanks{Email: anson@astro.as.utexas.edu}, 
	Jun Zhang$^{2,1}$, Paul R. Shapiro$^{1}$, and Yi Mao$^{3,4,1}$ \\
$^1$Department of Astronomy and Texas Cosmology Center, University of Texas, Austin, TX 78712, USA \\ $^2$Department of Physics, Shanghai Jiao Tong University, Shanghai 200240, China \\ $^3$CNRS, UPMC Univ Paris 06, Institut d'Astrophysique de Paris, UMR7095, 98 bis, boulevard Arago, F-75014, Paris, France\\
$^4$Sorbonne Universit\'es, Institut Lagrange de Paris (ILP), 98 bis, boulevard Arago, F-75014 Paris, France
}

\begin{document}

\maketitle

\begin{abstract}
The rise of large-scale structure in the universe depends upon the statistical distribution of initial density fluctuations generated by inflation.  While the simplest models of inflation predict an almost perfectly Gaussian distribution, more-general models predict primordial deviations from Gaussianity that observations might yet be sensitive enough to detect.  Recent measurements of the Cosmic Microwave Background (CMB) temperature anisotropy bispectrum by the \emph{Planck} collaboration have significantly tightened observational limits on the level of primordial non-Gaussianity (PNG) in the Universe, but they are still far from the level predicted by the simplest models of inflation.  Probing levels of PNG below CMB sensitivities will require other methods, such as searching for the statistical imprint of PNG on the clustering of galactic halos.  During the cosmic epoch of reionization (EoR), the first stars and galaxies released radiation into the intergalactic medium (IGM) that created ionized patches whose large-scale geometry and evolution reflected the underlying abundance and large-scale clustering of the star-forming galaxies.  This statistical connection between ionized patches in the IGM and galactic halos suggests that observations of reionization may provide another means of constraining PNG.  We employ the linear perturbation theory of reionization and semi-analytic models based on the excursion-set formalism to model the effects of PNG on the EoR.  We quantify the effects of PNG on the large-scale structure of reionization by deriving the ionized density bias, i.e. the ratio of the ionized atomic to total matter overdensities in Fourier space, at small wavenumber. Just as previous studies found that PNG creates a scale-dependent signature in the halo bias, so, too, we find a scale-dependent signature in the ionized density bias.  Our results, which differ significantly from previous attempts in the literature to characterize this PNG signature, will be applied elsewhere to predict its observable consequences, e.g. in the cosmic 21cm background.
\end{abstract}


\section{Introduction}
\label{SEC:introduction}

Observational probes of the epoch of reionization (EoR) are of great interest not only because they promise to yield new information on primordial galaxies, and the effects of their radiation backgrounds on the inter-galactic medium (IGM);  these probes may some day deliver a wealth of cosmological information as well.  The power spectrum of redshifted 21cm brightness temperature fluctuations from the EoR is a notable example with great potential for cosmological application.  Theoretical investigations to date have mainly focused on exploiting the separation of the matter power spectrum from the influence of astrophysical uncertainties in the EoR, which according to linear perturbation theory is made possible by anisotropy from peculiar velocity in the neutral hydrogen gas \citep{2005ApJ...624L..65B}.  The possibility of this clean separation has stimulated a broad literature on the topic of constraining cosmology with future 21cm surveys [e.g. see \citet{2006PhR...433..181F} and references therein.  Also see \citet{2012arXiv1211.2036S} for a recent re-examination of this technique].          

 One topic currently at the forefront of research in cosmology is the possibility that the initial conditions for structure formation were not perfectly Gaussian.  While the simplest inflationary models -- the canonical single-field slow-roll models -- predict an almost perfectly Gaussian distribution of initial fluctuations,  more general inflationary models predict significant deviations from Gaussianity that observations might yet be sensitive enough to detect.  There is therefore great interest in developing new ways to measure primordial non-Gaussianity (PNG) since its detection (or non-detection) would have profound implications for inflationary theory.  

Perfectly Gaussian initial conditions can be entirely characterized by the power spectrum of initial density fluctuations, while PNG is characterized by higher-order statistics.  The lowest-order statistic that can distinguish Gaussian from non-Gaussian primordial fluctuations is the 3-point correlation function, or equivalently, its Fourier transform -- the bispectrum.  In general since the bispectrum is zero for the Gaussian case, we can write it as

\begin{equation}
B_{\Phi}(k_1,k_2,k_3)=\fNL F(k_1,k_2,k_3)
\end{equation}
where $F(k_1,k_2,k_3)$ is a function only of the magnitudes of the wavevectors of any three perturbation modes, $\boldsymbol{k}_{1,2,3}$ (assuming translational and rotational invariance), $\fNL$ is a dimensionless parameter which indicates the level of departure from Gaussianity, called the ``non-linearity parameter" \citep{1994ApJ...430..447G,2000PhRvD..61f3504W,2001PhRvD..63f3002K,2004JCAP...08..009B}, and $\Phi$ is the Bardeen potential fluctuation in the matter-dominated epoch.  The primordial spectra generated by inflationary models vary considerably from one model to the next and can be quite complicated.  Observational efforts have therefore focused on phenomenological templates which capture the dominate shapes, or functional dependences on wavenumber, of the spectra generated by wide classes of inflationary models.  

An important example is the so-called ``local" template, in which $\Phi(\boldsymbol{r})$ is obtained from a quadratic transformation of the local Gaussian fluctuation field, $\phi(\boldsymbol{r})$, according to\footnote{There are two conventions for equation (\ref{EQ:localquadratic}) in the literature:  1) The CMB convention in which $\Phi$ is evaluated in the matter-dominated epoch, and 2) The large-scale structure convention in which $\Phi$ is linearly extrapolated to the present day.  These two conventions result in different normalizations of the non-linearity parameter, $f_{\mathrm{NL}}^{\mathrm{local}}(\mathrm{LSS}) \approx 1.3 f_{\mathrm{NL}}^{\mathrm{local}}(\mathrm{CMB}) $.  Here and throughout the rest of the paper, we use the CMB convention.  }

\begin{equation}
\Phi(\boldsymbol{r}) = \phi(\boldsymbol{r}) + \fNL^{\mathrm{local}} \left[  \phi(\boldsymbol{r})^2 - \langle \phi^2(\boldsymbol{r}) \rangle \right],
\label{EQ:localquadratic}
\end{equation}
where $< >$ refers to an average over all space \citep{1990NuPhB.335..197H,1991ASPC...15..339K,1990PhRvD..42.3936S,1994ApJ...430..447G,2000MNRAS.313..141V, 2000PhRvD..61f3504W,2001PhRvD..63f3002K}.  To first order in $\fNL^{\mathrm{local}}$, equation (\ref{EQ:localquadratic}) gives a primordial bispectrum of the following form\footnote{The bispectrum in equation (\ref{EQ:localquadraticbispectrum}) is more general than equation (\ref{EQ:localquadratic}), as it can also be generated in a number of models that do not involve the latter \citep[see footnote 34 in][for example]{2011ApJS..192...18K}. It is nonetheless customary to refer to this form of the bispectrum as the local template. }:

\begin{align}
B^{\mathrm{local}}_{\Phi}(k_1,k_2,k_3) = 2 \fNL^{\mathrm{local}} [ P_{\phi}(k_1)P_{\phi}(k_2) \nonumber \\ + P_{\phi}(k_1)P_{\phi}(k_3) + P_{\phi}(k_2)P_{\phi}(k_3) ],
\label{EQ:localquadraticbispectrum}
\end{align}
where $P_{\phi}(k)$ the power spectrum of the Gaussian potential, defined by $\langle\phi(\boldsymbol{k}) \phi(\boldsymbol{k}') \rangle \equiv (2 \pi)^3 P_{\phi}(k) \delta_D(\boldsymbol{k}+\boldsymbol{k}')$.  This local template is important in the phenomenology of PNG, since all single-field models of inflation with standard\footnote{By ``standard" assumptions, we mean that the curvature perturbation is initially in the Bunch-Davies state, and that there is no super-horizon evolution of the curvature perturbation from a non-attractor solution.  See e.g. \citet{2011PhRvD..83f3526A}, \citet{2011PhRvD..84f3514G}, \citet{2012PhRvD..86b3518G}, \citet{2012JCAP...09..007A}, \citet{2012JCAP...04..039D}, \citet{2012JCAP...10..055D}, \citet{2012JCAP...12..012C}, \citet{2013EL....10139001N}, and \citet{2013arXiv1301.5699C} for models which relax these assumptions.} assumptions predict $\fNL^{\mathrm{local}} = (5/12)(1-n_s)\approx 0.016$, where $n_s=0.96$ is the spectral index of the primordial power spectrum \citep{2003NuPhB.667..119A,2003JHEP...05..013M,2004JCAP...10..006C,2005JCAP...06..003S,2007JCAP...01..002C,2008JCAP...02..021C}.  Being able to detect or limit $\fNL^{\mathrm{local}}$ at such a small level would therefore be an important test of single-field models.  On the other hand, other more-general models of inflation exist which predict larger $\fNL^{\mathrm{local}}$, so it is also possible to exclude or restrict such models by limiting $\fNL^{\mathrm{local}}$ (Henceforth, we shall remove the label ``local" from $\fNL$ and mean $\fNL^{\mathrm{local}}$ unless otherwise specified).  

Observational constraints have been placed on the local template by finding the range of allowed amplitudes, expressed in terms of $\fNL$.  For example, CMB anisotropy measurements by the \emph{Wilkinson Microwave Anisotropy Probe} nine-year data analysis (WMAP9) find a $95\%$ confidence limit of $-3 < \fNL < 77$ \citep{2012arXiv1212.5225B}.  Newer CMB anisotropy results based upon the first 15 months of data from the \emph{Planck} satellite mission are reported to yield $\fNL=2.7\pm5.8$ [$68\%$ confidence limit error bars, \citet{2013arXiv1303.5084P}].  This result from the highly sensitive all-sky CMB anisotropy experiment of \emph{Planck} may be close to the limit which can be achieved by measurement of the CMB bispectrum alone \citep{2001PhRvD..63f3002K,2004PhRvD..70h3005B}.  To confirm these limits and probe $\fNL$ further, it is necessary to consider other methods.    

Correlation statistics of the CMB temperature anisotropy directly probe the initial fluctuations while they are still in the linear regime.  However, PNG also has potentially observable effects on the large-scale structure that develops as the initial fluctuations grow to the highly non-linear point of forming halos \citep[see e.g.][and references therein]{2010CQGra..27l4011D}.  For example, since non-Gaussianity to first order adds a skewness to the distribution function of filtered initial density fluctuations, it can significantly impact the abundance of the rarest and most massive halos \citep[e.g.][]{Matarrese:2000pb,2001MNRAS.325..412V,2007MNRAS.382.1261G,Lo-Verde:2008rt,2009MNRAS.398..321G,2010A&amp;A...514A..46V,2011JCAP...08..003L,2010PhRvD..81b3006D}, or of low-density cosmic voids \citep{2009JCAP...01..010K,2011PhRvD..83b3521D}, both of which originate from the tails of the density distribution.  PNG can also leave a strong signature in the halo bias, i.e. the ratio in Fourier space of the fractional halo number overdensity to the fractional matter overdensity, by introducing a scale-dependence which originates from coupling between large and small scales modes in the initial non-Gaussian distribution \citep{2008PhRvD..77l3514D, 2008ApJ...677L..77M,2008PhRvD..78l3507A}.  For the local template in equation (\ref{EQ:localquadraticbispectrum}), this halo bias has been found to depart significantly from the Gaussian expectation by a correction term that approaches the form $\Delta b(k) \propto \fNL (b_{\mathrm{G}} - 1)  / k^2$ in the small-$k$ limit, where $b_{\mathrm{G}}$ is the expected Gaussian bias  \citep[for recent analytical derivations, see e.g.][]{2011PhRvD..84f3512D,2012JCAP...03..032S,2012PhRvD..86f3526A,2012arXiv1206.3305D,2012arXiv1210.2495Y}.  Based upon this assumed $k^{-2}$ scale-dependence of $b(k)$, \citet{2013arXiv1303.1349G} recently constrained $\fNL$ to be in the range $-37 < \fNL < 25$ ($95\%$ limit in their most conservative analysis) using both the large-scale clustering of massive galaxies and the integrated Sachs-Wolfe effect [see also \citet{2008JCAP...08..031S}, \citet{2010JCAP...08..013X}, \citet{2010ApJ...717L..17X}, \citet{2011JCAP...08..033X} and \citet{2013MNRAS.428.1116R} for previous constraints based on the large-scale clustering of galaxies].     

Fluctuations in the density of neutral and ionized hydrogen in the IGM during the EoR created by the energy released by the first stars and galaxies presents another opportunity to observe the difference between Gaussian and non-Gaussian initial conditions.  During this cosmological phase, which ended at $z>6$, before the universe was a billion years old, expanding ionized patches of the IGM were created wherever galactic halos formed to fuel their growth, so the large-scale structure of this patchiness was correlated with that of the halos.  The connection between patchiness and halo clustering introduces a new kind of bias, the \emph{ionized density bias}, i.e. ratio of the fractional overdensity of intergalactic H II to the fractional total matter overdensity, which can also reflect the difference caused by PNG.  The theory of this ionized density bias, however, is dependent not only on the clustering of halos, but on the further complications of galaxy formation and radiative transfer which determine: (i) How much and what kind of ionizing radiation is released by galactic halos of different masses, (ii) How the radiation ionizes the surrounding IGM, and (iii) How that process feeds back on the ability of galactic halos to form stars and release more ionizing radiation.  For galaxies to form stars, they must collapse and accrete the baryonic component of the IGM along with the dark matter that dominates the halo mass, and then they must make these baryons gravitationally unstable and self-gravitating within the halos, by radiatively cooling the gas below the halo virial temperature.  A general picture of how this process unfolded in the $\Lambda$CDM universe is the following.       

The first stellar sources of reionization likely formed through radiative cooling from collisional excitation of rotational-vibrational energy levels of H$_2$ molecules, within halos with masses between $M\sim10^{5}-10^{8}~\Msun$, and virial temperatures $\Tvir \lesssim 10^{4}$ K - the so-called minihalos.  However, this early period of star formation ($z\gtrsim 15$) is thought to have been quenched by a corresponding rise in the UV background, since the H$_2$ molecules needed for efficient cooling in minihalos were easily photo-disociated by UV photons in the Lyman-Werner bands of $H_2$ between 11.2 and 13.6 eV \citep{2000ApJ...534...11H}.  Halos in this mass range would also have been suppressed as sources of ionizing radiation if they formed in places where the IGM was already ionized.  The gas pressure of the ionized IGM, which is heated to $\sim10^4$K by photoionization, opposes baryonic gravitational collapse into minihalos, a phenomenon sometimes referred to as ``Jeans filtering" \citep{1994ApJ...427...25S}, and pre-existing minihalos would also lose their baryonic content to photoevaporation by the ionizing radiation inside IGM H II regions \citep{2004MNRAS.348..753S}.  

Halos with $\Tvir \gtrsim10^4$ K, on the other hand, could radiatively cool their gas through collisional excitation of atomic hydrogen.  These are the so-called atomic cooling halos (ACHs).   This minimum $\Tvir$ for ACHs corresponds to a minimum halo-mass scale of roughly $\mmin \sim 10^8\Msun$.  In fact, like minihalos, even ACHs with masses below the Jeans-filtering scale ($M\sim 10^9~\Msun$) may have been susceptible to negative feedback from IGM photo-heating if they formed within already ionized regions.  These low-mass ACHs (``LMACHS") may also not have been massive enough to overcome the IGM pressure forces which act to prevent the accretion of inter-galactic gas -- the fuel of star formation [see e.g. \citet{2007MNRAS.376..534I} and references therein].  The precise boundary between LMACHs and the high-mass ACHs (``HMACHs") which are massive enough to be unaffected by Jeans filtering is still uncertain.  Regardless of the astrophysical uncertainties in the theory, reionization was likely driven to completion by galaxies in halos that were massive enough to be rare and highly biased at the relevant redshifts ($z\gtrsim 6$).  It is therefore natural to expect characteristic differences between EoR models with Gaussian and non-Gaussian initial conditions, and for these differences to lead to new observational signatures of PNG.

There has already been some work on the theory of reionization in the context of PNG.  \citet{2009MNRAS.394..133C} calculated the effects of the modified ACH abundance due to PNG on the reionization history and the mean integrated electron scattering optical depth, $\tau_{es}$, of the IGM.  \citet{2012MNRAS.420..441T} investigated how the modified ACH abundance can impact the number count of ionized bubbles observed in future maps of the brightness temperature of the redshifted 21cm background from the EoR, which they suggest as a potential probe of $\fNL$.      \citet{2012MNRAS.426L..21C} found that the effects of PNG on nonlinear biasing of minihalos significantly increases the root-mean-square (RMS) of fluctuations in the 21cm brightness temperature during the EoR.  If minihalos are able to maintain their reservoirs of neutral hydrogen throughout the EoR, \citet{2012MNRAS.426L..21C} claim, their contribution to the 21cm RMS may allow detection of $\fNL=\mathcal{O}(1)$ by next-generation radio telescopes like the Square Kilometer Array (SKA).  The possibility to detect PNG by measuring the power spectrum of brightness temperature fluctuations in the 21cm background from the EoR was considered by \citet{2011PhRvL.107m1304J} (henceforth JDFKS).  They reported a scale-dependent signature in the 21cm power spectrum for PNG which is absent in the Gaussian case.  This scale-dependent signature is related to that known already for the underlying galactic halos, as expected since the latter are the sources of reionization. 

JDFKS modeled reionization by semi-numeral simulations using the SimFast21 code \citep{2010MNRAS.406.2421S}, modified to take account of local PNG.  This method builds upon the analytical approximation introduced by \citet{2004ApJ...613....1F} in which reionization is statistically assumed to occur in some region when it has collapsed a large enough fraction of its mass to ionize all the atoms in that region.  The collapsed fraction in the model of \citet{2004ApJ...613....1F} is computed analytically by an application of the extended Press-Schechter approximation, sometimes referred to as the excursion-set model (ESM) \citep{Bond:1991sf,1993MNRAS.262..627L}.  We shall refer to this analytical approximation, which applies the excursion-set formalism to model reionization statistically, as the excursion-set model of reionization (ESMR).  The semi-numerical SimFast21 simulations go beyond the statistical approach of the ESMR by creating a 3D realization of the initial density fluctuations on a cubic mesh in a finite comoving volume prior to reionization, and extrapolating them forward in time by linear theory (the Zel'dovich approximation), to produce an evolving 3D map of the ionization field over time.      

JDFKS defined a statistical quantity they called the ``bias of ionized regions," which could be used to predict the 21cm power spectrum in the small-$k$ limit for a given reionization model in terms of the power spectrum of the underlying matter density fluctuations.  In that limit, the matter-density fluctuations are linear and the effect of PNG allowed by existing constraints is negligible.  The JDFKS bias quantity is $\hat{b}_x(k) = (1/\bar{x}_\mathrm{H})\sqrt{P_{xx}(k) / P_{\da \da}(k)}$, where $\bar{x}_\mathrm{H}$ is the mean neutral fraction, and $P_{x x}(k)$ and $P_{\delta \delta}(k)$ are the power spectra of the ionized fraction and matter density fluctuation fields respectively, i.e. $\langle \tilde{x}(\boldsymbol{k}) \tilde{x}(\boldsymbol{k'})\rangle = (2 \pi)^3 P_{x x} (k) \delta_D(\boldsymbol{k} + \boldsymbol{k'})$, and $\langle \tilde{\da} (\boldsymbol{k}) \tilde{\da}(\boldsymbol{k'})\rangle = (2 \pi)^3 P_{\delta \delta} (k) \delta_D(\boldsymbol{k} + \boldsymbol{k'})$, where $\delta_D$ is the Dirac Delta function.  Using their simulated maps of the ionization field for cases with and without local PNG, JDFKS found that $\hat{b}_x$ has a strong $k$-dependence in the non-Gaussian simulations, while $\hat{b}_x$ is $k$-independent in the Gaussian simulations on large enough scales.  Much like prior studies of the halo bias, JDFKS found they could fit $\hat{b}^{\mathrm{NG}}_x(k)$ from a given non-Gaussian simulation, as a function of $\hat{b}^{\mathrm{G}}_x$ from the corresponding Gaussian simulation, using a simple fitting formula in which $[\hat{b}^{\mathrm{NG}}_x(k)-\hat{b}^{\mathrm{G}}_x]$ scales as $k^{-2}$ in the small-$k$ limit.  Since this scale-dependent correction for PNG was found to be proportional to $\fNL$ as well, a measurement of the 21cm power spectrum at small $k$ might enable a determination of the value of $\fNL$.  Indeed, JDFKS used their fitting formula to predict that the SKA and Murchison Widefield Array (MWA) surveys could detect $\fNL \sim 50$ and $\sim100$ respectively.  \citet{2012arXiv1205.0563T} subsequently used the JDFKS fitting formula again to explore the constraining power of the cross-correlation between CMB temperature anisotropies and 21cm fluctuations on $\fNL$. 

We are interested here in the general problem of predicting the signatures of PNG in the observable properties of the EoR.  For this purpose, a more fundamental quantity is the \emph{ionized density bias}, the ratio of the ionized atomic to total matter overdensities in Fourier space, $\bHII(k)\equiv \tilde{\da}_{\rhoHII}(k)/\tilde{\da}(k)$, where $\tilde{\da}_{\rhoHII}(k)$ is the Fourier transform of the contrast in the ionized hydrogen mass density, $\da_{\rhoHII}\equiv \rhoHII/\bar{\rho}_{\mathrm{HII}}-1$.  Our goal here is to derive this ionized density bias, $\bHII^{\mathrm{NG}}$, for reionization models with PNG and relate it to the corresponding quantity for the Gaussian case, $\bHII^{\mathrm{G}}$.  Along the way, we will also derive the related bias parameter $\hat{b}_x$ defined by JDFKS for which they have reported the fitting formula described above.  As we shall discuss, we find significant differences between our derived expression and their fitting formula.

The methodology of our work is as follows:  (i) We extend the analytical excursion-set model of reionization (ESMR) of \citet{2004ApJ...613....1F} to include non-Gaussian initial conditions with general bispectra.  This extension is made possible by the derivation of the non-Gaussian collapsed fraction in \citet{2012arXiv1206.3305D} [see also \citet{2012PhRvD..86f3526A} for a similar but independent calculation].   (ii)  We use our extension of the ESMR to derive, for the first time, expressions for the scale-independent and -dependent contributions to the non-Gaussian ionized density bias, which apply to PNG with general bispectra.  (iii)  We test our derived expressions against more fundamental numerical calculations of the ionized density bias using the linear perturbation theory of reionization (LPTR) developed by \citet{2007MNRAS.375..324Z}, in which the linearized ionization rate and radiative transfer equations are solved in the long-wavelength limit for the perturbations to the ionized density field.  This approach is ideally suited to the problem at hand since it directly employs the relevant physics equations governing the ionization state of the IGM and is expected to be most accurate on the large scales of interest \citep{2007MNRAS.375..324Z}.  The LPTR provides a powerful and computationally cheaper alternative to fully non-linear cosmological radiative transfer simulations, which at the present day are prohibitively expensive in this context, since they would have to both resolve scales down to the smallest galactic halos responsible for reionization, $\emph{and}$ be large enough, approaching $\sim1$ Gpc$^3$ in volume, to capture the mode-coupling effects of PNG responsible for the scale-dependent halo bias.  (iv)  Finally, we derive the JDFKS bias parameter from first principles using our non-Gaussian extension of the ESMR.  The significant differences we find between our result and the JDFKS fitting formula, which we also confirm using the LPTR, will lead to our revision of their forecasted 21cm power spectrum constraints on $\fNL$ --  the topic of a follow-up paper to this work \citep{2013arXiv1305.0313M}.         

The remainder of this paper is organized as follows.  In \S \ref{SEC:ESMR}, we present basic definitions, our non-Gaussian extension of the analytical ESMR, and our expressions for the scale-independent and -dependent non-Gaussian ionized denisty bias.  In \S \ref{SEC:LPTR}, we summarize the LPTR formalism.  In \S \ref{SEC:results} we present numerical results from both the ESMR and the LPTR as well as comparisons between these two approaches.  In \S \ref{SEC:JDFKS} we derive our ESMR prediction for the JDFKS $\hat{b}_x$ parameter, and discuss our differences from their fitting formula.  Finally, we offer concluding remarks in \S \ref{SEC:conclusion}. 

In the plots presented below, we use a fiducial $\Lambda$CDM cosmology with parameters $\Omega_m=0.28$, $\Omega_{\Lambda} = 0.72$, $\Omega_b = 0.046$, $H_0 = 100 h~\mathrm{km~s^{-1}~Mpc^{-1}}$ (with $h=0.7$), $n_s=0.96$ and $\sigma_8=0.82$, consistent with WMAP7 constraints \citep{2011ApJS..192...18K}.  We also employ the linear matter power spectrum of \citet{1999ApJ...511....5E}.  Although the expressions we derive apply to PNG with general bispectra, we use the above local template to generate all plots in the remainder of this paper.  All distances are reported in comoving units unless otherwise stated.

\section{Analytical model of the ionized density bias}
\label{SEC:ESMR}

\subsection{Basic definitions}
Let $\rho_m(\boldsymbol{r},z)$ be the total matter density at location $\boldsymbol{r}$ and redshift $z$.  We define the matter density contrast, 

\begin{equation}
\delta(\boldsymbol{r},z) \equiv \frac{\rho_m(\boldsymbol{r},z)}{\bar{\rho}_m} -1,
\end{equation}  
where $\bar{\rho}_m$ denotes the mean matter density.  

Similarly, let $\rhoHI(\boldsymbol{r},z)$ and $\rhoHII(\boldsymbol{r},z)$ denote the mass densities of neutral and ionized hydrogen respectively, and let $\rhoH(\boldsymbol{r},z)=\rhoHI(\boldsymbol{r},z) + \rhoHII(\boldsymbol{r},z)$ denote the total hydrogen mass density.  We will from here on refer to $\rhoHII(\boldsymbol{r},z)$ as the \emph{ionized density}.  We define the ionized density contrast as

\begin{equation}
\drhoHII(\boldsymbol{r},z)\equiv \frac{\rhoHII(\boldsymbol{r},z)}{\bar{\rho}_{\mathrm{HII}}(z)} - 1,
\end{equation}
where $\bar{\rho}_{\mathrm{HII}}(z)$ is the mean ionized density at redshift $z$.  The central quantity that we will use in this work to quantify the impact of PNG on the large-scale structure of reionization is the \emph{ionized density bias},

\begin{equation}
\bHII(k,z) \equiv \frac{\tilde{\da}_{\rhoHII}(k,z)}{\tilde{\da}(k,z)},
\end{equation}
where $\tilde{\da}(k,z)$ and $\tilde{\da}_{\rhoHII}(k,z)$ are the Fourier transforms of the matter and ionized density contrasts respectively.  We now discuss how the ionized density bias is related to other quantities sometimes used in the literature.    

Let $x_i(\boldsymbol{r},z) \equiv \rhoHII(\boldsymbol{r},z)/\rhoH(\boldsymbol{r},z)$ denote the ionized fraction, and let $\delta_x(\boldsymbol{r},z)$ denote the ionized fraction contrast,

\begin{equation}
\delta_x(\boldsymbol{r},z)= \frac{x_i(\boldsymbol{r},z)}{\bar{x}_i(z)} - 1,
\label{EQ:dx}
\end{equation}
where $\bar{x}_i(z)$ is the spatially averaged ionized fraction at redshift $z$.  On large scales ($k \ll 1$), the ionized density bias is related to the \emph{ionized fraction bias}, 

\begin{equation}
b_x(k,z)\equiv \frac{\tilde{\da}_x(k,z)}{\tilde{\da}(k,z)},
\label{EQ:bxdef}
\end{equation}
through the definition of the ionized fraction smoothed on scale $R$, $x_i(R,\boldsymbol{r},z) \equiv \int \dd^3 \boldsymbol{r'} W(|\boldsymbol{r'} - \boldsymbol{r}|,R)~\rho_{\mathrm{HII}}(\boldsymbol{r},z)/\rho_{\mathrm{H}}(\boldsymbol{r},z)$, where $W(|\boldsymbol{r'} - \boldsymbol{r}|,R)$ is a spherically symmetric filter with characteristic cut-off scale $R$.  If the smoothing scale is large, then the above equation for $x_i(R,\boldsymbol{r},z)$ yields the leading-order relation, $\tilde{\delta}_x(k,z) + \tilde{\delta}(k,z) = \tilde{\delta}_{\rho \mathrm{HII}}(k,z)$, in the limit of $k \ll 1$, where we have assumed that the total hydrogen density fluctuations faithfully trace dark matter fluctuations on large scales.  Hence, the ionized density bias is related to the ionized fraction bias for small $k$ through 

\begin{equation}
\label{EQ:bHII2bx}
b_x(k,z) = \bHII(k,z) - 1.
\end{equation}  
Equation (\ref{EQ:bHII2bx}) can be used to transform the results for $\bHII$ presented in this paper, which are always restricted to the small-$k$ regime, to $b_x$ as needed.  

Consider a spherical region with radius $R$.  Let $\fcoll(\boldsymbol{r},\mmin,R,z)$ be the fraction of mass within that region contained in halos with masses above some threshold $\mmin$, i.e. the collapsed fraction.  At times, we will consider the contrast in the collapsed fraction, 

\begin{equation}
\dcoll(\boldsymbol{r},\mmin,R,z) \equiv \frac{\fcoll(\boldsymbol{r},\mmin,R,z)}{\bar{f}_{\mathrm{coll}}(\mmin,z)} - 1,
\end{equation} 
where $\bar{f}_{\mathrm{coll}}(\mmin,z)$ is the mean collapsed fraction at $z$, and a corresponding bias parameter,

\begin{equation}
\label{EQ:bfcolldef}
\bcoll(k,\mmin,z) \equiv \frac{\tilde{\da}_{\mathrm{fcoll}}(k,\mmin,z)}{\tilde{\da}(k,z)},
\end{equation} 
where the dependence on $R$ has been dropped under the assumption of $k\ll 2 \pi/R$.

\subsection{Generalizing the excursion-set model of reionization (ESMR) to include PNG}

In this section we extend the ESMR of \citet{2004ApJ...613....1F} for Gaussian initial conditions to include PNG with general bispectra, and use it to derive expressions for the Gaussian and non-Gaussian ionized density bias parameters.  These expressions will provide analytical expectations against which to compare the numerical LPTR results presented in \S \ref{SEC:results}.

\subsubsection{General Principles of the ESMR}

The basic postulate of the ESMR is that the local ionized fraction within a spherical volume with radius $R$ is proportional to the number of ionizing photons produced within that volume or, equivalently, the local collapsed fraction of mass in luminous sources,  

\begin{equation}
x_i(\mmin,R,z) = \zeta_{\mathrm{ESMR}}~\fcoll(\mmin,R,z).
\label{EQ:ESMR}
\end{equation}
Here, the $\zeta_{\mathrm{ESMR}}$ parameter accounts for the efficiency of this mass in releasing ionizing photons into the IGM (note that the condition of a fully ionized volume implies $\fcoll \geq \zeta_{\mathrm{ESMR}}^{-1}$).  Rather than try to compute this parameter from first principles, we will treat it as a free parameter.  When numerically evaluating our results in \S \ref{SEC:results}, we will fix the value of $\zeta_{\mathrm{ESMR}}$ by specifying the electron scattering optical depth for the model.  For simplicity, we assume that ACHs provide the only sources of ionizing radiation throughout reionization (see \S \ref{SEC:conclusion} for a discussion of the limitations of this assumption).  The minimum $\Tvir \gtrsim 10^4$ K criterion for ACHs roughly corresponds to a minimum halo mass of

\begin{align}
\label{EQ:Mmin}
\mmin ~\approx ~&1.3\times 10^{7}~\Msun \left( \frac{\Tvir}{10^4\mathrm{K}}\right)^{3/2} \left(\frac{1+z}{21} \right)^{-3/2} \nonumber \\ & \left( \frac{\Omega_m}{ 0.3}\right)^{-1/2} \left(\frac{h}{0.7} \right)^{-1} \left( \frac{\mumol}{1.22} \right)^{-3/2},
\end{align}
where $\mumol$ is the mean molecular weight of the gas \citep[see e.g.][]{2003ApJ...595....1H}.  
 
\subsubsection{Gaussian initial conditions}

Consider a spherical region with initial comoving radius $R$, containing mass $M\approx \bar{\rho}_m 4 \pi R^3 / 3$.  In models with Gaussian initial conditions, the excursion set expression for the fraction of mass within this comoving volume that will end up in ACHs at redshift $z$ is \footnote{Strictly speaking, equation (\ref{EQ:fcollgm}) applies in the case of Gaussian initial conditions \emph{and} a sharp $k$-space filtering function. Under these conditions, the stochastic evolution of excursion set trajectories is Markovian, and fully tractable analytically.  The use of the coordinate space top-hat filter necessitates more sophisticated analytical techniques resulting in correction terms to equation (\ref{EQ:fcollgm}).  For simplicity, we neglect these terms here.}

\begin{equation}
\fcollgm(\mmin,R,z) = \mathrm{erfc}\left[ \frac{\delta_c - \delta_R(z)}{ \sqrt{2 [\Smin(z) - S_R(z)]}} \right], 
\label{EQ:fcollgm}
\end{equation}
where $\delta_c\approx 1.686$ is the critical density in the spherical collapse model (in an Einstein-de Sitter universe),
\begin{equation}
 \da_R(z) = D(z) \int{\dd^3 \boldsymbol{r'}~W\left( |\boldsymbol{r}-\boldsymbol{r'} |,R\right) \da(\boldsymbol{r}',z=0)}
\label{EQ:smootheddelta}
\end{equation}
is the density contrast, smoothed on scale $R$ with the coordinate-space top-hat filter function, linearly extrapolated\footnote{In contrast to \cite{2012arXiv1206.3305D}, we do not adopt the usual convention in which the density field is linearly extrapolated all the way to the present day.  The convention we adopt in the current work, where the density field is linearly extrapolated to the epoch of interest, makes the redshift dependence in many of the summarized expressions more transparent.  } to the epoch $z$, with variance  

\begin{equation}
S_R(z) \equiv \sigma^2_R = \langle \da^2_R \rangle = \int \frac{\dd^3 \boldsymbol{k}}{(2 \pi)^3} \MM^2_R(k,z) P_{\Phi}(k),
\end{equation}
and $\Smin(z)$ is the variance of density fluctuations smoothed on scale $R_{\mathrm{min}} = (3 \mmin / 4 \pi \bar{\rho}_m)^{1/3}$.  Above, $D(z)$ is the linear growth factor normalized to unity at $z=0$, and we have written $S_R$ in terms of the power spectrum of the primordial potential fluctuation, $P_{\Phi}$, and the factor $\MM_R(k,z)$

\begin{equation}
\mathcal{M}_R(k,z) \equiv \frac{2}{3} \frac{k^2 \mathcal{T}(k) D(z)}{\Omega_m H_0^2 } g(0) \tilde{W}(k,R),
\end{equation}
between the potential and the smoothed density contrast in the cosmological Poisson equation in the synchronous-comoving gauge.  Here, $g(0)=(1+z_i)^{-1}D^{-1}(z_i)$, where $z_i$ corresponds to the initial epoch, i.e. the limit of large redshift, is the linear growth factor of the potential normalized to unity in the matter dominated epoch, evaluated at the present day [$g(0) \approx 0.76$ in our fiducial cosmology], and $\mathcal{T}(k)$ is the matter transfer function normalized to unity on large scales.

\subsubsection{Extension to non-Gaussian initial conditions}   

PNG complicates the analytical calculation of the collapsed fraction considerably because the additional correlations quantified in the three-point correlation function (and in general higher-order correlation functions) make the excursion set random walks non-Markovian.  \citet{2012arXiv1206.3305D} applied the non-Markovian extension\footnote{See also \citet{2012MNRAS.419..132P,2012MNRAS.420.1429P,2012MNRAS.420..369M,2012MNRAS.423L.102M,2012arXiv1205.3401M} for other works on non-Markovian extensions of the excursion-set formalism.} of the excursion set formalism of \citet{2010ApJ...711..907M,2010ApJ...717..515M,2010ApJ...717..526M} to calculate the collapsed fraction perturbatively [see also \citet{2012PhRvD..86f3526A} for a similar but independent calculation].  We adopt equation (33) of \citet{2012arXiv1206.3305D} for the collapsed fraction of ACHs,  

\begin{equation}
\fcoll^{\mathrm{NG}}(\mmin,R,z) = \fcollgm + \Delta \fcoll^{\mathrm{NG}},
\label{EQ:fcollPNGdef}
\end{equation}
where the non-Gaussian correction is\footnote{Here we use the leading-order non-Gaussian correction to the collapsed fraction of \citet{2012arXiv1206.3305D}.  They showed that this expression is consistent with the previously obtained non-Gaussian halo bias \citep{2011PhRvD..84f3512D}, which is in good agreement with results from N-body simulations at low redshifts \citep{2011PhRvD..84f1301D}.  \citet{2012arXiv1206.3305D} also obtained a next-to-leading order non-Gaussian correction which alters the collapsed fraction by a few percent.  For simplicity, we neglect this correction term here.} 

\begin{align}
\Delta f^{\mathrm{NG}}_{\mathrm{coll}}&(\mmin,R,z) = \frac{\Afd}{3}  \left(  \frac{\delta_c-\delta_R}{\Smin-S_R}   - \frac{1}{\da_c-\da_R} \right) \frac{\pd \fcollgm}{\pd \Smin} \nonumber \\ & + \frac{\Bfd}{S_R} \left[ \da_c - ( \da_c -\da_R) \coth \left( \frac{\da_c^2-\da_c \da_R }{ S_R} \right)  \right] \frac{\pd \fcollgm}{\pd \Smin}  \nonumber \\ & + \Cfd \cdot \frac{\Smin-S_R}{S_R^2 (\da_c - \da_R)} \biggl\{ \da_R^2- S_R \nonumber \\  &  - 2 (\da_c^2-\da_c\da_R) \left[\coth \left( \frac{ \da_c^2-\da_c \da_R }{ S_R} \right)  - 1 \right] \biggr\} \frac{\pd \fcollgm}{\pd \Smin},
\label{EQ:fcollPNGlong}
\end{align}
with

\begin{equation}
 \frac{\pd \fcollgm}{\pd \Smin} = \frac{(\da_c - \da_R)}{\sqrt{2 \pi} (\Smin - S_R)^{3/2}}  \exp\left[ -\frac{(\da_c - \da_R)^2}{2 (\Smin - S_R)} \right].
\end{equation}
The mode coupling effects of PNG between small and large scales are encoded in the functions,

\begin{align}
\label{EQ:Afunc}
& \mathcal{A} \equiv \langle \da^3_{\mathrm{min}} \rangle -  \langle \da^3_R \rangle + 3~\langle \da^2_R \da_{\mathrm{min}} \rangle - 3~\langle \da_R \da^2_{\mathrm{min}} \rangle \\ \nonumber \\
\label{EQ:Bfunc}
& \mathcal{B} \equiv  \langle \da^3_R \rangle +  \langle \da_R \da^2_{\mathrm{min}}\rangle - 2~\langle \da^2_R \da_{\mathrm{min}} \rangle  \\ \nonumber \\
\label{EQ:Cfunc}
& \mathcal{C} \equiv \langle \da^2_R \da_{\mathrm{min}} \rangle -  \langle \da^3_R \rangle,
\end{align}
(where $\damin \equiv \delta_{R =R_{\mathrm{min}}}$) containing the three-point correlation functions\footnote{It is technically the \emph{connected} two- and three-point correlation functions that appear in the above ESMR expressions.  However, since $\langle \delta \rangle =0$, the connected two- and three-point correlation functions are equal to the full two- and three-point correlation functions \citep[see e.g.][]{2002PhR...372....1C}.  We therefore drop the distinction here.} of the smoothed, linearly extrapolated density field,
\begin{align}
\langle \da_{R_1} \da_{R_2} \da_{R_3} \rangle  =  & \int \frac{\dd^3 \mathbf{k}_1}{(2 \pi)^3} \frac{\dd^3 \mathbf{k}_2}{(2 \pi)^3} B_{\Phi}(k_1,k_2,k_3) \\ & \MM_{R_1}(k_1,z) \MM_{R_2}(k_2,z) \MM_{R_3}(k_3,z), \nonumber 
\end{align}
where $k_3 = \sqrt{k_1^2+k_2^2+2 \boldsymbol{k_1}\cdot \boldsymbol{k_2}}$.  Our non-Gaussian extension of the ESMR is achieved by plugging equation (\ref{EQ:fcollPNGdef}) into equation (\ref{EQ:ESMR}).

\subsection{Ionized density bias from the ESMR}

 \subsubsection{Gaussian initial conditions}
 \label{SEC:ESMRGauss}
 
 We compute the ionized density bias in essentially the same way as \citet{2012arXiv1206.3305D} computed the halo bias\footnote{For Gaussian initial conditions, see also \citet{2005ApJ...630..643M} and \citet{2006ApJ...647..840A}.  \citet{2005ApJ...630..643M} calculated the bias of the overdensity of H II bubble counts with respect to the underlying matter density field in the standard ESMR approach.  \citet{2006ApJ...647..840A} calculated both the collapsed fraction bias and the ionized fraction bias -- the latter with their own model of reionization.}.  The first step is to write down the ionized fraction contrast, equation (\ref{EQ:dx}).  Assuming Gaussian initial conditions, we find    

\begin{equation}
\delta_x = \frac{\fcollgm}{\avgfcollgm} -1,
\end{equation}
where the subscript, $0$, in the denominator denotes the global mean collapsed fraction, i.e. evaluation in the limit of $\da_R\rightarrow 0$ and $S_R\rightarrow 0$.  We then plug in equation (\ref{EQ:fcollgm}) and Taylor expand $\delta_x$ in $\delta_R$ about $\delta_R = 0$ in the limit of $S_R \rightarrow 0$.   This procedure yields the leading-order relationship between the ionized fraction contrast and the initial matter contrast in the unevolved density field -- the so-called Lagrangian ionized fraction bias.  In order to relate this to the Eulerian bias defined in equation (\ref{EQ:bxdef}), which relates the ionized fraction contrast to the evolved matter density contrast, we adopt the spherical collapse model as in \citet{1996MNRAS.282..347M}.  In this model, the initial mass contained within the unevolved (i.e. Lagrangian) region is conserved, so the fraction of ionized mass in the evolved (i.e. Eulerian) region at $z$ is equal to the fraction of the initial mass that will end up ionized at $z$.  Moreover, in the large-scale limit, the evolved matter contrast is to first-order equal to the linearly extrapolated initial contrast.  The large-scale Lagrangian and Eulerian $b_x$ are therefore equivalent in this framework.  As a final step, we convert $b_x$ to $\bHII$ using equation (\ref{EQ:bHII2bx}) to obtain

\begin{equation}
\bHII^{\mathrm{G}}=1+ \frac{2} {\da_c} \frac{\pd \ln \avgfcollgm}{\pd \ln \Smin}.
\label{EQ:ESMRbHII}
\end{equation}
Hence, the ESMR predicts that the large-scale ionized density bias is scale-independent in models with Gaussian initial conditions, with an amplitude given by equation (\ref{EQ:ESMRbHII}).  

\subsubsection{Scale-dependent ionized density bias from non-Gaussian initial conditions}
\label{SEC:PNGsources}

Here we shall derive the ionized density bias $\bHII^{\mathrm{NG}}$ for non-Gaussian initial conditions.  For comparison with the Gaussian bias, we will write $\bHII^{\mathrm{NG}}$ as follows:

\begin{equation}
\bHII^{\mathrm{NG}} = \bHII^{\mathrm{G}} + \Delta \bHII^{(i)} + \Delta \bHII^{(d)}.
\label{EQ:bcfPNG_total}
\end{equation} 
where the labels ``i" and ``d" refer to scale-independent and scale-dependent corrections respectively.  Before proceeding we make some key simplifications.  Since our goal here is to compute the large-scale ionized density bias, where we will ultimately take the limit of $|\da_R| \ll 1$ and $S_R \ll 1$, we can recast equation (\ref{EQ:fcollPNGlong}) in the much simplified form,

\begin{align}
\label{EQ:fcollPNG}
& \Delta f^{\mathrm{NG}}_{\mathrm{coll}}  =  \frac{\langle \damin^3 \rangle - 3 \langle \damin^2 \da_R\rangle}{3} \\ & \times \left(  \frac{\delta_c-\delta_R}{\Smin-S_R}   - \frac{1}{\da_c-\da_R} \right) \frac{\pd \fcollgm}{\pd \Smin} \nonumber \\ & + \frac{\langle \damin^2 \da_R \rangle}{S_R} \left[ \da_c - ( \da_c -\da_R) \coth \left( \frac{\da_c^2-\da_c \da_R }{ S_R} \right)  \right] \frac{\pd \fcollgm}{\pd \Smin}.  \nonumber \end{align}
Additionally, we will require the mean non-Gaussian collapsed fraction, for which $\da_R \rightarrow 0$ and $S_R \rightarrow 0$.  In this limit, equations (\ref{EQ:fcollPNGdef}) and (\ref{EQ:fcollPNG}) simplify to

\begin{equation}
\label{EQ:avgfcollsum}
\avgfcoll^{\mathrm{NG}}=\avgfcollgm + \Delta \avgfcoll^{\mathrm{NG}}
\end{equation}
and

\begin{align}
\label{EQ:fcollPNGavg}
\Delta \avgfcoll^{\mathrm{NG}} =  \frac{\langle \damin^3 \rangle}{3}  \left(  \frac{\delta_c}{\Smin}   - \frac{1}{\da_c} \right) \frac{\pd \avgfcollgm}{\pd \Smin},
\end{align}
respectively.  

As before, we write down the ionized fraction contrast, 

\begin{equation}
\delta_x= \left(\frac{\fcollgm}{\avgfcollgm} - 1 \right) - \frac{\fcollgm~\Delta \avgfcoll^{\mathrm{NG}} }{\left(\avgfcollgm \right)^{2} } + \frac{\Delta \fcoll^{\mathrm{NG}}}{\avgfcollgm}.
\label{EQ:deltax}
\end{equation}
Like the analogous expression for the halo bias [see equation (57) of \citet{2012arXiv1206.3305D}], this equation contains both scale-independent and -dependent contributions to the ionized density bias.  We give technical details of our derivation in the appendix and summarize the main results here. The first term on the right-hand side of equation (\ref{EQ:deltax}) gives the Gaussian term from the last section, equation (\ref{EQ:ESMRbHII}).  The terms from equation (\ref{EQ:fcollPNG}) with $\langle \damin^3 \rangle$ contribute a scale-independent correction from PNG to the bias, whereas one of the terms with $\langle \damin^2 \da_R \rangle$ yields a non-zero scale-dependent correction.  The scale-independent correction is given by

\begin{align} 
 \Delta \bHII^{(i)} =      -\frac{\mathcal{S}^{(3)}_{\mathrm{min}} }{6} \Smin \left( \bHII^{\mathrm{G}} -1 \right) \biggl[ \frac{3\da_c \Smin - \da_c^3}{\Smin^2} \nonumber \\   +  \left(\bHII^{\mathrm{G}} -1 \right) \left( \frac{\da_c^2}{\Smin}  -1 \right) \biggr], 
 \label{EQ:bcfPNG_SI}
\end{align}
where $\mathcal{S}^{(3)}_{\mathrm{min}} \equiv \langle \damin^3 \rangle / \Smin^2$ denotes the skewness of density fluctuations smoothed on the $\mmin$ scale.  The scale-dependent correction is given by

\begin{equation}
\Delta \bHII^{(d)}(k) = 2 \delta_c \left( \bHII^{\mathrm{G}} -1 \right) \frac{\mathcal{F}^{(3)}_{\mathrm{min}}(k)}{\MM_{\mathrm{min}}(k)},
\label{EQ:SDcorrection}
\end{equation}
where the form factor is 

\begin{align}
\label{EQ:F3formfactor}
\mathcal{F}^{(3)}_{R}(k)=  & \frac{1}{ 4 S_{R} P_{\phi}(k)}   \int \frac{\dd^3 \mathbf{k}_1}{(2 \pi)^3} \\ & \times  \MM_{R}(k_1,z) \MM_{R}(q,z)  B_{\Phi}(k,k_1,q), \nonumber
\end{align}
with $q=\sqrt{k^2+k_1^2+2\boldsymbol{k}\cdot \boldsymbol{k_1}}$ (as before, we use the shorthand notation $\mathcal{F}^{(3)}_{\mathrm{min}}\equiv \mathcal{F}^{(3)}_{R=R_{\mathrm{min}}}$ and $\MM_{\mathrm{min}}\equiv \MM_{R=R_{\mathrm{min}}}$).  The scale-dependence in equation (\ref{EQ:SDcorrection}) is encapsulated in the factor $\mathcal{F}^{(3)}_{\mathrm{min}}(k)/ \MM_{\mathrm{min}}(k)$, which applies to PNG with general bispectra.  We note that a useful simplification of equation (\ref{EQ:SDcorrection}) can be made in the $k \lesssim 0.1~\Mpc^{-1}$ regime of the local template, where $\mathcal{F}^{(3)}_{\mathrm{min}}(k)$ is approximately\footnote{The mass thresholds corresponding to $T_{\mathrm{vir}}\sim10^{4}\mathrm{K}$ are around $\mmin=10^{8}~\Msun$.  For the corresponding smoothing scales, $\mathcal{F}^{(3)}_{\mathrm{min}}(k)$ deviates from its asymptotic value of $\fNL$ by a maximum of a few percent at $k\sim 0.1~\Mpc^{-1}$.} constant and equal to $\fNL$, and the smoothing kernel, $\tilde{W}(k,R)$, in $\mathcal{M}_{\mathrm{min}}(k)$ can be set to unity.  In this case, the scale-dependent ionized density bias takes the simplified form\footnote{As we discuss in \S \ref{SEC:JDFKS} below, the ionized density bias in equations (\ref{EQ:bcfPNG_total}), (\ref{EQ:bcfPNG_SI}), and (\ref{EQ:HybridSD})  is \emph{not} equivalent to the ``bias of ionized regions" defined by JDFKS, and therefore equation (\ref{EQ:HybridSD}) should not be compared directly to their equation (4).  In fact, we will show in \S \ref{SEC:JDFKS} that equation (\ref{EQ:HybridSD}) implies a very different expression for that quantity. },

\begin{equation}
\Delta \bHII^{(d)}(k,z) = 3 \fNL  \left[ \bHII^{\mathrm{G}}(z) - 1 \right]  \frac{ \da_c \Omega_m H_0^2}{ g(0) D(z) k^2 \mathcal{T}(k)}, 
\label{EQ:HybridSD}
\end{equation}
which makes explicit the $1/k^2$ scaling in the limit of $k\ll1$ for the local template. 

\section{Linear perturbation theory of reionization (LPTR)}
\label{SEC:LPTR}

As we have shown in the last section, the ESMR predicts that the large-scale ionized density bias is scale-independent in Gaussian models, and acquires both scale-independent and -dependent corrections in the case with PNG, in a manner analogous to the halo bias.  In $\S$ \ref{SEC:results}, we will check these predictions against a more general method, the LPTR of \citet{2007MNRAS.375..324Z}, which starts with the fundamental equations governing the ionization state of the IGM -- the ionization rate and radiative transfer equations.  Here we summarize the formalism of \citet{2007MNRAS.375..324Z} and our extension of it to include PNG, and in the next section present numerical results from both the ESMR and LPTR.         

\subsection{Fundamental equations}

\ctable[ caption = LPTR notation conventions ]{c  c  c }{

\label{TAB:LPTRnotation}

}{
\hline\hline
 \rule{0pt}{3ex }Quantity &  \citet{2007MNRAS.375..324Z} & This paper  \vspace{0.1cm}   \\ \hline
\\ [-1ex]
Coordinate vector & $\boldsymbol{x}$ &$\boldsymbol{r}$ \vspace{0.1cm}\\   
 Conformal time & $\tau$ & $\eta$  \vspace{0.1cm} \\
$\ln \nu-\ln \nu_0$ & $\mu$ & $x_\nu$ \vspace{0.1cm} \\
 Source emissivity & $S$ & $j$ \vspace{0.1cm} \\
 Secondary ionization & $\kappa$ & $\betaSI$  \\ 
 boost factor & & \vspace{0.1cm} \\
Mean photon density & $f_\gamma$ & $\bar{\xi}_\gamma$   \\
in units of $\bnH$ & &  \vspace{0.1cm} \\ 
Mean source emissivity & $f_s$ & $\bar{\xi}_s$ \\
in units of $\bnH$ & &  \vspace{0.1cm} \\
$\ln a$ & $\omega$ & $y$ \vspace{0.1cm} \\
Source spectrum & $\beta$ & $s$  \\
power-law index & & \\
[1.5ex]
\hline
}

The main physical quantities which appear in the LPTR formalism are as follows\footnote{Our notation is different from the original notation of \citet{2007MNRAS.375..324Z}.  See Table \ref{TAB:LPTRnotation} for the conversions.  These conversions are intended to bring the notation of the LPTR equations into closer correspondence with standard notation in the astrophysical literature on radiative transfer and photoionized nebulae.}:  

\begin{itemize}

\item{ The total comoving number density of hydrogen atoms (ionized and neutral) and of ionized hydrogen, $\nH=\nH(\boldsymbol{r},\eta)$ and $\nHII = \nHII(\boldsymbol{r},\eta)$ respectively.  Here, $\boldsymbol{r}$ is the comoving spatial coordinate vector and $\eta$ is the conformal time.}
\item{$\ngam= \ngam(\boldsymbol{r},\eta,x_\nu,\Omega)$, the comoving number density of photons per unit solid angle, $\dd^2 \boldsymbol{\Omega}$, around the propagation direction $\boldsymbol{\Omega}$, per unit frequency parameter, $x_\nu\equiv \ln \nu - \ln \nu_0$, where $\nu$ is the photon frequency, and $\nu_0=13.6\mathrm{eV}/(2 \pi \hbar)$ is the ionization threshold of hydrogen.  }
\item{The differential emissivity of the ionizing photon sources:  the number of photons emmited per unit comoving volume, per unit conformal time, per unit $x_\nu$, per unit solid angle.  Following \citet{2007MNRAS.375..324Z}, we write the emissivity as $j/4\pi$, where $j=j(\boldsymbol{r},\eta,x_\nu,\boldsymbol{\Omega})$.    }

\end{itemize} 
The local ionization state of the IGM is governed by the ionization rate equation, a continuity equation which takes into account photoionization and local recombination rates:

\begin{align}
\frac{\pd \nHII}{\pd \eta} & + \nabla \cdot (\nHII \boldsymbol{u}) \nonumber \\
= & \left( \nH - \nHII \right) \int_0^{\infty} \dd x_\nu \int \dd^2 \boldsymbol{\Omega} ~ \ngam \frac{\sigma(x_\nu)}{a^2(\eta)} \betaSI(x_\nu,x_i) \nonumber \\  &- \frac{\alphaB \nHII^2}{a^2(\eta)},
\label{EQ:ionizationbalance}
\end{align}
where $\boldsymbol{u}$ is the comoving velocity of a volume-element of ionized hydrogen, $\sigma(x_\nu)$ is the photoionization cross section, $\alpha_B=2.6\times 10^{-13}\mathrm{cm}^3 /\mathrm{s}$ is the case-B recombination coefficient at an IGM temperature $T=10^{4} \mathrm{K}$, and $a(\eta)$ is the cosmological scale factor.  In the photoionization term of equation (\ref{EQ:ionizationbalance}), the factor $\betaSI(x_\nu,x_i)$ accounts for secondary ionizations due to energetic free electrons produced when X-ray photons ionize hydrogen.  In what follows, we adopt a soft source spectrum which is dominated by UV photons [see text surrounding equation (\ref{EQ:spectrum})], such that secondary ionizations are negligible, and $\betaSI$ may be set to unity.  

The radiation field that drives the photoionization term in equation (\ref{EQ:ionizationbalance}) evolves according to the radiative transfer equation, 

\begin{align}
\frac{\pd \ngam}{\pd \eta} + \boldsymbol{\Omega} \cdot \nabla \ngam - H(\eta) a(\eta) \frac{\pd \ngam}{ \pd x_\nu} \nonumber \\ =
\frac{j}{4 \pi} - \left( \nH - \nHII \right) \ngam \frac{\sigma(x_\nu)}{a^2(\eta)},
\label{EQ:radiativetransfer}
\end{align}
where $H(\eta)$ is the Hubble parameter.  As in \citet{2007MNRAS.375..324Z}, we write the main physical quantities in equations (\ref{EQ:ionizationbalance}) and (\ref{EQ:radiativetransfer}) in terms of spatial averages and linear perturbations in the following way:

\begin{align}
\nHII  & = \bnH \left[ \bar{x}_i(\eta) + \DHII(\boldsymbol{r},\eta) \right] \nonumber \\ \nonumber \\
\nH & =\bnH\left[ 1+\delta(\boldsymbol{r},\eta)\right] \nonumber \\  \nonumber \\
\ngam & = \bnH\left[ \bar{\xi}_{\gamma}(\eta,x_\nu) + \Delta_{\gamma}(\boldsymbol{r},\eta,x_\nu,\boldsymbol{\Omega}) \right] \nonumber \\  \nonumber \\
j&=\bnH \left[ \bar{\xi}_s(\eta,x_\nu) + \Delta_s(\boldsymbol{r},\eta,x_\nu,\boldsymbol{\Omega})\right].
\label{EQ:pertdefs}
\end{align}            
Before going into more detail about the equations governing the spatial averages and linear perturbations, we describe our models of the source emissivity.  

\subsection{The source emissivity}

To facilitate comparison with the analytical expectations derived from the ESMR, we will assume in our LPTR calculations that ACHs provide the only sources of ionizing radiation, so only halos above $\mmin$ given by equation (\ref{EQ:Mmin}) contribute to the source emissivity.  

In what follows, we consider two source models.  In the first model, henceforth referred to as ``source-model A," we assume that the number of photons produced in some time interval $\Delta \eta$ is proportional to the change in the number of hydrogen atoms in collapsed halos in that time interval.  In this model, photon-production is fueled only by \emph{newly collapsed hydrogen}, which may be a reasonable approximation if either internal or external feedback mechanisms quickly act to limit continuous star formation, and/or if the steep rise in abundance of ACHs results in the effective dominance of reionization by newly collapsed halos.  In the second model, ``source-model B," we assume that the \emph{rate} of photon production at a time $\eta$ is proportional to the number of collapsed hydrogen atoms at that time.  In this model, photon-production is continuously fueled by hydrogen once it collapses into halos.  

Our motivation for considering the two scenarios above is that source-model A is more similar to the ansatz adopted in the ESMR, where the number of ionizing photons produced in some region is assumed proportional to the number of collapsed baryons in that region.  On the other hand, radiative transfer simulations often assume that the photon production rate is proportional to the collapse baryon number, as in source-model B.  It is therefore useful to compare results using both assumptions.  Physically, these models represent two limiting cases of source lifetimes.  In source-model A the lifetimes are assumed to be much shorter than the duration of reionization, whereas in source-model B they are assumed to be much longer.  
  
Consider the emissivity $j$ smoothed in coordinate space over large scales using a smoothing kernel with characteristic scale $R$ (the Fourier space linear perturbation equations in \S \ref{SEC:linearperturbations} are rendered independent of $R$ in the large-scale limit where $k\ll 2\pi/R$).  If we assume that ACHs emit on average $\gamma^A(x_\nu)$ ionizing photons per unit $x_\nu$, per collapsed hydrogen atom, then we may write the smoothed emissivity in source-model A as

\begin{equation}
j^A_R(\mathbf{r},\eta, x_\nu) = \gamma^A(x_\nu)  \frac{\pd}{\pd \eta} \left[ n_H(\mathbf{r},\eta, R ) \fcoll(\mmin,R,\da_R,\eta) \right].
\label{EQ:emissivityA}
 \end{equation}
 In contrast, the smoothed emissivity in source-model B can be written as 
 
 \begin{equation}
j^B_R(\mathbf{r},\eta, x_\nu) = \gamma^B(x_\nu)   n_H(\mathbf{r},\eta ,R) \fcoll(\mmin,R,\da_R,\eta),
\label{EQ:emissivityB}
 \end{equation} 
 where $\gamma^B(x_\nu)$ is the number of photons emitted per unit $x_\nu$, per unit $\eta$, per collapsed hydrogen atom.  Following  \citet{2007MNRAS.375..324Z}, we parameterize the source spectrum with a power law in $\nu$, 
 
 \begin{equation}
 \gamma^{A,B}(x_\nu)~\dd x_\nu = -\zeta^{A,B}_{\mathrm{LPTR}}(1+ s) \exp\left[ \left( s+1\right) x_\nu \right] \dd x_\nu,
 \label{EQ:spectrum}
 \end{equation}
where $\zeta^A_{\mathrm{LPTR}}$ is the total number of photons per collapsed hydrogen atom, per unit $x_\nu$, and $\zeta^B_{\mathrm{LPTR}}$ is the total number of photons, per unit $\eta$, per collapsed hydrogen atom, per unit $x_\nu$.  The normalization of equation (\ref{EQ:spectrum}) has been chosen so that $\int_0^{\infty} \gamma^{A,B}(x_\nu)~\dd x_\nu = \zeta^{A,B}_{\mathrm{LPTR}}$ under the condition $s < -1$.  In our numerical results presented in \S \ref{SEC:results}, we fix $\zeta^{A,B}_{\mathrm{LPTR}}$ by fixing the value of the electron-scattering optical depth. The power law index, $s$, can be used to shift the spectrum towards the soft (UV) or hard (X-Ray) photons.  In this work, we restrict ourselves to a soft source spectrum, with $s = -3$ [see \citet{2007MNRAS.375..324Z} for LPTR calculations with a hard spectrum in models with Gaussian initial conditions].      
 
Finally, we note that the LPTR can be sourced by any model of the collapsed fraction, including those extracted from N-body simulations.  However, for the sake of comparison, we use excursion-set equations (\ref{EQ:fcollgm}) and (\ref{EQ:fcollPNGdef}) in the Gaussian and non-Gaussian models respectively, so the statistics of the halo sources in the ESMR and LPTR calculations are exactly the same.  However, we emphasize that the LPTR is otherwise independent of the ESMR in how it models reionization; the former incorporates large-scale physics of radiative transfer, photoionization, and recombinations in the IGM, while the latter follows from the ansatz in equation (\ref{EQ:ESMR}). 

\subsection{The equations of the spatial averages}
\label{SEC:thespatialaverages}

The spatial averages of equation (\ref{EQ:ionizationbalance}) and (\ref{EQ:radiativetransfer}) are

\begin{align}
\label{EQ:avgionizationbalance}
\frac{\pd \bar{x}_i }{\pd \eta} = 4 \pi (1- \bar{x}_i) \int \dd x_\nu \frac{\sigma \bnH}{a^2} \langle\betaSI \rangle \bar{\xi}_{\gamma} \clumpone \\ - \frac{\alphaB \bnH}{a^2} \bar{x}^2_i \clumpHII \nonumber
\end{align}
and 

\begin{equation}
\frac{\pd \bar{\xi}_{\gamma}}{\pd \eta} = \frac{\bar{\xi}_s}{4 \pi} + H a \frac{\pd \bar{\xi}_{\gamma}}{\pd x_\nu}  - \frac{\sigma \bnH}{a^2} \left( 1 - \bar{x}_i \right) \bar{\xi}_{\gamma} \clumptwo
\label{EQ:avgradiativetransfer}
\end{equation}
respectively, where the quantity $\clumpHII \equiv \langle \nHII^2 \rangle / \langle \nHII \rangle^2$ is the clumping factor for hydrogen recombination, and

\begin{equation}
\clumpone \equiv \frac{\langle \nHI n_{\gamma} \betaSI \rangle }{ \langle \nHI \rangle \langle n_{\gamma} \rangle \langle \betaSI \rangle}
\end{equation}
and
\begin{equation}
\clumptwo \equiv \frac{ \langle \nHI n_{\gamma} \rangle }{ \langle \nHI \rangle \langle n_{\gamma} \rangle }
\end{equation}
are the photoionization clumping factors.  In the case of a soft source spectrum, in which secondary ionizations are negligible, $\clumpone \approx \clumptwo$.  From here on we will drop the distinction and denote both photoionization factors with $\clumpGH$.  These clumping factors cannot be calculated analytically, since they are sensitive to non-linear density fluctuations on small scales, and detailed feedback effects of the radiation background.  We therefore employ a few simple models for the clumping factors, motivated by numerical results previously reported in the literature, which we describe in \S \ref{SEC:illustrativemodels}.

Note that the source emissivity appears in the above spatially averaged equations through the function $\bar{\xi}_s$.  In appendix \ref{APP:emissivity}, we show that $\bar{\xi}_s$ is given by

\begin{equation}
\bar{\xi}^{\mathrm{A}}_s = \gamma^A(x_\nu) \frac{\pd \avgfcoll }{ \pd \eta}
\label{EQ:fsA}
\end{equation}
in source-model A, and

\begin{equation}
\bar{\xi}^{\mathrm{B}}_s =  \gamma^B(x_\nu) \avgfcoll
\label{EQ:fsB}
\end{equation}
in source-model B.  For Gaussian initial conditions, we insert equation (\ref{EQ:fcollgm}) with $\da_R=0$ and $S_R=0$ into the above equations.  For non-Gaussian initial conditions, we use (\ref{EQ:avgfcollsum}) along with equation (\ref{EQ:fcollPNGavg}).

\subsection{The linear perturbation equations}
\label{SEC:linearperturbations}

Following \citet{2007MNRAS.375..324Z}, we substitute the conformal time with the variable $y\equiv \ln a(\eta)$, and define $\hat{\sigma}(\eta,x_\nu)\equiv \sigma(x_\nu) \bnH / (H a^3)$ and $\TalphaB \equiv \alphaB \bnH / (H a^3)$ for notational convenience.  Under these substitutions, $\hat{\sigma}$ represents the probability that a photon propagating through a neutral Universe with frequency parameter $x_\nu$ directly ionizes a single atom within a Hubble time.  Similarly, $\TalphaB$ is the average number of times a proton in a fully ionized universe recombines within a Hubble time.  We take the Fourier transforms of equation (\ref{EQ:ionizationbalance}) and (\ref{EQ:radiativetransfer}) and keep only terms that are first-order in the Fourier transforms of $\delta$, $\Delta_s$, $\DHII$, and $\Delta_{\gamma}$ (from here on denoted by $\Tdelta$, $\TDs$, $\TDHII$, and $\TDgam$ respectively).  Noting that the dark matter density contrast is to first order proportional to the growth factor, $D(\eta)$, and that the peculiar velocity is proportional to the gradient of the gravitational potential, we obtain 

\begin{equation}
\frac{\pd \TDHII}{\pd y} = F_2 \Tdelta - F_1 \TDHII + \int_0^{\infty} \dd x_\nu \langle \betaSI \rangle \int \dd^2 \boldsymbol{\Omega} \TDgam (1-\bar{x}_i) \hat{\sigma}
\label{EQ:FOionizationbalance}
\end{equation}
and

\begin{equation}
\frac{\pd \TDgam}{\pd y} = \frac{\pd \TDgam }{\pd x_\nu} - F_3 \TDgam + \frac{\TDs}{4 \pi H a} +  \hat{\sigma} \bar{\xi}_{\gamma} \left( \TDHII - \Tdelta\right),
\label{EQ:FOradiativetransfer}
\end{equation}
where we have defined the following auxiliary functions:

\begin{align}
F_1  =  & 2 \TalphaB \bar{x}_i \\  & + 4 \pi \int_0^{\infty} \dd x_\nu \hat{\sigma} \bar{\xi}_{\gamma} \left[ \langle \betaSI \rangle - (1-\bar{x}_i) \left. \frac{\pd \betaSI}{\pd \phi} \right|_{\phi=\bar{x}_i} \right], \nonumber
\end{align}

\begin{align}
F_2= & \frac{\dd \ln D}{ \dd y} \bar{x}_i \\ & + 4 \pi \int_0^{\infty}\dd x_\nu \hat{\sigma} \bar{\xi}_{\gamma} \left[ \langle \betaSI \rangle - (1 - \bar{x}_i) \bar{x}_i \left. \frac{\pd \betaSI}{\pd \phi} \right|_{\phi=\bar{x}_i} \right], \nonumber
\end{align}

\begin{equation}
F_3=(1-\bar{x}_i) \hat{\sigma} - \frac{i \boldsymbol{k} \cdot \boldsymbol{\Omega}}{H a},
\end{equation}

The source emissivity appears in the above equations through the linear source fluctuation, $\tilde{\Delta}_s(k,\eta)$.  In appendix \ref{APP:emissivity}, we show that

\begin{equation}
\tilde{\Delta}^{\mathrm{A}}_s(k,\eta) =  \frac{\tilde{\delta}(\boldsymbol{k},\eta)}{D(\eta)} \gamma^A(x_\nu) \frac{\pd}{\pd \eta} \left\{ D(\eta) \avgfcoll \left[ 1 + \bcoll(k,\eta) \right] \right\}.
\label{EQ:DsA}
\end{equation}  
in source-model A, and
  
 \begin{equation}
\tilde{\Delta}^{\mathrm{B}}_s(k,\eta) =  \tilde{\delta}(\boldsymbol{k},\eta) \gamma^B(x_\nu) \avgfcoll \left[ 1 + \bcoll(k,\eta) \right]
\label{EQ:DsB}
\end{equation}  
in source-model B, where the collapsed fraction bias, $\bcoll$, is obtained as described in the last paragraph of appendix \ref{APP:emissivity}.

 The procedure for numerically solving equations (\ref{EQ:FOionizationbalance}) and (\ref{EQ:FOradiativetransfer}) is detailed in \citet{2007MNRAS.375..324Z}.  First, equations (\ref{EQ:avgionizationbalance}) and (\ref{EQ:avgradiativetransfer}) are solved for the global reionization history, which serves as input for the linear perturbation equations.   In summary, the linear coordinate transformation in equation (12) of \citet{2007MNRAS.375..324Z} is used to change equation (\ref{EQ:avgradiativetransfer}) into a first-order ordinary differential equation.  The resulting equation can be solved for $\bar{\xi}_{\gamma}$, for some initial $\bar{x}_i$.  The solution for $\bar{\xi}_{\gamma}$ is then used to find a new solution of equation (\ref{EQ:avgionizationbalance}) for $\bar{x}_i$.  This procedure is repeated until $\bar{\xi}_{\gamma}$ and $\bar{x}_i$ converge, which tends to happen after about ten iterations.  A similar procedure is then applied to solve for the linear perturbations.  The coordinate transformation in equation (12) of \citet{2007MNRAS.375..324Z} is again applied to change equation (\ref{EQ:FOradiativetransfer}) into a first-order ordinary differential equation.  The resulting equation is solved for $\TDgam$, which is then integrated over $\boldsymbol{\Omega}$ to obtain the monopole perturbation of the radiation field, under the assumption that the monopole of $\TDs$ constitutes the main contribution.  The monopole perturbation of the radiation field is then used as input for equation (\ref{EQ:FOionizationbalance}), and the solutions are iterated until convergence is achieved.

\section{Results}
\label{SEC:results}

\subsection{Illustrative models of reionization}
\label{SEC:illustrativemodels}

\ctable[ caption = Illustrative models of reionization ]{c  c  c c c}{

\tnote[a]{Here, $\zeta_\mathrm{LPTR}^\mathrm{A}$ gives the number of photons per collapsed hydrogen atom, per unit $x_\nu$, while $\zeta_\mathrm{LPTR}^\mathrm{B}$ gives the number of photons per collapsed hydrogen atom, per $H_0^{-1}$, per unit $x_\nu$.}

\label{TAB:reionizationmodels}

}{
\hline\hline
 \rule{0pt}{3ex }Model & $\clumpHII$ & $\clumpGH$ & Emissivity & Efficiency \\ & & & & Parameter\tmark[a] \vspace{0.1cm} \\ \hline
\\ [-1ex]
 LPTR1 & Eq. \ref{EQ:clumpHII} & 1  & Eq. \ref{EQ:emissivityA} & $\zeta_{\mathrm{LPTR}}^{\mathrm{A}} = 70.3$ \vspace{0.1cm}\\   
 LPTR2 & 2 & 1  &  Eq. \ref{EQ:emissivityA} & $ \zeta_{\mathrm{LPTR}}^{\mathrm{A}}= 54.2$              
 \vspace{0.1cm} \\ 
  LPTR3 & 10 &  1 & Eq. \ref{EQ:emissivityA} & $\zeta_{\mathrm{LPTR}}^{\mathrm{A}} = 87.5$
 \vspace{0.1cm} \\
 LPTR4 & Eq. \ref{EQ:clumpHII} & 1  &  Eq. \ref{EQ:emissivityB} & $\zeta_{\mathrm{LPTR}}^{\mathrm{B}} = 11800$ \vspace{0.1cm} \\
 ESMR & N/A & N/A & N/A  & $\zeta_{\mathrm{ESMR}}=50.2$ \\
[1.5ex]
\hline
}

Here we describe the five reionization models which are evaluated in this section.  The parameters of these models are summarized in Table \ref{TAB:reionizationmodels}.  We consider four models using the LPTR, in which we must specify the recombination and photoionization clumping factors, in addition to the source prescription, i.e. source-model A [eq. (\ref{EQ:emissivityA})] or source-model B [eq. (\ref{EQ:emissivityB})].  In the first model, LPTR1, we use source-model A and $\clumpHII$ varies with redshift according to the fitting formula in \citet{2007MNRAS.376..534I},       

\begin{equation}
\clumpHII = 26.2917 \exp\left(-0.1822 ~ z + 0.003505 ~z^2\right),
\label{EQ:clumpHII}
\end{equation}
obtained from a high-resolution, small-scale N-body simulation, which resolved the mass scale of minihalos (for which, e.g., $\clumpHII= 3.8$, $6$, and $10$ at $z=15$, $10$, and $6$ respectively).  They computed the clumping factor from the density field of dark matter \emph{outside} of halos; a procedure which assumes as a first-order approximation that the ionized density fluctuations faithfully trace dark matter fluctuations in the IGM.  However, the above fit does not account for feedback effects, such as photoionization heating, which act to suppress the clumping by increasing the Jeans scale inside ionized regions.  

Most recently, \cite{2012arXiv1209.2489F} confirmed the suppression of the recombination clumping factor by photoionization heating \citep[also see][]{2009MNRAS.394.1812P,2011ApJ...743...82M,2012ApJ...747..100S}.  They found that $\clumpHII$ rises from $\sim1$ at $z=10$ to $\sim 3$ at $z=6$.  However, they did not resolve the mass scale of minihalos which dominates the small-scale structure responsible for the recombination clumping factor.  Nevertheless, as an opposite extreme, we adopt a constant $\clumpHII = 2$ in LPTR2, which may be more representative of the photo-heated IGM.  On the other hand, LPTR3, with $\clumpHII =10$, serves as an upper bracket to the range of recombination rates considered in this work.  The LPTR4 model is used to illustrate differences between sourcing photons with the differential and cumulative collapsed fraction [see text surrounding equations (\ref{EQ:emissivityA}) and (\ref{EQ:emissivityB})].  It has the same clumping factors as LPTR1, but uses source-model B.

The photoionization clumping factor required by these models is more uncertain.  \citet{2010ApJ...724..244A} find from small-scale reionization simulations that $\clumpGH=1$ should be a reasonable approximation to the actual value, which they find to be suppressed relative to $\clumpHII$ \citep[also see][]{2007ApJ...657...15K}.  For simplicity, we set $\clumpGH=1$ for all of the LPTR models in Table \ref{TAB:reionizationmodels}.  Finally, we also evaluate the ESMR.   

For a given reionization history, the electron scattering optical depth is  

\begin{equation}
\tau_{es} = \sigma_T \int_{z_{\mathrm{rec}}}^{0} \bar{n}_e(z) \frac{\dd t}{ \dd z} \dd z,
\label{EQ:tau_es}
\end{equation}
where $\sigma_T$ is the Thompson cross section, $\bar{n}_e(z)=\bar{x}_i(z) \bnH(z)$ is the cosmic-average free electron density, and $z_{\mathrm{rec}}$ is the redshift of the recombination epoch ($z_{\mathrm{rec}}\approx10^3$).  For all of the above models, we set the normalizations (i.e. by tuning efficiency parameters, either $\zeta_{\mathrm{LPTR}}$ or $\zeta_{\mathrm{ESMR}}$) by fixing $\tau_{es}=0.08$ \emph{in the case with Gaussian initial conditions}.  When we consider cases with non-zero $\fNL$, the normalization values in the last column of Table \ref{TAB:reionizationmodels} stay fixed, while the $\tau_{es}$ values vary fractionally by at most a couple of percent for the illustrative $\fNL = \pm 50$ models adopted below, as we show in the next section.

\subsection{Global reionization histories}

\begin{figure}
\begin{center}
\resizebox{8.5cm}{!}{\includegraphics{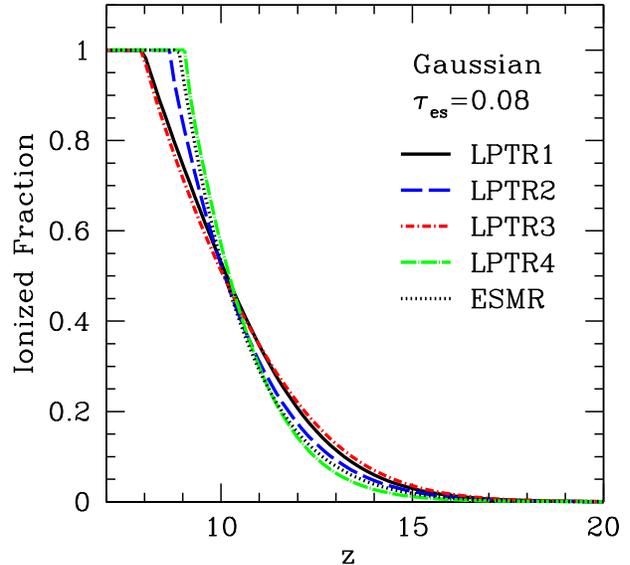}}
\end{center}
\caption{Global reionization histories in models with Gaussian initial conditions (see Table \ref{TAB:reionizationmodels} for model parameters).  The efficiency parameters in the LPTR models and the ESMR are fixed by setting the electron scattering optical depth to $\tau_{es}=0.08$.    }
\label{FIG:Greionhist}
 \end{figure}

  \begin{figure}
\begin{center}
\resizebox{8.5cm}{!}{\includegraphics{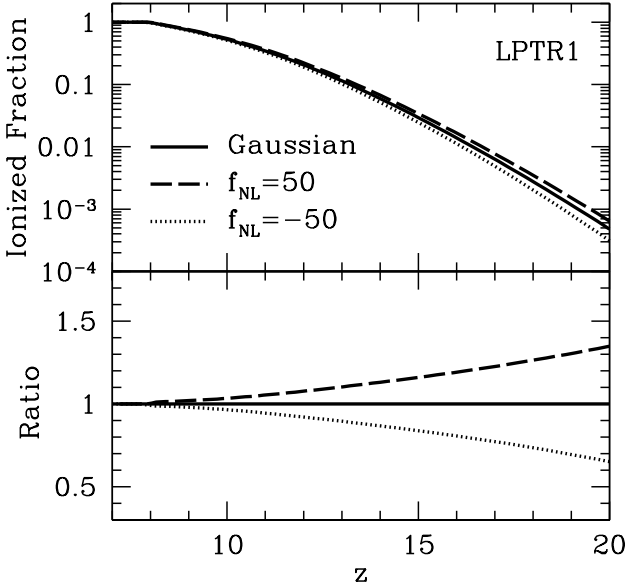}}
\end{center}
\caption{The effects of local PNG on the global reionization history in LPTR1.  Note that the source efficiency parameter is fixed to $\zeta_{\mathrm{LPTR}}^{\mathrm{A}}=70.3$ across all cases shown.  Top panel:  mean ionized fraction as a function of redshift.  Bottom panel:  ratio of non-Gaussian to Gaussian models.       }
\label{FIG:fNLreionhist}
 \end{figure}
 
  \begin{figure}
\begin{center}
\resizebox{8.8cm}{!}{\includegraphics{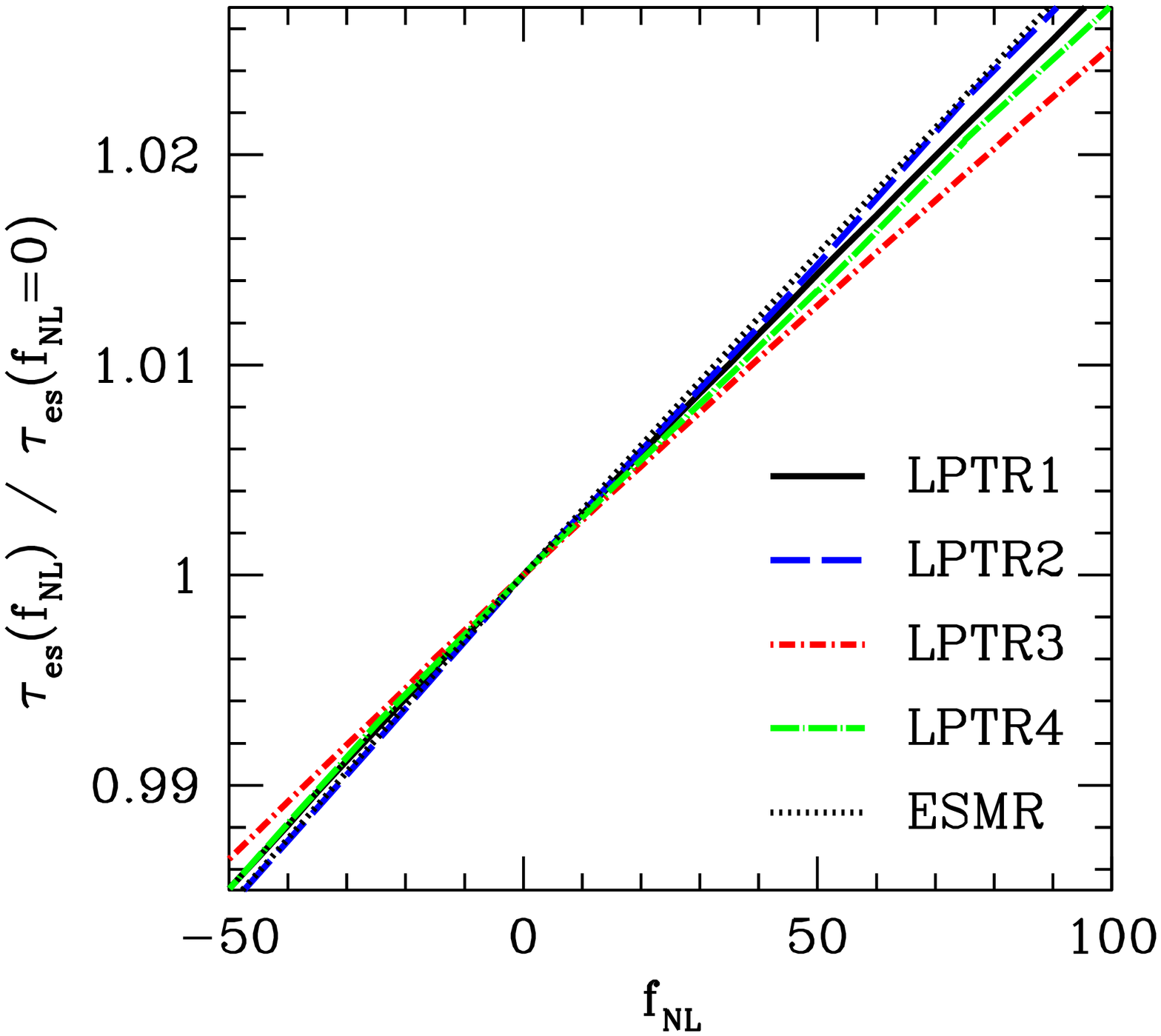}}
\end{center}
\caption{The effects of local PNG on the electron-scattering optical depth, $\tau_{es}$, for a range of $\fNL$ values.  We show the ratios of non-Gaussian to Gaussian models ($\tau_{es}$ is fixed to $0.08$ in the Gaussian models).   }
\label{FIG:fNLtau_es}
 \end{figure}

Before presenting our main results for the ionized density bias, it is instructive to consider the global reionization histories in the models of Table \ref{TAB:reionizationmodels}.  A reionization history is obtained in the LPTR by solving for the mean ionized fraction in equations (\ref{EQ:avgionizationbalance}) and (\ref{EQ:avgradiativetransfer}).  In Figure \ref{FIG:Greionhist}, we show the mean ionized fraction as a function of redshift for the Gaussian LPTR models of Table \ref{TAB:reionizationmodels}.  We also plot the mean ionized fraction from the ESMR, which is just given by $\bar{x}_i = \zeta_{\mathrm{ESMR}} \avgfcollgm$.  We note that reionization ends between $z\sim8-9$ in all of the models shown, which may be at odds which recent observations [see e.g. \citet{2013arXiv1301.1228R} and references therein].  We stress, however, that these models are employed below for illustrative purposes to show the effects of PNG on the large-scale structure of reionization.   
  
Figure \ref{FIG:fNLreionhist} illustrates how PNG affects the reionization history through its impact on the halo abundance.  Although we display only the LPTR1 case for clarity, the results using other models are quantitatively similar.  The top panel compares the mean ionized fraction in the Gaussian and non-Gaussian models -- the latter with $\fNL=\pm 50$ (we choose these $\fNL$ values for illustrative purposes).  The normalization of the source emissivity is $\zeta_{\mathrm{LPTR}}^{\mathrm{A}} =70.3$ in all three cases.   The bottom panel shows the ratio of the non-Gaussian to Gaussian $\bar{x}_i$.  The case with $\fNL=50(-50)$ has an ionized fraction that is $\sim16 \%$ higher(lower) than the Gaussian case at $z\sim15$, when the universe is only a few percent ionized.  However, the differences drop to $\sim 3(1)\%$ when the universe is about $50(95)\%$ ionized at $z\sim10(8.1)$.  Despite the differences early on, all three cases finish reionization at approximately the same redshift, $z\approx8$.  The largest effects of PNG in our models occur at higher redshifts during the early stages of reionization when, due to the imposed $\mmin$ for ACHs [equation (\ref{EQ:Mmin})], the collapsed fraction above $\mmin$ is dominated by halos that are rare at those redshifts.  For example, $\mmin = 4(5.9)(10)\times10^7~\Msun$ at $z=20(15)(10)$, corresponding to a peak height of $\nu \equiv \delta_c/\sqrt{\Smin} = 4.5(3.6)(2.6)$.       
 
  \begin{figure}
\begin{center}
\resizebox{8.7cm}{!}{\includegraphics{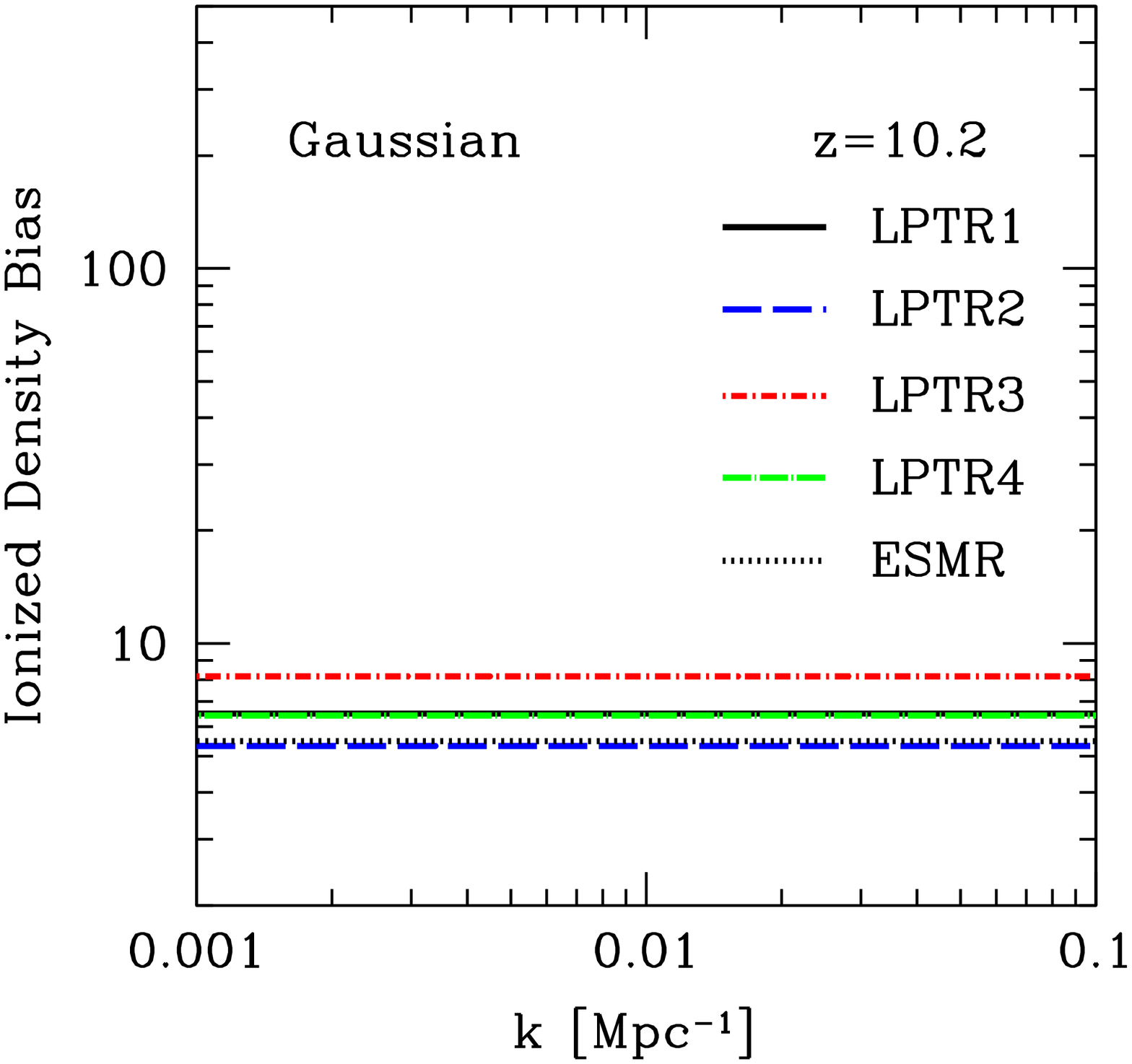}}
\resizebox{8.5cm}{!}{\includegraphics{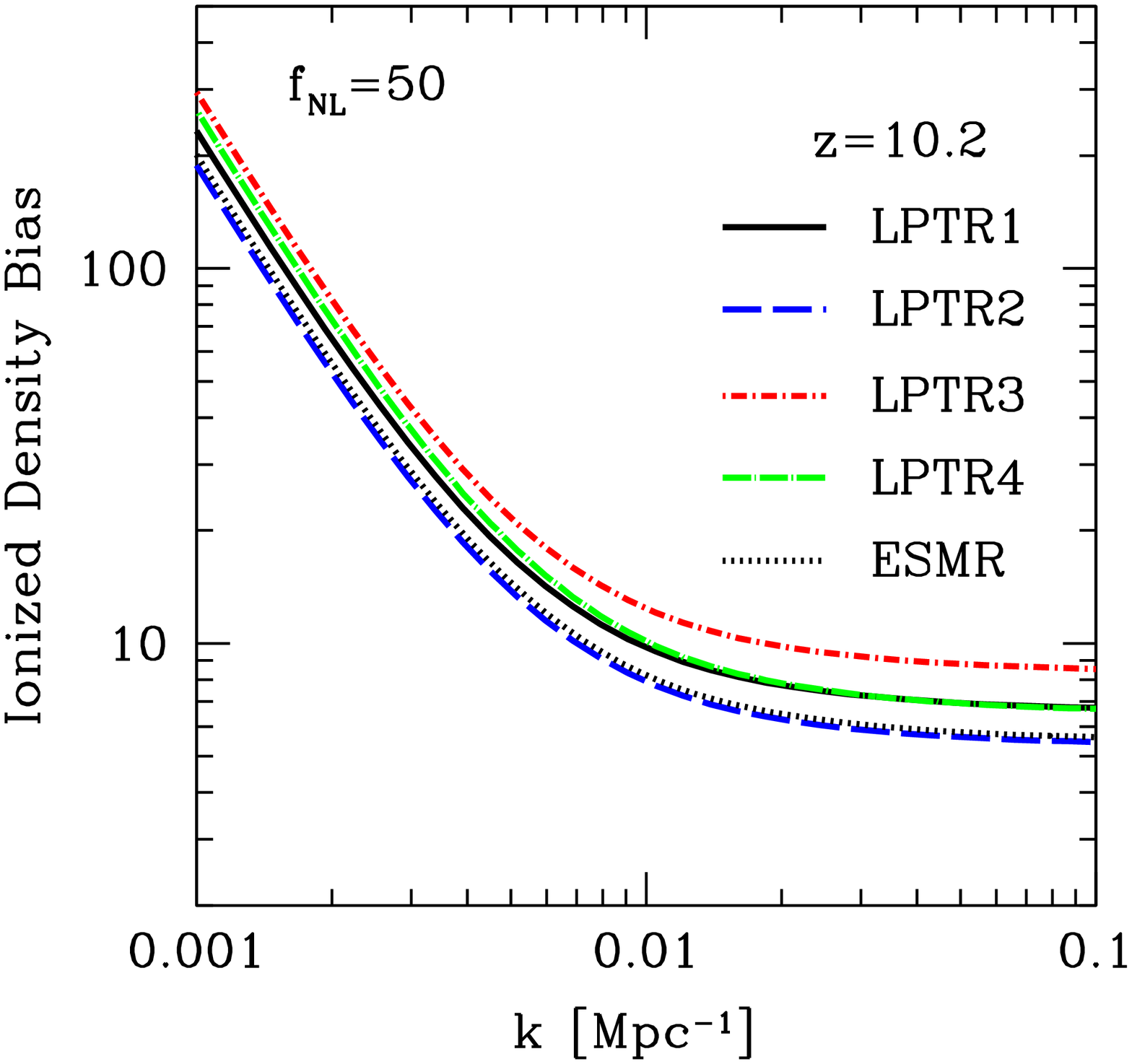}}
\end{center}
\caption{The scale-dependence of the ionized density bias.  Top panel: for models with Gaussian initial conditions.  Bottom panel:  the scale-dependent signature of local PNG with $\fNL=50$.  We show results at a fixed redshift of $z=10.2$, which corresponds to a global ionized fraction of $\approx 50\%$ in all models shown.  }
\label{FIG:HIIbiasCompareA}
 \end{figure}

 \begin{figure}
\begin{center}
\resizebox{8.5cm}{!}{\includegraphics{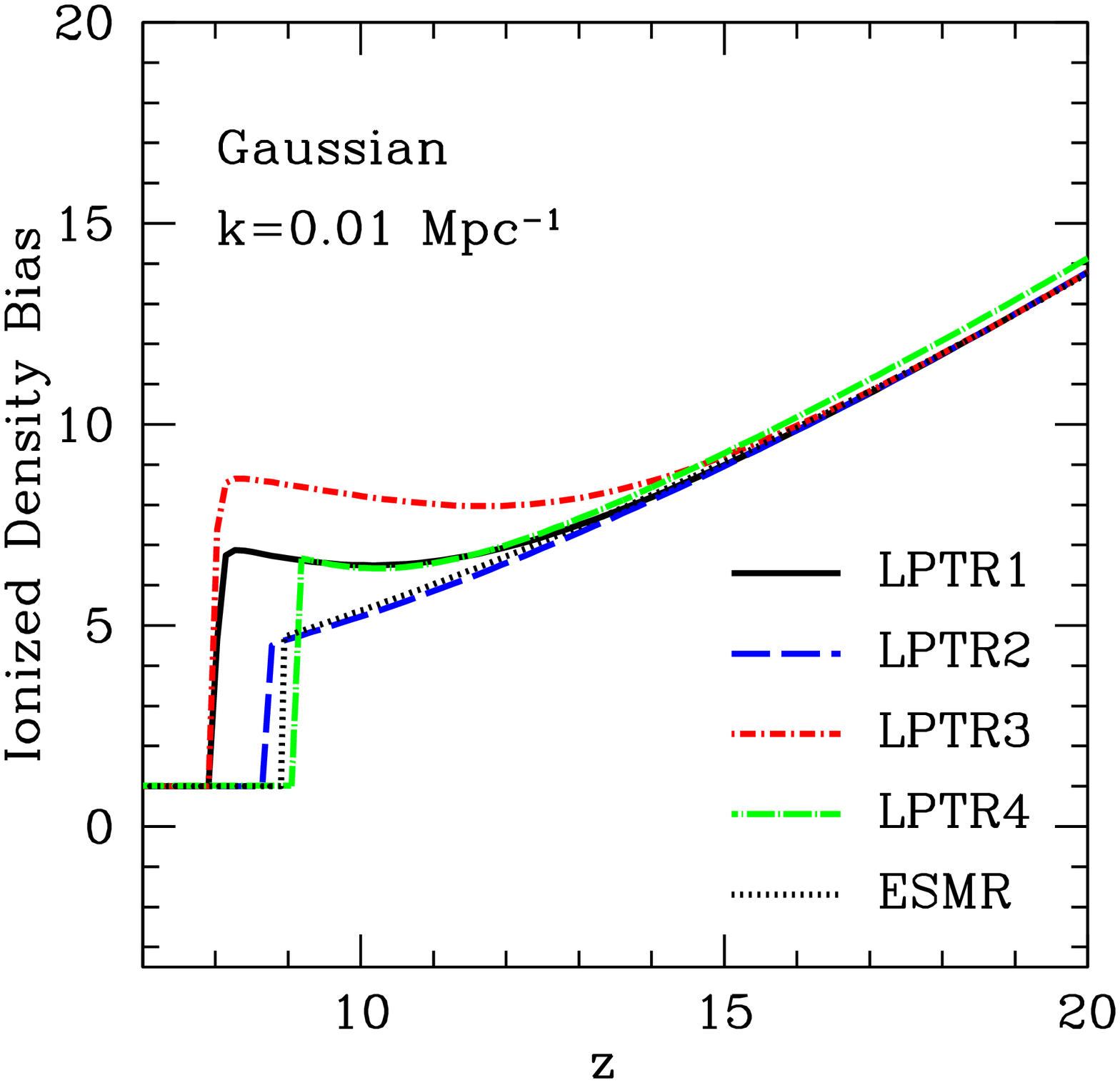}}
\resizebox{8.5cm}{!}{\includegraphics{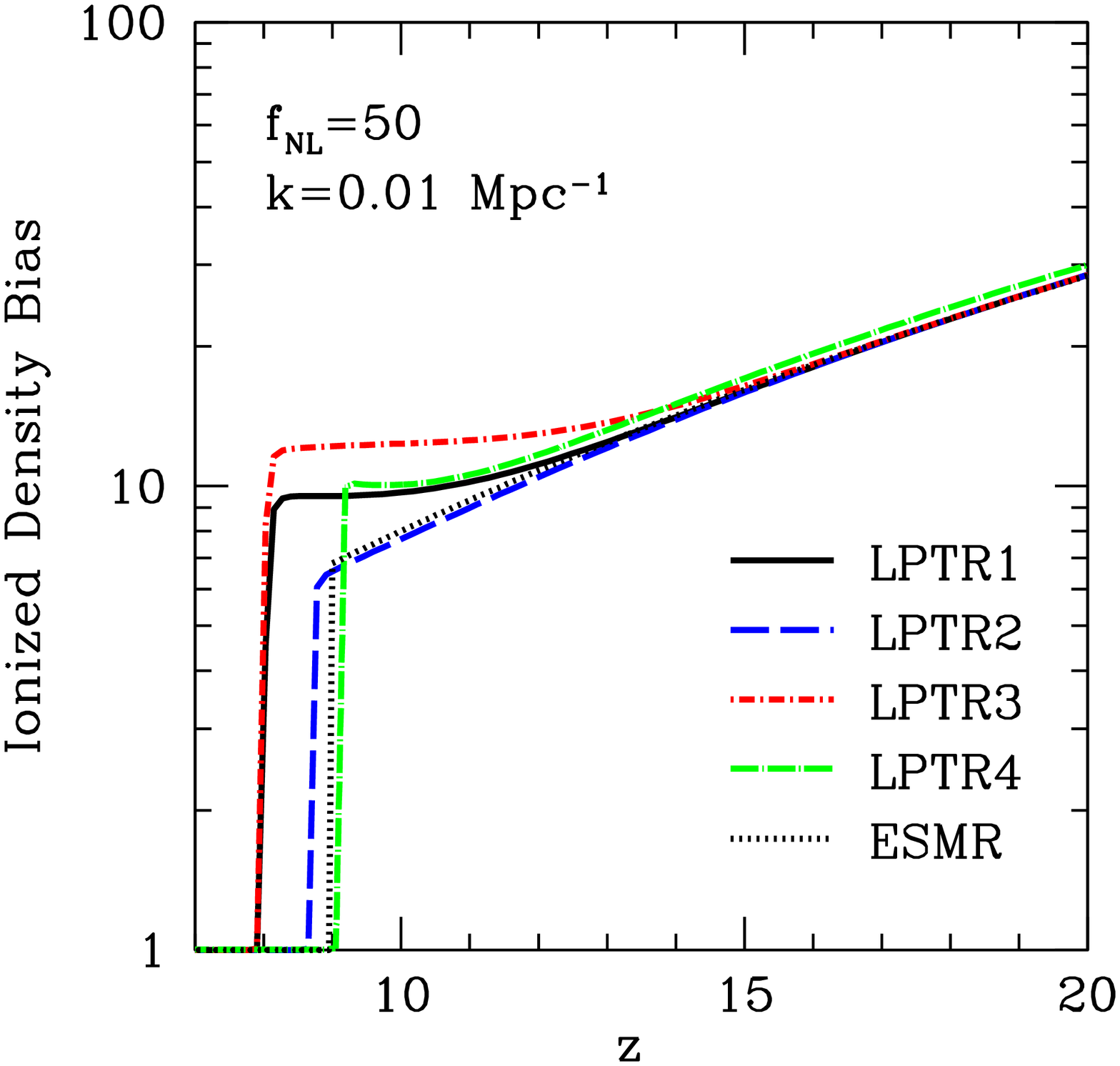}}
\end{center}
\caption{The redshift evolution of the ionized density bias for a fixed scale of $k=0.01~\Mpc^{-1}$.  }
\label{FIG:HIIbiasCompareB}
 \end{figure}
 
\begin{figure}
\begin{center}
\resizebox{8.5cm}{!}{\includegraphics{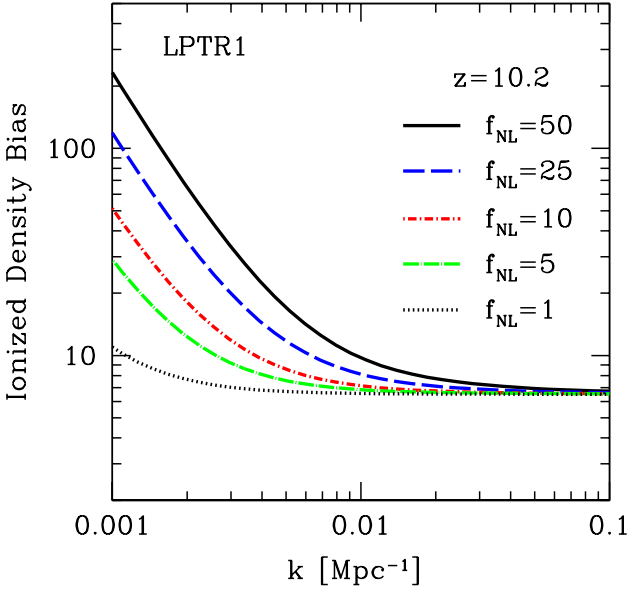}}
\end{center}
\caption{The ionized density bias in LPTR1 at a fixed $z=10.2$ for a range of local $\fNL$ values.  }
\label{FIG:HIIbiasfNLrange}
 \end{figure}
 
 \begin{figure}
\begin{center}
\resizebox{8.6cm}{!}{\includegraphics{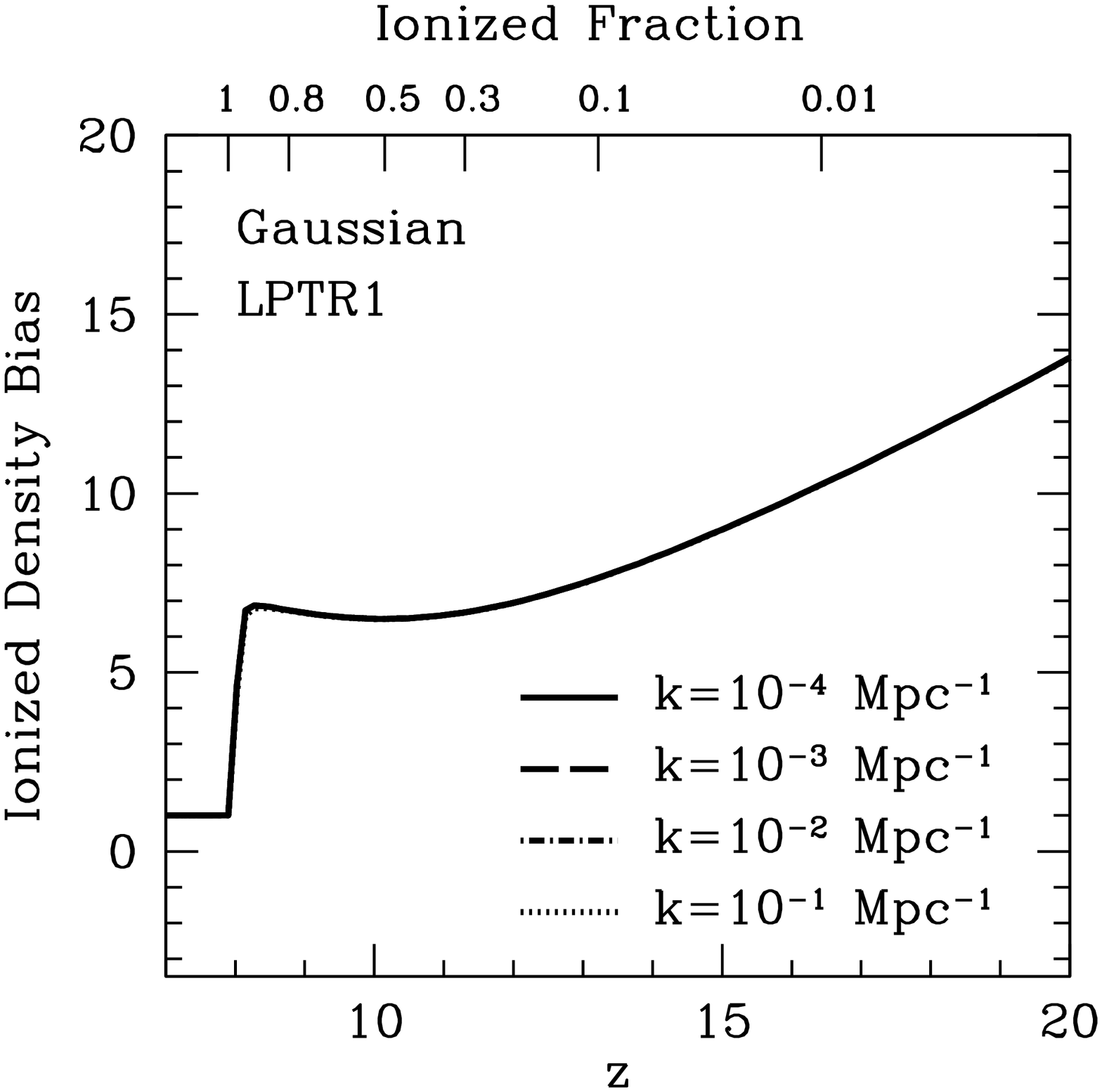}}
\resizebox{8.6cm}{!}{\includegraphics{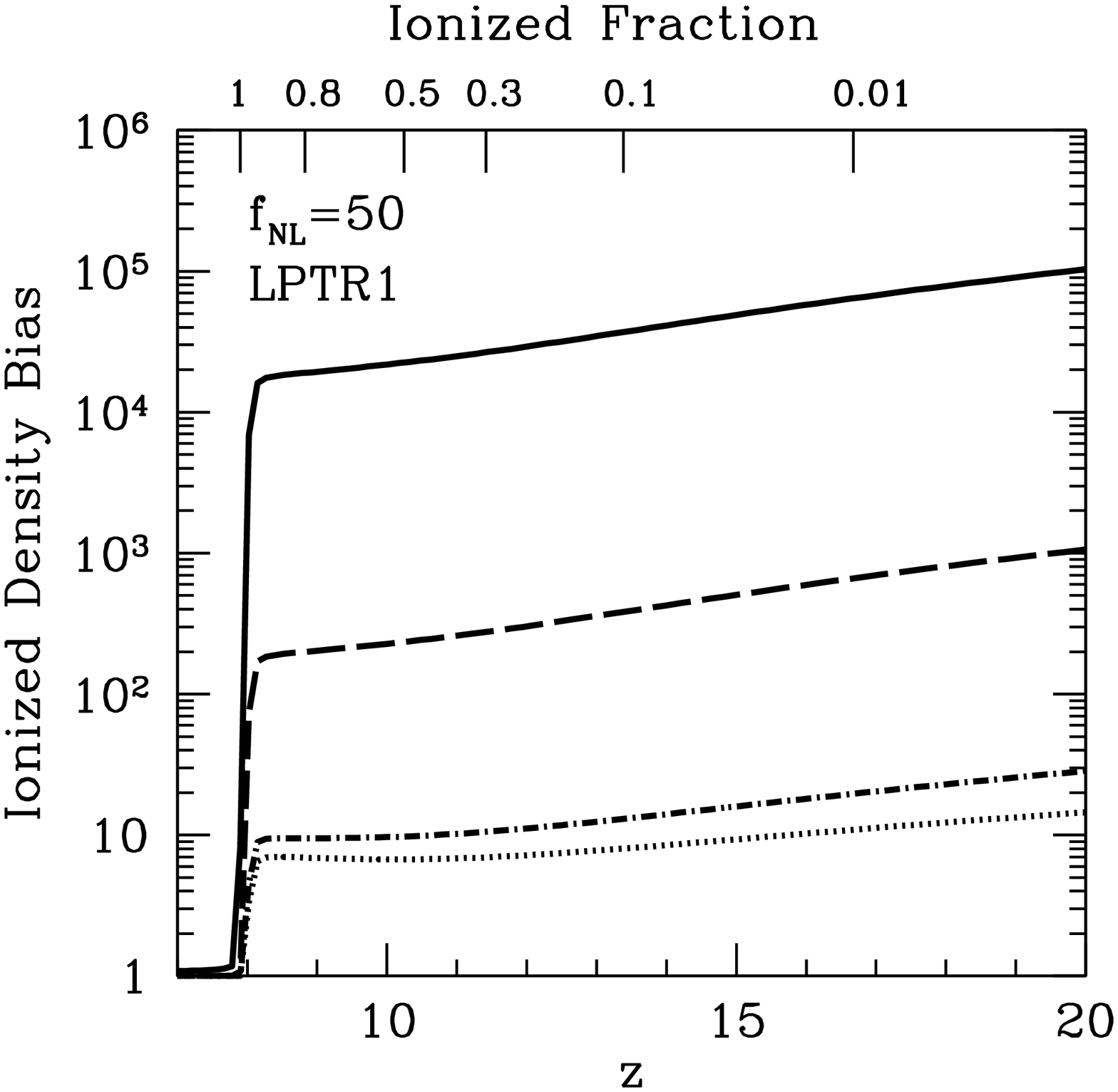}}
\end{center}
\caption{Evolution of the ionized density bias across a range of scales. Top panel: in the LPTR1 model with Gaussian initial conditions.  All curves here are degenerate, indicating that the Gaussian ionized density bias remains to a good approximation scale-independent until the end of reionization.  Bottom panel: for local PNG with $\fNL=50$.  The line styles here match those in the top panel.  The scale-dependent imprint of the non-Gaussian source bias on the ionized density bias is preserved until the end of reionization in the LPTR. }     
\label{FIG:LPT1_HIIbias}
\end{figure}

Since PNG alters the reionization history for a fixed source emissivity, it also alters the optical depth according to equation (\ref{EQ:tau_es}).  Figure \ref{FIG:fNLtau_es} shows the effects of PNG on $\tau_{es}$ for a range of $\fNL$ values.  For comparison, models with $\fNL\sim100$ can boost the optical depth by up to only $\sim 3\%$, which is somewhat larger than the results of \citet{2009MNRAS.394..133C}.  They used an independent analytical model to calculate a $\sim1\%$ enhancement for local $\fNL\sim100$.  
 
 \subsection{Large-scale ionized density bias}
   
Figure \ref{FIG:HIIbiasCompareA} shows the ionized density bias as a function of wavenumber for the reionization models in Table \ref{TAB:reionizationmodels}, with Gaussian (top) and non-Gaussian ($\fNL=50$, bottom) initial conditions\footnote{The comoving particle horizon is $R\sim4900~\Mpc$ at $z\sim10$, which corresponds to a wavenumber of $k\sim10^{-3}~\Mpc^{-1}$.  While we have adopted the Newtonian approximation in this work, we note that general relativistic effects on the source statistics may come into play on scales approaching $k\sim10^{-3}~\Mpc^{-1}$ [for detailed studies on galaxy bias in the context of general relativity, see \citet{2009PhRvD..80h3514Y,2010PhRvD..82h3508Y,2011JCAP...04..011B,2011PhRvD..84d3516C,2011PhRvD..84f3505B,2011JCAP...10..031B,2012PhRvD..85d1301B,2012PhRvD..85b3504J}].  On the other hand, the mean free path of UV photons during reionization is always much smaller than the horizon, so we do not anticipate additional general relativistic corrections to the LPTR itself to be significant.  From a practical viewpoint, such large scales are usually disregarded when forecasting constraints on cosmological parameters from the EoR 21cm power spectrum, because foreground contamination is most problematic on those scales \citep[e.g.][]{2008PhRvD..78b3529M}.  So general relativistic corrections are unlikely to influence the application of our results to the EoR 21cm power spectrum. }.  We show results at a single redshift, $z=10.2$, corresponding to $\bar{x}_i \approx 50\%$ for all models.  Figure \ref{FIG:HIIbiasCompareB} shows the redshift evolution of the ionized density bias for a fixed wavenumber, $k=0.01~\Mpc^{-1}$.  The steep decline of $\bHII$ to unity at lower redshifts marks the end of reionization, after which fluctuations in the IGM ionized density field faithfully trace matter fluctuations (see below for an explanation of the steepness).  Figure \ref{FIG:HIIbiasfNLrange} shows the ionized density bias for a range of $\fNL$ values in the LPTR1 model.  

Figures \ref{FIG:HIIbiasCompareA} and \ref{FIG:HIIbiasCompareB} show that the ESMR matches the LPTR2 results particularly well throughout reionization.  This is perhaps not surprising, since LPTR2 is the model in Table \ref{TAB:reionizationmodels} for which recombinations play the least significant role ($\clumpHII=2$).  The ESMR also does well at matching the evolution of $\bHII$ in the other LPTR models at early times.  While there are, as one might expect, significant differences in the evolution of $\bHII$ between the ESMR and LPTR models at later times, two basic predictions of the ESMR are corroborated by the LPTR:  1) The Gaussian ionized density bias is scale-independent on large scales.  2) Local PNG introduces through the clustering of sources a strong scale-dependent signature in the ionized density bias. 

 \begin{figure}
\begin{center}
\resizebox{8.7cm}{!}{\includegraphics{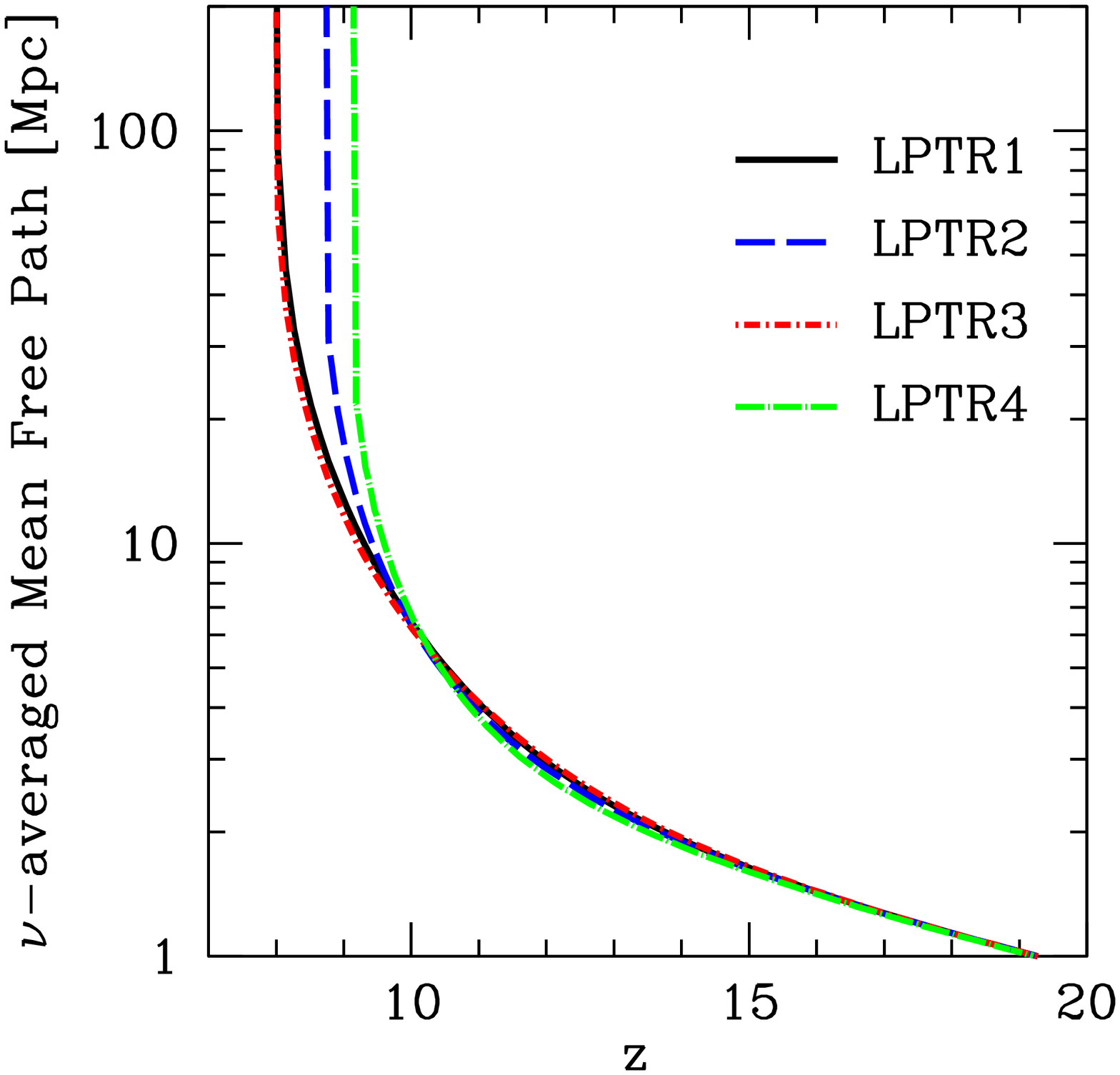}}
\end{center}
\caption{The mean free path of photons through the IGM averaged over frequency $\nu$ [see equations (\ref{EQ:MFPmu}) and (\ref{EQ:MFP})].  The $\nu$-averaged mean free path is the characteristic scale above which fluctuations in the ionized density field correspond to fluctuations in the source distribution.  Below this scale, $\bHII$ is suppressed towards unity, since ionized density fluctuations trace matter fluctuations inside of fully ionized regions.  }
\label{FIG:MFP}
 \end{figure} 
 
By construction, the ESMR $\bHII$ is determined by the source bias, so the similarities in shape between the ESMR and LPTR results in Figure \ref{FIG:HIIbiasCompareA} confirm, at least at $\bar{x}_i \sim 50\%$, that there is a strong connection between the scale-dependence of the source and ionized density bias parameters.  However, we may ask whether this connection is preserved throughout reionization.  The answer is not clear a priori, since the LPTR tracks the propagation of ionizing radiation, which can, especially towards the end of reionization, act over long distances to smooth out the patchiness in the ionized density field, and diminish its correspondence with the local clustering of sources.  We answer this question by plotting the ionized density bias for a range of scales ($k=10^{-4}, 10^{-3}, 10^{-2}$ and $0. 1~\Mpc^{-1}$) as a function of redshift.  The top and bottom panels of Figure \ref{FIG:LPT1_HIIbias} show the Gaussian and non-Gaussian LPTR1 results respectively.  Although we focus here on LPTR1 for clarity, we have checked that the results are similar for the other LPTR models in Table \ref{TAB:reionizationmodels}.  

In the top panel of Figure \ref{FIG:LPT1_HIIbias}, all four curves are so similar that they are indistinguishable in the plot, indicating that the Gaussian $\bHII$ is to a very good approximation scale-independent in the LPTR for $k \lesssim 10^{-1}~\mathrm{Mpc}^{-1}$ effectively throughout all of reionization.  In the bottom panel, the different amplitudes of the curves illustrate the scale-dependence of $\bHII$ due to PNG.  The important point is that the curves decline at approximately the same rate until the end of reionization, indicating a preservation of their relative amplitudes. This is especially true for the top two curves, corresponding to the largest scales, with $k=10^{-4}$ and $10^{-3}~\Mpc^{-1}$.  The small change with time in the relative amplitude between the bottom two curves, with $k=10^{-2}$ and $0.1~\Mpc^{-1}$, results from the scale-independent non-Gaussian correction to the bias, which is only noticeable at larger $k$, where the scale-dependent term is less dominant.

As \citet{2007MNRAS.375..324Z} point out, the preservation of the halo clustering imprint on the ionized density bias can be understood in terms of a characteristic scale -- the mean free path of ionizing photons through the IGM -- which for a given $x_\nu$ is  

\begin{equation}
\lambda(x_\nu) = a^2 / [\bnH \sigma(x_\nu) (1-\bar{x}_i)].
\label{EQ:MFPmu}  
\end{equation}
The relevant scale is the $\nu$-averaged mean free path, 

\begin{equation}
\langle \lambda \rangle_{\nu} = \frac{\int \dd x_\nu~\bar{\xi}_\gamma \lambda(x_\nu)}{\int \dd x_\nu \bar{\xi}_\gamma},
\label{EQ:MFP}
\end{equation}
which corresponds to the average distance that a photon travels before photoionizing a neutral atom.  Ionizing radiation is unable to smooth out the patchiness of the ionized density field on scales much larger than $\langle \lambda \rangle_\nu$.  On these scales, the ionized density field traces large-scale fluctuations in the sources, so $\bHII$ is expected to have approximately the same scale-dependence as the source bias (though with a different amplitude).  On the other hand, for scales less than $\langle \lambda \rangle_\nu$, but larger than the Jeans scale, the ionized density field typically traces the matter density fluctuations, which suppresses the ionized density bias towards $\bHII \sim 1$.  As reionization proceeds, $\langle \lambda \rangle_\nu$ increases, so the scale below which $\bHII$ is suppressed increases until $\bHII=1$ for all scales at the end of reionization.

Figure \ref{FIG:MFP} shows $\langle \lambda \rangle_\nu$ in our four LPTR models.  For clarity, we show only the Gaussian models, since the non-Gaussian results are essentially identical. It is now clear why the halo clustering imprint is so well preserved in the LPTR throughout reionization, and why $\bHII$ drops so rapidly to unity at the end of reionization.  The value of $\langle \lambda \rangle_\nu$ is well below the scales of interest ($k<0.1~\Mpc^{-1}$ corresponding to $\gtrsim 60 ~\Mpc$) until the very end of reionization, at which point it is a very steeply rising\footnote{We note that the LPTR neglects absorption of photons by Lyman-limit systems, which would cap the steep rise of the mean free path at the end of reionization (see  Figure \ref{FIG:MFP}).}.  In fact, $\langle \lambda \rangle_\nu$ increases so rapidly at the end of reionization that $\bHII$ drops to unity at approximately the same redshift for the four wavenumbers considered in Figure \ref{FIG:LPT1_HIIbias}.  In the next section, we explore the relationship between Gaussian and non-Gaussian $\bHII$ implied by the source-clustering imprint.

\subsection{Analytical mapping between Gaussian and non-Gaussian ionized density bias parameters}

In the last section we saw the ESMR prediction of a strong scale-dependent imprint in $\bHII$ from the source bias confirmed.  On the other hand, the ESMR does not capture the details of the imprint, i.e. the bias amplitude and its evolution, in all of our LPTR models.  Indeed, we should not expect it to; after all, the simple ESMR developed in \S \ref{SEC:ESMR} is a one parameter model, whereas each LPTR model contains four parameters (counting the source spectrum slope, $s$, which we have not varied here).  One approach we could take is to expand upon the ESMR by re-parameterizing it to better match the LPTR results.  While this may be a fruitful topic of future investigation, it is perhaps more useful at this stage to test whether the ESMR at least gives the correct relationship between the Gaussian and non-Gaussian $\bHII$.  To this end, for each LPTR model in Table \ref{TAB:reionizationmodels}, we consider the accuracy of a ``hybrid" calculation, which takes the ESMR equations (\ref{EQ:bcfPNG_SI}) and (\ref{EQ:SDcorrection}) for the non-Gaussian $\bHII$, and inserts the numerical values of the Gaussian $\bHII$ from the particular LPTR model.  We then compare the results of this calculation to results from the full LPTR calculation.     

\begin{figure}
\begin{center}
\resizebox{8.3cm}{!}{\includegraphics{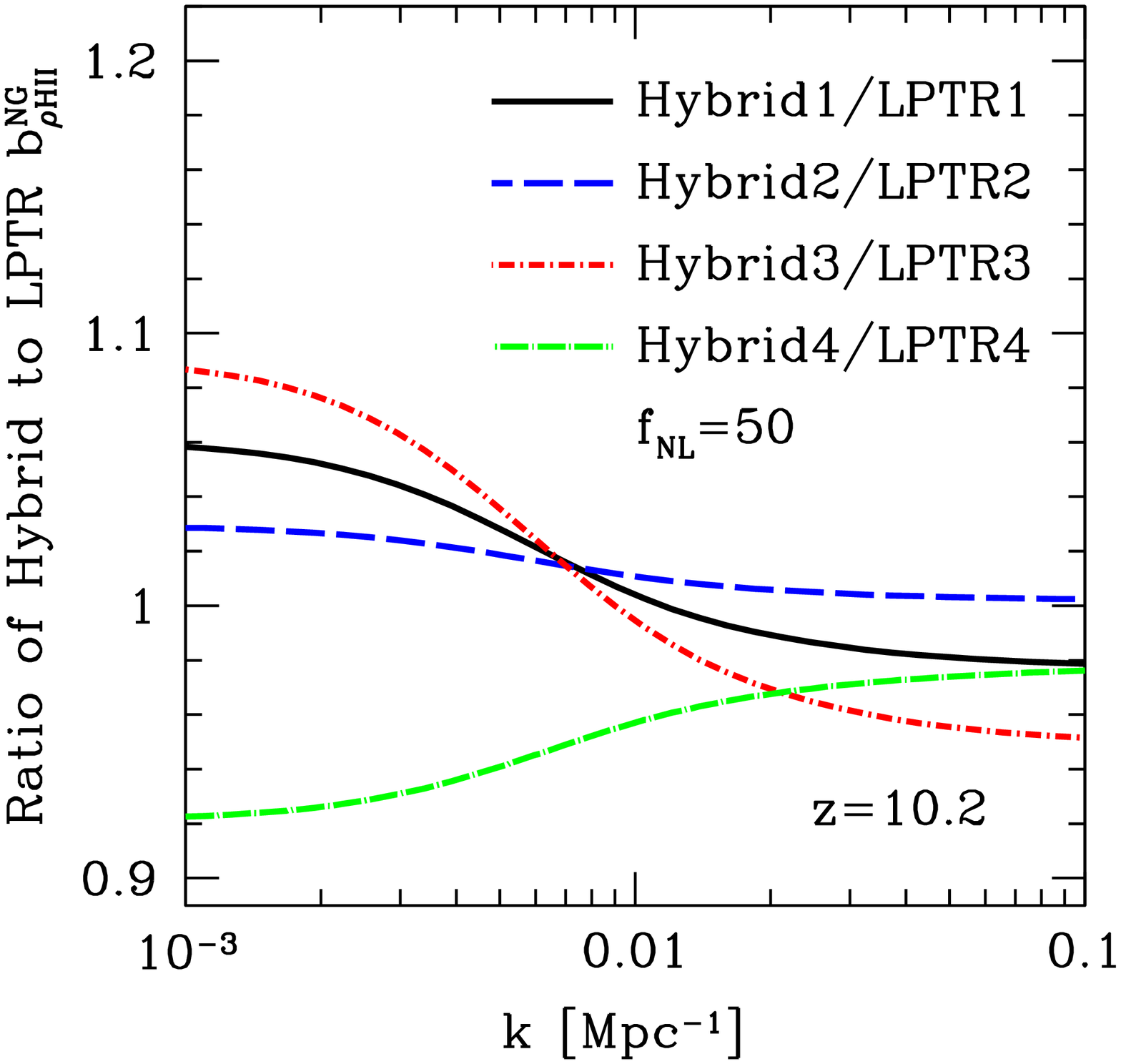}}
\resizebox{8.3cm}{!}{\includegraphics{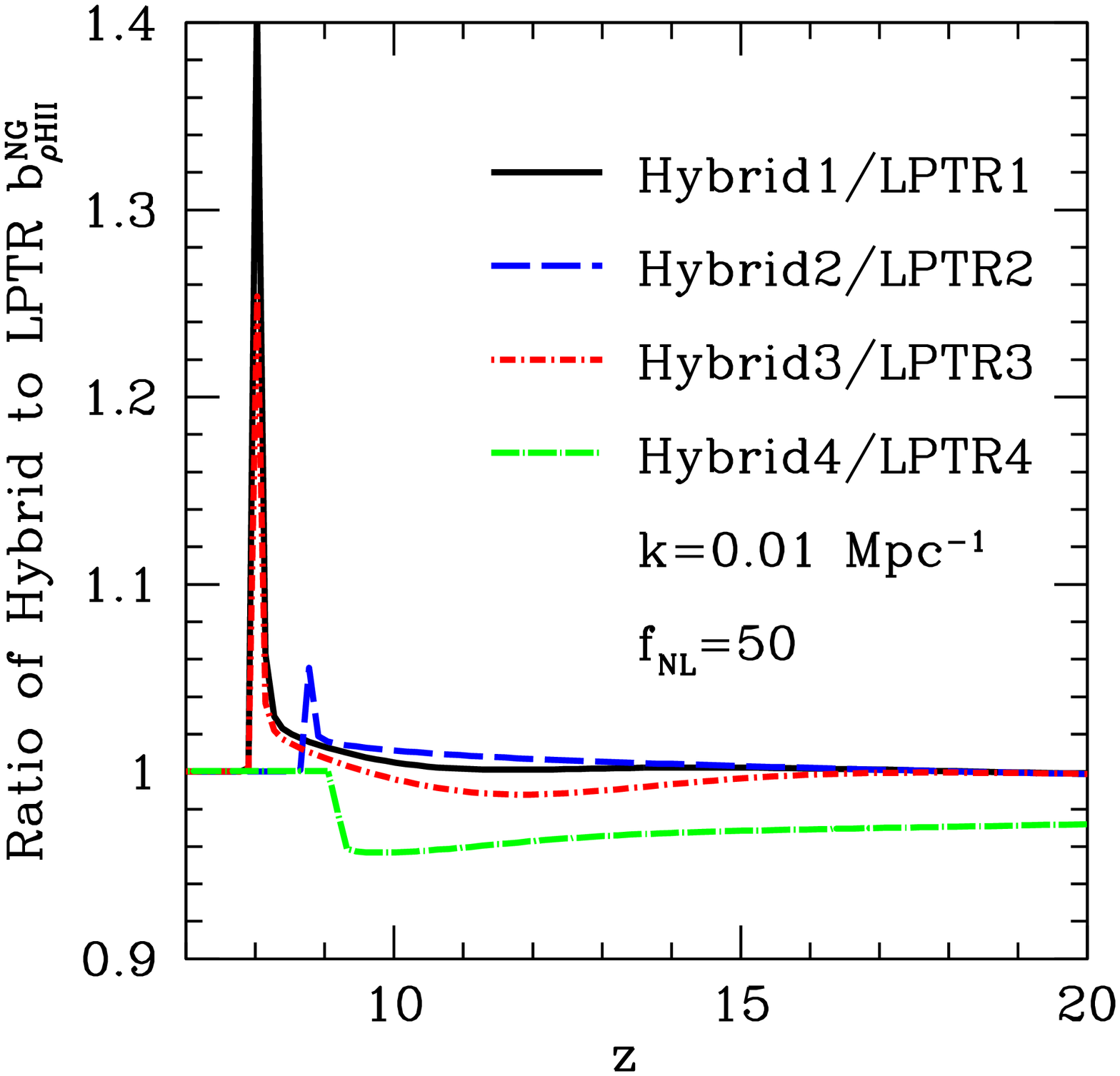}}
\end{center}
\caption{Testing how accurately the ESMR predicts the relationship between Gaussian and non-Gaussian ionized density bias parameters in our LPTR models.  In the ``hybrid" calculation, for a given LPTR model, we take the Gaussian bias, $\bHII^{\mathrm{G}}$, and insert it into the ESMR expressions [equations (\ref{EQ:bcfPNG_SI}) and (\ref{EQ:SDcorrection})] to obtain the non-Gaussian, $\bHII^{\mathrm{NG}}$.  These results are then compared to the full LPTR calculation, which provides the exact $\bHII^{\mathrm{NG}}$.  Above, we show the ratio of hybrid to full LPTR $\bHII^{\mathrm{NG}}$ as a function of $k$ for fixed $z=10.2$ (top), and as a function of $z$ for fixed $k=0.01~\Mpc^{-1}$ (bottom).}
\label{FIG:HybridCompare}
 \end{figure}

Figure \ref{FIG:HybridCompare} shows the ratios of $\bHII$ obtained from the hybrid and full LPTR calculations.  The top panel shows the ratios as a function of wavenumber for a fixed $z=10.2$, while the bottom panel shows them as a function of redshift for a fixed $k=0.01~\Mpc^{-1}$.  Equations (\ref{EQ:bcfPNG_SI}) and (\ref{EQ:SDcorrection}) best capture the relationship between Gaussian and non-Gaussian $\bHII$ in the LPTR2 model.  In that case, the hybrid result is within $5 \%$ of the full result until the end of reionization.  Perhaps more importantly, the flatness of the dashed curve in the top panel of Figure \ref{FIG:HybridCompare} shows that the scale-dependence of $\bHII$ is well approximated by the ESMR.  For the other LPTR models, the hybrid calculations work best at earlier times, reproducing the LPTR results to within a few percent for $z>12$ (corresponding to $\bar{x}_i \sim 10-20 \%$) after which the discrepancies tend to rise.  Nonetheless, the hybrid calculations are accurate to within $\lesssim 20 \%$ across $10^{-3} \lesssim k \lesssim 0.1~\Mpc^{-1}$ up until the end of reionization for all LPTR models.  We highlight the mild mismatch in the scale-dependence of $\bHII$ between the hybrid and full LPTR calculations.  This mismatch is most obvious in LPTR3; the model in which recombinations are most significant.   In fact, the functional dependence of $\bHII$ on $k$ appears to depend mildly on $\clumpHII$.  This effect may prove to be minimal, if the lower values of $\clumpHII$ ($\sim 1-3$) suggested by recent studies are confirmed.

Given the success of equations (\ref{EQ:bcfPNG_SI}) and (\ref{EQ:SDcorrection}) in mapping to a reasonable accuracy the Gaussian to non-Gaussian $\bHII$ from the range of LPTR models considered here, we hope that these equations will work equally well when applied to more detailed cosmological radiative transfer simulations.  Equations (\ref{EQ:bcfPNG_SI}) and (\ref{EQ:SDcorrection}) should be similarly tested against full simulation results once large enough boxes become available.

\section{Comparison with previous work}
\label{SEC:JDFKS}

In this section we compare our results to the results of JDFKS.  We begin by summarizing important details of the JDFKS SimFast21 reionization simulations [see \citet{2010MNRAS.406.2421S} for details on the SimFast21 code methodology].  Their initial density fields, both Gaussian and non-Gaussian, were generated on a regular cubic mesh with $N=2048^3$ cells in a (3 Gpc)$^3$ volume, corresponding to a halo mass-resolution of $M\approx 5\times10^{11}\Msun$.  In order to account for mass in smaller halos, down to their ACH mass-threshold of $\mmin=10^8\Msun$, JDFKS employed an analytical sub-grid model in which equation (\ref{EQ:fcollgm}) -- the excursion-set expression for the \emph{Gaussian} collapsed fraction -- is used to fill in the unresolved collapsed mass in each mesh cell \citep{2011PhRvL.107m1304J}.  At $z=7.5$, the redshift at which their analysis was based, the $10^8\Msun$ threshold corresponds to a peak height of $\nu \equiv \delta_c/\sqrt{S} \approx 2$, while the $5\times10^{11}\Msun$ mass-resolution limit corresponds to $\nu \approx 4.7$.  Their reionization simulations were therefore sourced mainly by unresolved sub-grid halos, since these were by far more abundant than halos resolved directly by their mesh.  

As we point out in $\S$ \ref{SEC:introduction}, JDFKS did not directly calculate the ionized density bias. Rather, they calculated a quantity they called the ``bias of ionized regions," whose definition has caused some confusion in the literature.  They defined this bias parameter to be $\hat{b}_x \equiv \sqrt{\mathscr{P}_{x x} / \mathscr{P}_{\delta \delta}}$ (Note that we use the ``hat" notation to distinguish this quantity from the distinct but related ionized fraction bias, $b_x$.  See below for the relationship between $\hat{b}_x$ and $b_x$).  Here, $\mathscr{P}_{\delta \delta} \equiv \hat{T}_b^2 \bar{x}^2_{\mathrm{H}} P_{\delta \delta} $ and $\mathscr{P}_{x x} \equiv \hat{T}_b^2 P_{x x} $, where $\hat{T}_b$ is related to the spatially averaged 21cm brightness temperature ($\bar{T}_b$) by $\hat{T}_b=\bar{T}_b/\bar{x}_{\mathrm{H}}$ under the usual assumption that the spin temperature is much greater than the CMB temperature, as expected during reionization.  We emphasize the important distinction between $\hat{b}_x$, used in JDFKS, and the ionized fraction bias, $b_x$, defined in equation (\ref{EQ:bxdef}).  These quantities are related as follows: $\hat{b}_x = \sqrt{\mathscr{P}_{x x} / \mathscr{P}_{\delta \delta} }= (1/ \bar{x}_{\mathrm{H}}) \sqrt{ P_{x x} / P_{\delta \delta} }= (\bar{x}_i / \bar{x}_{\mathrm{H}}) \sqrt{ P_{\delta_x \delta_x} / P_{\delta \delta}}$, which yields

\begin{equation}
\hat{b}_x = \frac{\bar{x}_i}{\bar{x}_{\mathrm{H}}} b_x.
\end{equation}  
Using equation (\ref{EQ:bHII2bx}), we also find a simple relationship between $\hat{b}_x$ and the ionized density bias, $\bHII$, which applies on large scales: 

\begin{equation}
\hat{b}_x(k,z) = \frac{\bar{x}_i(z)}{ \bar{x}_{\mathrm{H}}(z)} \left[\bHII(k,z) - 1\right].
\label{EQ:bxtransformation}
\end{equation} 

 \begin{figure}
\begin{center}
\resizebox{8.3cm}{!}{\includegraphics{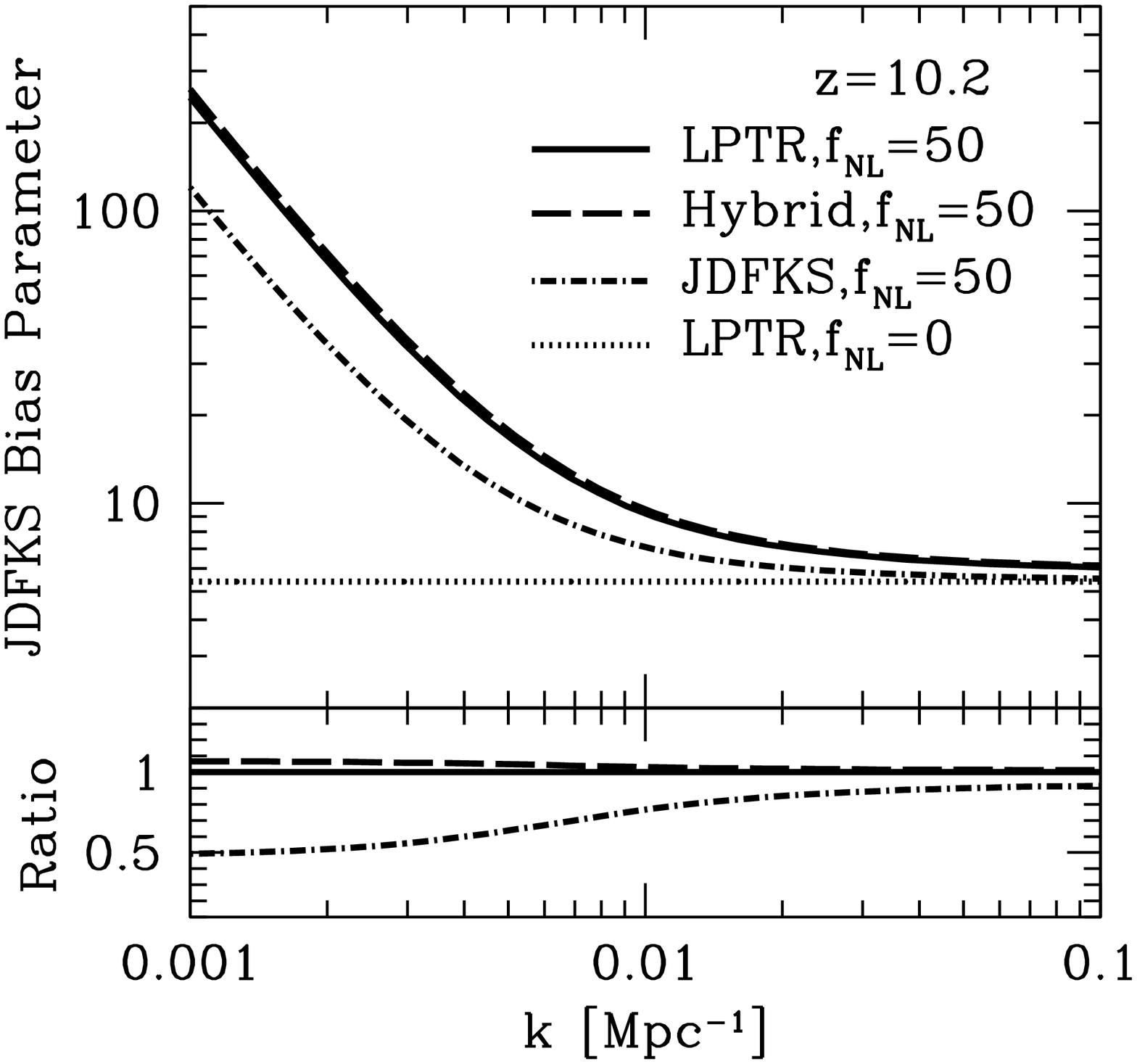}}
\resizebox{8.3cm}{!}{\includegraphics{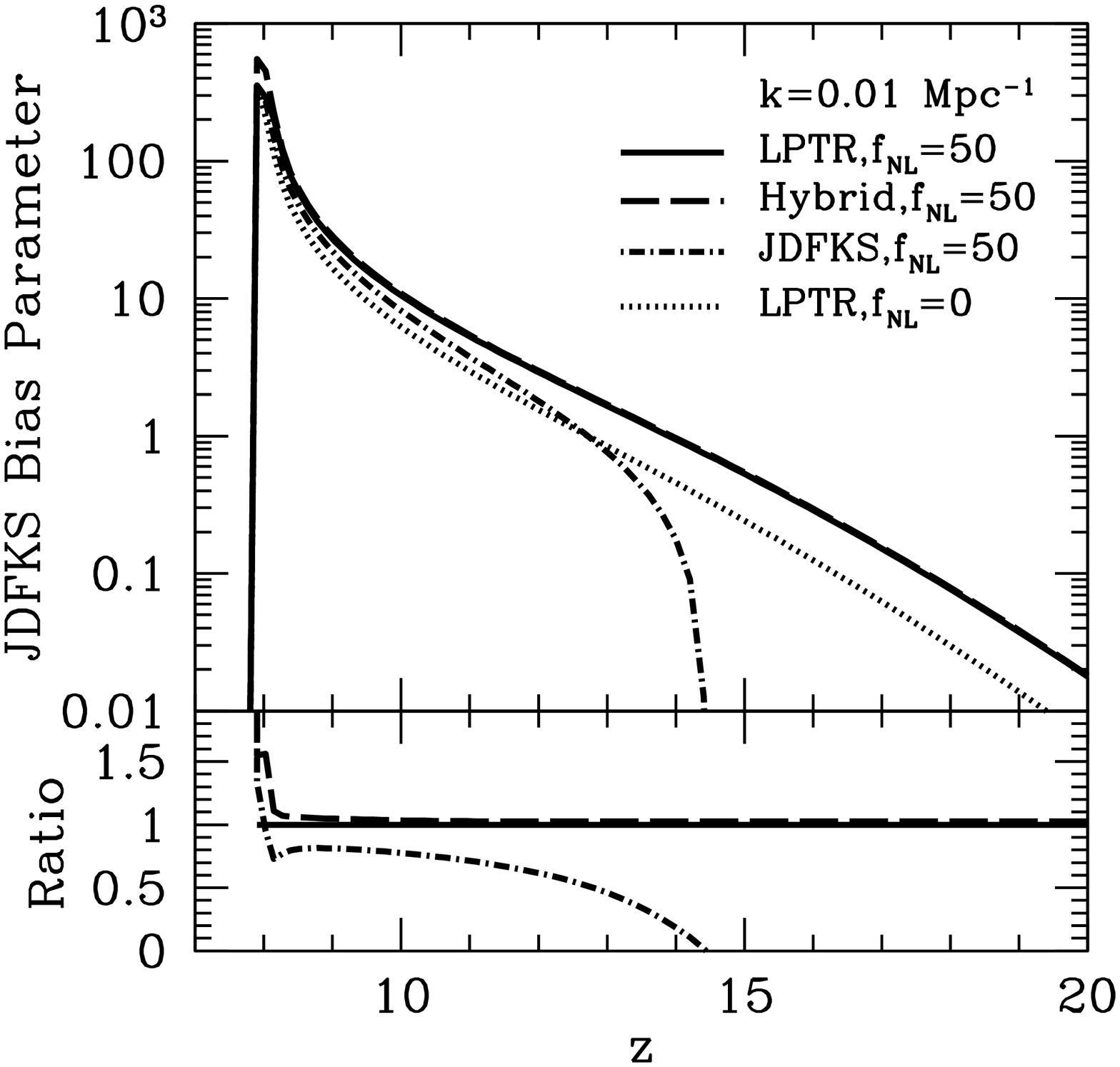}}
\end{center}
\caption{Testing the accuracy of our ESMR prediction [eq. (\ref{EQ:hatbx})] versus the JDFKS fitting formula [eq. (\ref{EQ:JoudakiBias})] in predicting the relationship between Gaussian and non-Gaussian $\hat{b}_x = \sqrt{\mathscr{P}_{xx} / \mathscr{P}_{\da \da}}$ (The JDFKS bias parameter) in our LPTR calculations.  To obtain the curve labeled ``hybrid", we take $\hat{b}_x^{\mathrm{G}}$ from LPTR1 and insert it into equation (\ref{EQ:hatbx}).  The curve labeled ``JDFKS" is obtained in a similar way, but using (\ref{EQ:JoudakiBias}).  For clarity, we only show results from the LPTR1 model in Table \ref{TAB:reionizationmodels}.       }
\label{FIG:JDFKScompare}
 \end{figure}

JDFKS report that the $\hat{b}_x$ computed from their non-Gaussian simulations are to a good approximation related to the $\hat{b}_x$ from corresponding Gaussian simulations by the following proposed fitting function\footnote{Here we restore a factor of $g(0)$ in the denominator of the scale-dependent term.  JDFKS did not normalize $D(0)=1$, so $g(0)$ does not appear in equation (4) of their paper; it is absorbed into their definition of $D(z)$ \citep{JoudakiEmail}.  In contrast, we choose to normalize $D(0)=1$, so $g(0)$ must appear in the denominator of our equation (\ref{EQ:JoudakiBias}).}:  

\begin{equation}
\hat{b}^{\mathrm{NG}}_x(k,z) = \hat{b}_x^{\mathrm{G}}(z)+ 3 \fNL  \left[\hat{b}^{\mathrm{G}}_x(z) - 1 \right]   \frac{\bar{\delta}_x \Omega_m H_0^2}{g(0) D(z) k^2 T(k)}. 
\label{EQ:JoudakiBias}
\end{equation}
The quantity $\bar{\delta}_x$ is analogous to the critical overdensity for collapse which appears in the well-known non-Gaussian halo bias formula \citep{2008ApJ...677L..77M,2008PhRvD..77l3514D,2008PhRvD..78l3507A}.  The condition $\fcoll(\mmin,R,z) \geq \zeta_{\mathrm{ESMR}}^{-1}$ for a fully ionized region in the ESMR along with equation (\ref{EQ:fcollgm}) implies (for Gaussian initial conditions) a minimum linearly extrapolated overdensity of

\begin{equation}
\delta_x \geq \da_c - \sqrt{2} K( \zeta_{\mathrm{ESMR}}) \sqrt{\Smin - S_R}
\label{EQ:dxbarrier}
\end{equation}
for a region with Lagrangian radius $R$ to self-ionize, where $K( \zeta_{\mathrm{ESMR}})\equiv \mathrm{erf}^{-1}(1- \zeta_{\mathrm{ESMR}}^{-1})$.  JDFKS defined $\bar{\delta}_x$ to be the ``average" $\delta_x$.   For illustrative purposes, they calculated $\bar{\delta}_x = 1.1$ by averaging equation (\ref{EQ:dxbarrier}) over the fraction of volume filled by H II bubbles.  However, they set $\bar{\delta}_x=1$ for convenience when they used equation (\ref{EQ:JoudakiBias}) to forecast 21cm power spectrum constraints on $\fNL$.   

We note that equation (\ref{EQ:JoudakiBias}) was not derived from analytical theory.  Rather, it was postulated by JDFKS based on their intuition from previous literature on the halo bias.  Here, we derive a formula for $\hat{b}_x^{\mathrm{NG}}$ from first principles in the ESMR, and compare it directly to equation (\ref{EQ:JoudakiBias}).  We then compare the accuracies of our result and equation (\ref{EQ:JoudakiBias}) in capturing the relationship between $\hat{b}_x^{\mathrm{NG}}$ and $\hat{b}_x^{\mathrm{G}}$ when both quantities are computed numerically with the LPTR.  

To derive our analogue of the JDFKS fitting formula from the ESMR, we use equation (\ref{EQ:bxtransformation}) to convert equation (\ref{EQ:bcfPNG_total}) to $\hat{b}_x^{\mathrm{NG}}$.  Note that PNG alters the mean ionized fraction relative to the Gaussian case at a fixed redshift.  We will therefore use the super-scripts ``NG" and ``G" on the mean ionized and neutral fractions to distinguish the two cases.  Taking this caveat into account, equation (\ref{EQ:bcfPNG_total}) implies

\begin{align}
 \hat{b}_x^{\mathrm{NG}} = \frac{\bar{x}_i^{\mathrm{NG}}  }{ \bar{x}_H^{\mathrm{NG}}  } \frac{ \bar{x}_H^{\mathrm{G}} }{ \bar{x}_i^{\mathrm{G}} } \Biggl\{ \hat{b}_x^{\mathrm{G}}  -\frac{\mathcal{S}^{(3)}_{\mathrm{min}} }{6} \Smin  \hat{b}_x^{\mathrm{G}}  \biggl[ \frac{3\da_c \Smin - \da_c^3}{\Smin^2} \nonumber \\   +  \left(  \frac{ \bar{x}_H^{\mathrm{G}} }{ \bar{x}_i^{\mathrm{G}} } \hat{b}_x^{\mathrm{G}} \right) \left( \frac{\da_c^2}{\Smin}  -1 \right) \biggr] +  2 \delta_c \hat{b}_x^{\mathrm{G}}  \frac{\mathcal{F}^{(3)}_{\mathrm{min}}(k) }{\mathcal{M}_{\mathrm{min}}(k)} \Biggr\}. 
\label{EQ:hatbx}
\end{align}
Note that this expression applies to PNG with general bispectra.  In order to compare this result to the JDFKS formula [equation (\ref{EQ:JoudakiBias})] we restrict equation (\ref{EQ:hatbx}) to the local template, where we can make the substitution,

\begin{equation}
 2 \delta_c \hat{b}_x^{\mathrm{G}}  \frac{\mathcal{F}^{(3)}_{\mathrm{min}}(k) }{\mathcal{M}_{\mathrm{min}}(k)} \rightarrow 3 \fNL\hat{b}_x^{\mathrm{G}} ~\delta_c  \frac{ \Omega_m H_0^2}{g(0) D(z) k^2 \mathcal{T}(k)}.
\end{equation}
Applying this substitution to equation (\ref{EQ:hatbx}) yields

\begin{align}
\hat{b}_x^{\mathrm{NG}} = \frac{\bar{x}_i^{\mathrm{NG}}  }{ \bar{x}_H^{\mathrm{NG}}  } \frac{ \bar{x}_H^{\mathrm{G}} }{ \bar{x}_i^{\mathrm{G}} } \Biggl\{ \hat{b}_x^{\mathrm{G}}  -\frac{\mathcal{S}^{(3)}_{\mathrm{min}} }{6} \Smin  \hat{b}_x^{\mathrm{G}}  \biggl[ \frac{3\da_c \Smin - \da_c^3}{\Smin^2} \nonumber \\   +  \left(  \frac{ \bar{x}_H^{\mathrm{G}} }{ \bar{x}_i^{\mathrm{G}} } \hat{b}_x^{\mathrm{G}} \right) \left( \frac{\da_c^2}{\Smin}  -1 \right) \biggr] + 3 \fNL\hat{b}_x^{\mathrm{G}}  \frac{\delta_c \Omega_m H_0^2}{g(0) D(z) k^2 \mathcal{T}(k)}  \Biggr\}. 
\label{EQ:hatbxlocal}
\end{align}

Our analytical result differs from the JDFKS fitting formula in several important ways: (i) The second term in equation (\ref{EQ:hatbxlocal}) is the scale-independent correction to $\hat{b}_x$ due to PNG, which JDFKS neglected.  For the local template, this correction contributes up to a few percent to $\hat{b}_x$ for $k \sim 0.1$ Mpc$^{-1}$.  As an example, in the ESMR model at $z=10.2$ with $\fNL=50$ ($\bar{x}_i^{\mathrm{NG}}=0.51$, $\bar{x}^{\mathrm{G}}_i=0.48$, $\hat{b}^{\mathrm{G}}_x$=4.15), we find the fractional contribution of the scale-independent term to be $2.5$, $1.6$ and $0.06$ percent at $k=10^{-1}$, $10^{-2}$, and $10^{-3}$ Mpc$^{-1}$ respectively.  (ii) The critical density for halo collapse, $\delta_c\approx1.686$, appears in the last term of (\ref{EQ:hatbx}) (the scale-dependent term) and not $\bar{\delta}_x\approx 1$ as assumed by JDFKS.  (iii) Equation (\ref{EQ:hatbx}) has an overall factor involving mean ionized and neutral fractions which is absent in the JDFKS formula.  (iv) It is $\hat{b}_x^{\mathrm{G}}$ that sets the amplitude of the scale-dependent term in our formula, and not $(\hat{b}_x^{\mathrm{G}}-1)$ as assumed by JDFKS.  They assumed the latter because the non-Gaussian scale-dependent Eulerian halo bias is proportional to $(b^{\mathrm{G}}-1)$, where $b^{\mathrm{G}}$ is the Gaussian (Eulerian) halo bias.  This $(b^{\mathrm{G}}-1)$ form in the halo bias comes from the spherical collapse model, in which the Lagrangian and Eulerian halo bias parameters are related by $b^{\mathrm{G}}=b^{\mathrm{G}}_{\mathrm{L}}+1$ \citep{1996MNRAS.282..347M}.  However, as we discussed in \S \ref{SEC:ESMRGauss}, the Lagrangian and Eulerian ionized fraction bias parameters are equivalent, and since $\hat{b}_x = (\bar{x}_i / \bar{x}_H) b_x$, so are the Lagrangian and Eulerian $\hat{b}_x$.  The factor of $(\hat{b}_x^{\mathrm{G}}-1)$ should therefore not appear in the formula.      

We now explore the numerical differences between equations (\ref{EQ:JoudakiBias}) and (\ref{EQ:hatbx}).  Figure \ref{FIG:JDFKScompare} shows the comparison between three calculations:  (i) The purely numerical LPTR1 calculation (solid), using equation (\ref{EQ:bxtransformation}) to convert $\bHII^{\mathrm{NG}}$ to $\hat{b}^{\mathrm{NG}}_x$, (ii) Our ESMR prediction (dashed), equation (\ref{EQ:hatbx}),  and (iii) The JDFKS fitting formula (dot-dashed), equation (\ref{EQ:JoudakiBias}).  In calculations ii and iii, we insert the numerical values of $\hat{b}_x^{\mathrm{G}}$ from the LPTR1 calculation into equations (\ref{EQ:hatbx}) and (\ref{EQ:JoudakiBias}) respectively (like the ``hybrid" calculations described in the last section), since our goal here is to determine how well these equations capture the mapping between $\hat{b}_x^{\mathrm{G}}$ and $\hat{b}_x^{\mathrm{NG}}$.  The top panel shows $\hat{b}_x$ as a function of $k$ at a fixed redshift of $z=10.2$, corresponding to a mean ionized fraction of $\approx 50 \%$.  The bottom panel shows the redshift evolution for a fixed $k=0.01~\Mpc^{-1}$.  While the ESMR prediction matches the LPTR results well, the JDFKS fitting function always yields a significantly lower amplitude of $\hat{b}^{\mathrm{NG}}_x$.  The differences can be quantified in terms of $\fNL$, which sets the amplitude of the scale-dependent term.  As a simple, ``back-of-the-envelope" illustration, if we assume that $\fNL=50$ is the ``true" model, and that the LPTR1 curve in the top panel of Figure \ref{FIG:JDFKScompare} represents the ``measured" $\hat{b}^{\mathrm{NG}}_x$ at $\bar{x}_i \sim 0.5$, then an observer using the JDFKS fitting formula would falsely infer an $\fNL\sim 100$, assuming that $\hat{b}^{\mathrm{G}}_x$ is known exactly.  The bottom panel of Figure \ref{FIG:JDFKScompare} shows that the JDFKS formula also does not reproduce the redshift evolution of $\hat{b}^{\mathrm{NG}}_x$.  At high redshifts, the JDFKS formula becomes negative due to $\hat{b}^{\mathrm{G}}_x$ dropping below unity at those epochs.  This behavior of the JDFKS fitting formula is not observed in our results.

\section{Summary and Discussion}
\label{SEC:conclusion}

We have presented a first-principals investigation on the effects of primordial non-Gaussianity on the large-scale structure of reionization.  We employed two methods that are independent in how they model reionization:  1) An extension of the analytical excursion-set model of reionization (ESMR) of \citet{2004ApJ...613....1F} to include PNG, which allowed us to derive analytical expressions for the ionized density bias.  2)  The linear perturbation theory of reionization (LPTR) of \citet{2007MNRAS.375..324Z}, which has the advantage that it directly solves the ionization rate and radiative transfer equations, allowing us to explore a range of recombination clumping factors, and two distinct models of photon-production rates in galactic sources.  Our main results can be summarized as follows:    

\begin{itemize}

\item{Equations (\ref{EQ:fcollPNGdef}) and (\ref{EQ:fcollPNGlong}) inserted into equation (\ref{EQ:ESMR}) constitute our extension of the analytical ESMR of \citet{2004ApJ...613....1F} to include non-Gaussian initial conditions with general bispectra.  }

\item{In our non-Gaussian extension of the ESMR, equation (\ref{EQ:ESMRbHII}) gives the Gaussian ionized density bias, while equations (\ref{EQ:bcfPNG_SI}) and (\ref{EQ:SDcorrection}) give the non-Gaussian scale-independent and -dependent ionized density bias corrections.  These expressions follow from the assumption that, on large scales, fluctuations in the ionized density field follow fluctuations in the source distribution.  The ESMR predicts that the ionized density bias is scale-independent in models with Gaussian initial conditions, while for non-Gaussian initial conditions the bias acquires a scale-dependent correction which scales as $1/k^2$ in the local template for small $k$.    }

\item{Numerical calculations using the LPTR confirm that the ionized density bias in the small-$k$ limit ($k<0.1~\Mpc^{-1}$) is scale-independent in models with Gaussian initial conditions, and strongly scale-dependent for local PNG, reflecting the impact of PNG on the source bias.  Moreover, the imprint of the source bias on the ionized density bias persists throughout the reionization epoch.  We attribute this characteristic of our models to the mean free path of UV photons through the IGM, which remains small relative to the scales on which PNG affects the halo bias, until just before reionization ends.   }

\item{While the simple one-parameter ESMR model developed in this work cannot capture the detailed evolution of even the Gaussian ionized density bias computed in the LPTR, we found that equations (\ref{EQ:bcfPNG_SI}) and (\ref{EQ:SDcorrection}) provide a reasonably accurate map between Gaussian and non-Gaussian ionized density bias parameters.  As shown in Figure \ref{FIG:HybridCompare}, these formulae work best at earlier times, corresponding to lower mean ionized fractions, and in models with lower recombination rates.}

\item{Equation (\ref{EQ:hatbx}) gives our prediction for the ``bias of ionized" regions defined by JDFKS, which is related to the ionized density bias used in this work by equation (\ref{EQ:bxtransformation}).  This result can be compared directly to the JDFKS fitting formula given in equation (\ref{EQ:JoudakiBias}), which they used to forecast constraints on $\fNL$ by future measurements of the EoR 21cm power spectrum.  We note significant differences in the forms of these expressions.   In a companion paper, we use our result to revisit the topic of constraining PNG with the EoR 21cm power spectrum  \citep{2013arXiv1305.0313M}.}
 
 \end{itemize}

There are several scenarios one could imagine which might require extensions to both our analytical ESMR predictions and our LPTR calculations.  For example, all of our models assumed that atomic cooling halos were the only photon-sources during reionization.  These were not the first halos to form stars, however, and there was likely a modest contribution to reionization from minihalos at earlier times before feedback mechanisms effectively shut them down \citep[see e.g.][]{2012ApJ...756L..16A}.  We also did not take into account the possible ``self-regulation" effects of reionization on the star formation rates of lower mass atomic cooling halos, below the Jeans filtering scale, $M \sim 10^{9}\Msun$.  These feedback effects can lead to a more complicated reionization history and possibly a more complicated evolution and scale dependence of the ionized density bias \cite[for studies on the role of feedback during reionization, see e.g.][]{1994ApJ...427...25S,2000ApJ...534..507C,2003ApJ...595....1H,2003ApJ...588L..69W,2004ApJ...610....1O,2005ApJ...634....1F,2006ApJ...649..570K,2007MNRAS.376..534I,2007MNRAS.379.1647W,2007ApJ...659..890W,2008MNRAS.390.1071M,2011MNRAS.412.2781K,2012ApJ...747..126A,2012ApJ...756L..16A,2013arXiv1301.6781S}.

We have also assumed in the case with no feedback that the bias of ionizing sources is equal to the halo bias.  This assumption is often adopted in the literature, but its accuracy is uncertain.  One situation in which this assumption might not hold is if mergers play a significant role in driving star formation during reionization (see e.g. \citet{2013arXiv1302.1363L} for recent observations at lower redshifts).  In this case, the source bias might be quite different from the halo bias, since the bias of merging halos is different from the general halo population.  This difference would likely be reflected in the ionized density bias, and it is not clear whether our current ESMR expressions would apply in such a scenario [see \citet{2007MNRAS.374...72C} for an extension of the Gaussian ESMR to include the effects of halo mergers].  Scenarios in which the source bias is not simply related to the halo bias should be explored to gauge their impact on the ionized density bias.  

We have considered a range of reionization models in this work, but there is an important characteristic in common with all of these models.  The mean free path of typical photons through the IGM is short relative to the large scales of interest, until the end of reionization.  This characteristic is a consequence of the soft source spectrum we adopted.  As noted above, this feature is crucial in preserving the imprint of the source bias on the large-scale ionized density bias.  It is therefore important to consider scenarios in which this condition might not hold.  One such scenario is that of a hard source spectrum, i.e. a significant contribution from X-ray photons, which have a significantly larger mean free path relative to UV photons, and may therefore act to suppress the amplitude of the ionized density bias on larger scales, approaching the scales of interest for the effects of PNG.  Further numerical work should be devoted towards exploring more detailed reionization models and their implications for the analytical results presented here.  

Finally, we have discussed one observable effect of the scale-dependent ionized density bias: a scale-dependent signature in the EoR 21cm power spectrum.  However, the scale-dependent ionized density bias can also have an impact on secondary temperature and polarization anisotropies in the CMB sourced by the EoR.  Extensions of this work may also involve observational signatures beyond the redshifted 21cm background.

\emph{Note:}  during the preparation of this manuscript, a paper appeared on the ArXiv preprint archive on the topic of constraining PNG with the EoR 21cm power spectrum \citep{2013arXiv1303.4387C}.  \citet{2013arXiv1303.4387C} calculated the ionized fraction bias (and not the ionized density bias considered in our work) based on the reionization model of \citet{2006ApJ...647..840A}.  We note that the approach of \citet{2013arXiv1303.4387C} does not yield an analytical expression for the non-Gaussian ionized fraction bias, in contrast to the ESMR approach taken here.  Shortly after this manuscript was submitted, a paper by \citet{2013arXiv1304.8049L} appeared on the archive which expanded upon the work of JDFKS with more-detailed semi-numerical simulations of reionization with PNG.  They also considered the effects of foreground subtraction in their forecasts of constraints on $\fNL$ from the EoR 21cm power spectrum.

\section*{Acknowledgments}
A.D. thanks Vincent Desjacques, Donghui Jeong, and Eiichiro Komatsu for useful discussions, and Steve Furlanetto for helpful comments on this manuscript.  The authors also thank Shahab Joudaki and Mario Santos for additional information on the work in JDFKS, and the referee of this manuscript for helpful comments. This work was supported in part by U.S. NSF grants AST-0708176 and AST-1009799, and NASA grants NNX07AH09G and NNX11AE09G.  YM was supported by the French state funds managed by the ANR within the Investissements d'Avenir programme under reference ANR-11-IDEX-0004-02.   

\bibliographystyle{mn2e}
\bibliography{HIIbias}

\begin{thebibliography}{}

\bibitem[\protect\citeauthoryear{{Acquaviva}, {Bartolo}, {Matarrese} \&
  {Riotto}}{{Acquaviva} et~al.}{2003}]{2003NuPhB.667..119A}
{Acquaviva} V.,  {Bartolo} N.,  {Matarrese} S.,    {Riotto} A.,  2003, Nuclear
  Physics B, 667, 119

\bibitem[\protect\citeauthoryear{{Ade}, {Aghanim}, {Armitage-Caplan}, {Arnaud},
  {Ashdown}, {Atrio-Barandela}, {Aumont}, {Baccigalupi}, {Banday} \& et
  al.}{{Ade} et~al.}{2013}]{2013arXiv1303.5084P}
{Ade} P.~A.~R.,  {Aghanim} N.,  {Armitage-Caplan} C.,  {Arnaud} M.,  {Ashdown}
  M.,  {Atrio-Barandela} F.,  {Aumont} J.,  {Baccigalupi} C.,  {Banday} A.~J.,
    et al. 2013, ArXiv e-prints

\bibitem[\protect\citeauthoryear{{Adshead}, {Baxter}, {Dodelson} \&
  {Lidz}}{{Adshead} et~al.}{2012}]{2012PhRvD..86f3526A}
{Adshead} P.,  {Baxter} E.~J.,  {Dodelson} S.,    {Lidz} A.,  2012, \prd, 86,
  063526

\bibitem[\protect\citeauthoryear{{Afshordi} \& {Tolley}}{{Afshordi} \&
  {Tolley}}{2008}]{2008PhRvD..78l3507A}
{Afshordi} N.,  {Tolley} A.~J.,  2008, \prd, 78, 123507

\bibitem[\protect\citeauthoryear{{Agullo} \& {Parker}}{{Agullo} \&
  {Parker}}{2011}]{2011PhRvD..83f3526A}
{Agullo} I.,  {Parker} L.,  2011, \prd, 83, 063526

\bibitem[\protect\citeauthoryear{{Agullo} \& {Shandera}}{{Agullo} \&
  {Shandera}}{2012}]{2012JCAP...09..007A}
{Agullo} I.,  {Shandera} S.,  2012, \jcap, 9, 7

\bibitem[\protect\citeauthoryear{{Ahn}, {Iliev}, {Shapiro}, {Mellema}, {Koda}
  \& {Mao}}{{Ahn} et~al.}{2012}]{2012ApJ...756L..16A}
{Ahn} K.,  {Iliev} I.~T.,  {Shapiro} P.~R.,  {Mellema} G.,  {Koda} J.,    {Mao}
  Y.,  2012, \apjl, 756, L16

\bibitem[\protect\citeauthoryear{{Alvarez} \& {Abel}}{{Alvarez} \&
  {Abel}}{2012}]{2012ApJ...747..126A}
{Alvarez} M.~A.,  {Abel} T.,  2012, \apj, 747, 126

\bibitem[\protect\citeauthoryear{{Alvarez}, {Komatsu}, {Dor{\'e}} \&
  {Shapiro}}{{Alvarez} et~al.}{2006}]{2006ApJ...647..840A}
{Alvarez} M.~A.,  {Komatsu} E.,  {Dor{\'e}} O.,    {Shapiro} P.~R.,  2006,
  \apj, 647, 840

\bibitem[\protect\citeauthoryear{{Aubert} \& {Teyssier}}{{Aubert} \&
  {Teyssier}}{2010}]{2010ApJ...724..244A}
{Aubert} D.,  {Teyssier} R.,  2010, \apj, 724, 244

\bibitem[\protect\citeauthoryear{{Babich}, {Creminelli} \&
  {Zaldarriaga}}{{Babich} et~al.}{2004}]{2004JCAP...08..009B}
{Babich} D.,  {Creminelli} P.,    {Zaldarriaga} M.,  2004, \jcap, 8, 9

\bibitem[\protect\citeauthoryear{{Babich} \& {Zaldarriaga}}{{Babich} \&
  {Zaldarriaga}}{2004}]{2004PhRvD..70h3005B}
{Babich} D.,  {Zaldarriaga} M.,  2004, \prd, 70, 083005

\bibitem[\protect\citeauthoryear{{Baldauf}, {Seljak}, {Senatore} \&
  {Zaldarriaga}}{{Baldauf} et~al.}{2011}]{2011JCAP...10..031B}
{Baldauf} T.,  {Seljak} U.,  {Senatore} L.,    {Zaldarriaga} M.,  2011, \jcap,
  10, 31

\bibitem[\protect\citeauthoryear{{Barkana} \& {Loeb}}{{Barkana} \&
  {Loeb}}{2005}]{2005ApJ...624L..65B}
{Barkana} R.,  {Loeb} A.,  2005, \apjl, 624, L65

\bibitem[\protect\citeauthoryear{{Bartolo}, {Matarrese} \& {Riotto}}{{Bartolo}
  et~al.}{2011}]{2011JCAP...04..011B}
{Bartolo} N.,  {Matarrese} S.,    {Riotto} A.,  2011, \jcap, 4, 11

\bibitem[\protect\citeauthoryear{{Bennett}, {Larson}, {Weiland}, {Jarosik},
  {Hinshaw}, {Odegard}, {Smith}, {Hill} \& {et al.}}{{Bennett}
  et~al.}{2012}]{2012arXiv1212.5225B}
{Bennett} C.~L.,  {Larson} D.,  {Weiland} J.~L.,  {Jarosik} N.,  {Hinshaw} G.,
  {Odegard} N.,  {Smith} K.~M.,  {Hill} R.~S.,    {et al.} 2012, ArXiv e-prints

\bibitem[\protect\citeauthoryear{{Bond}, {Cole}, {Efstathiou} \&
  {Kaiser}}{{Bond} et~al.}{1991}]{Bond:1991sf}
{Bond} J.~R.,  {Cole} S.,  {Efstathiou} G.,    {Kaiser} N.,  1991, \apj, 379,
  440

\bibitem[\protect\citeauthoryear{{Bonvin} \& {Durrer}}{{Bonvin} \&
  {Durrer}}{2011}]{2011PhRvD..84f3505B}
{Bonvin} C.,  {Durrer} R.,  2011, \prd, 84, 063505

\bibitem[\protect\citeauthoryear{{Bruni}, {Crittenden}, {Koyama}, {Maartens},
  {Pitrou} \& {Wands}}{{Bruni} et~al.}{2012}]{2012PhRvD..85d1301B}
{Bruni} M.,  {Crittenden} R.,  {Koyama} K.,  {Maartens} R.,  {Pitrou} C.,
  {Wands} D.,  2012, \prd, 85, 041301

\bibitem[\protect\citeauthoryear{{Carney}, {Fischler}, {Paban} \&
  {Sivanandam}}{{Carney} et~al.}{2012}]{2012JCAP...12..012C}
{Carney} D.,  {Fischler} W.,  {Paban} S.,    {Sivanandam} N.,  2012, \jcap, 12,
  12

\bibitem[\protect\citeauthoryear{{Challinor} \& {Lewis}}{{Challinor} \&
  {Lewis}}{2011}]{2011PhRvD..84d3516C}
{Challinor} A.,  {Lewis} A.,  2011, \prd, 84, 043516

\bibitem[\protect\citeauthoryear{{Chen}, {Firouzjahi}, {Namjoo} \&
  {Sasaki}}{{Chen} et~al.}{2013}]{2013arXiv1301.5699C}
{Chen} X.,  {Firouzjahi} H.,  {Namjoo} M.~H.,    {Sasaki} M.,  2013, ArXiv
  e-prints

\bibitem[\protect\citeauthoryear{{Chen}, {Huang}, {Kachru} \& {Shiu}}{{Chen}
  et~al.}{2007}]{2007JCAP...01..002C}
{Chen} X.,  {Huang} M.-x.,  {Kachru} S.,    {Shiu} G.,  2007, \jcap, 1, 2

\bibitem[\protect\citeauthoryear{{Cheung}, {Fitzpatrick}, {Kaplan} \&
  {Senatore}}{{Cheung} et~al.}{2008}]{2008JCAP...02..021C}
{Cheung} C.,  {Fitzpatrick} A.~L.,  {Kaplan} J.,    {Senatore} L.,  2008,
  \jcap, 2, 21

\bibitem[\protect\citeauthoryear{{Chiu} \& {Ostriker}}{{Chiu} \&
  {Ostriker}}{2000}]{2000ApJ...534..507C}
{Chiu} W.~A.,  {Ostriker} J.~P.,  2000, \apj, 534, 507

\bibitem[\protect\citeauthoryear{{Chongchitnan}}{{Chongchitnan}}{2013}]{2013ar%
Xiv1303.4387C}
{Chongchitnan} S.,  2013, ArXiv e-prints

\bibitem[\protect\citeauthoryear{{Chongchitnan} \& {Silk}}{{Chongchitnan} \&
  {Silk}}{2012}]{2012MNRAS.426L..21C}
{Chongchitnan} S.,  {Silk} J.,  2012, \mnras, 426, L21

\bibitem[\protect\citeauthoryear{{Cohn} \& {Chang}}{{Cohn} \&
  {Chang}}{2007}]{2007MNRAS.374...72C}
{Cohn} J.~D.,  {Chang} T.-C.,  2007, \mnras, 374, 72

\bibitem[\protect\citeauthoryear{{Cooray} \& {Sheth}}{{Cooray} \&
  {Sheth}}{2002}]{2002PhR...372....1C}
{Cooray} A.,  {Sheth} R.,  2002, \physrep, 372, 1

\bibitem[\protect\citeauthoryear{{Creminelli} \& {Zaldarriaga}}{{Creminelli} \&
  {Zaldarriaga}}{2004}]{2004JCAP...10..006C}
{Creminelli} P.,  {Zaldarriaga} M.,  2004, \jcap, 10, 6

\bibitem[\protect\citeauthoryear{{Crociani}, {Moscardini}, {Viel} \&
  {Matarrese}}{{Crociani} et~al.}{2009}]{2009MNRAS.394..133C}
{Crociani} D.,  {Moscardini} L.,  {Viel} M.,    {Matarrese} S.,  2009, \mnras,
  394, 133

\bibitem[\protect\citeauthoryear{{Dalal}, {Dor{\'e}}, {Huterer} \&
  {Shirokov}}{{Dalal} et~al.}{2008}]{2008PhRvD..77l3514D}
{Dalal} N.,  {Dor{\'e}} O.,  {Huterer} D.,    {Shirokov} A.,  2008, \prd, 77,
  123514

\bibitem[\protect\citeauthoryear{{D'Aloisio}, {Zhang}, {Jeong} \&
  {Shapiro}}{{D'Aloisio} et~al.}{2012}]{2012arXiv1206.3305D}
{D'Aloisio} A.,  {Zhang} J.,  {Jeong} D.,    {Shapiro} P.~R.,  2012, ArXiv
  e-prints

\bibitem[\protect\citeauthoryear{{D'Amico}, {Musso}, {Nore{\~n}a} \&
  {Paranjape}}{{D'Amico} et~al.}{2011}]{2011PhRvD..83b3521D}
{D'Amico} G.,  {Musso} M.,  {Nore{\~n}a} J.,    {Paranjape} A.,  2011, \prd,
  83, 023521

\bibitem[\protect\citeauthoryear{{Desjacques}, {Jeong} \&
  {Schmidt}}{{Desjacques} et~al.}{2011a}]{2011PhRvD..84f1301D}
{Desjacques} V.,  {Jeong} D.,    {Schmidt} F.,  2011a, \prd, 84, 061301

\bibitem[\protect\citeauthoryear{{Desjacques}, {Jeong} \&
  {Schmidt}}{{Desjacques} et~al.}{2011b}]{2011PhRvD..84f3512D}
{Desjacques} V.,  {Jeong} D.,    {Schmidt} F.,  2011b, \prd, 84, 063512

\bibitem[\protect\citeauthoryear{{Desjacques} \& {Seljak}}{{Desjacques} \&
  {Seljak}}{2010a}]{2010CQGra..27l4011D}
{Desjacques} V.,  {Seljak} U.,  2010a, Classical and Quantum Gravity, 27,
  124011

\bibitem[\protect\citeauthoryear{{Desjacques} \& {Seljak}}{{Desjacques} \&
  {Seljak}}{2010b}]{2010PhRvD..81b3006D}
{Desjacques} V.,  {Seljak} U.,  2010b, \prd, 81, 023006

\bibitem[\protect\citeauthoryear{{Dey}, {Kovetz} \& {Paban}}{{Dey}
  et~al.}{2012}]{2012JCAP...10..055D}
{Dey} A.,  {Kovetz} E.,    {Paban} S.,  2012, \jcap, 10, 55

\bibitem[\protect\citeauthoryear{{Dey} \& {Paban}}{{Dey} \&
  {Paban}}{2012}]{2012JCAP...04..039D}
{Dey} A.,  {Paban} S.,  2012, \jcap, 4, 39

\bibitem[\protect\citeauthoryear{{Eisenstein} \& {Hu}}{{Eisenstein} \&
  {Hu}}{1999}]{1999ApJ...511....5E}
{Eisenstein} D.~J.,  {Hu} W.,  1999, \apj, 511, 5

\bibitem[\protect\citeauthoryear{{Finlator}, {Oh}, {{\"O}zel} \&
  {Dav{\'e}}}{{Finlator} et~al.}{2012}]{2012arXiv1209.2489F}
{Finlator} K.,  {Oh} S.~P.,  {{\"O}zel} F.,    {Dav{\'e}} R.,  2012, ArXiv
  e-prints

\bibitem[\protect\citeauthoryear{{Furlanetto} \& {Loeb}}{{Furlanetto} \&
  {Loeb}}{2005}]{2005ApJ...634....1F}
{Furlanetto} S.~R.,  {Loeb} A.,  2005, \apj, 634, 1

\bibitem[\protect\citeauthoryear{{Furlanetto}, {Oh} \& {Briggs}}{{Furlanetto}
  et~al.}{2006}]{2006PhR...433..181F}
{Furlanetto} S.~R.,  {Oh} S.~P.,    {Briggs} F.~H.,  2006, \physrep, 433, 181

\bibitem[\protect\citeauthoryear{{Furlanetto}, {Zaldarriaga} \&
  {Hernquist}}{{Furlanetto} et~al.}{2004}]{2004ApJ...613....1F}
{Furlanetto} S.~R.,  {Zaldarriaga} M.,    {Hernquist} L.,  2004, \apj, 613, 1

\bibitem[\protect\citeauthoryear{{Ganc}}{{Ganc}}{2011}]{2011PhRvD..84f3514G}
{Ganc} J.,  2011, \prd, 84, 063514

\bibitem[\protect\citeauthoryear{{Ganc} \& {Komatsu}}{{Ganc} \&
  {Komatsu}}{2012}]{2012PhRvD..86b3518G}
{Ganc} J.,  {Komatsu} E.,  2012, \prd, 86, 023518

\bibitem[\protect\citeauthoryear{{Gangui}, {Lucchin}, {Matarrese} \&
  {Mollerach}}{{Gangui} et~al.}{1994}]{1994ApJ...430..447G}
{Gangui} A.,  {Lucchin} F.,  {Matarrese} S.,    {Mollerach} S.,  1994, \apj,
  430, 447

\bibitem[\protect\citeauthoryear{{Giannantonio}, {Ross}, {Percival},
  {Crittenden}, {Bacher}, {Kilbinger}, {Nichol} \& {Weller}}{{Giannantonio}
  et~al.}{2013}]{2013arXiv1303.1349G}
{Giannantonio} T.,  {Ross} A.~J.,  {Percival} W.~J.,  {Crittenden} R.,
  {Bacher} D.,  {Kilbinger} M.,  {Nichol} R.,    {Weller} J.,  2013, ArXiv
  e-prints

\bibitem[\protect\citeauthoryear{{Grossi}, {Dolag}, {Branchini}, {Matarrese} \&
  {Moscardini}}{{Grossi} et~al.}{2007}]{2007MNRAS.382.1261G}
{Grossi} M.,  {Dolag} K.,  {Branchini} E.,  {Matarrese} S.,    {Moscardini} L.,
   2007, \mnras, 382, 1261

\bibitem[\protect\citeauthoryear{{Grossi}, {Verde}, {Carbone}, {Dolag},
  {Branchini}, {Iannuzzi}, {Matarrese} \& {Moscardini}}{{Grossi}
  et~al.}{2009}]{2009MNRAS.398..321G}
{Grossi} M.,  {Verde} L.,  {Carbone} C.,  {Dolag} K.,  {Branchini} E.,
  {Iannuzzi} F.,  {Matarrese} S.,    {Moscardini} L.,  2009, \mnras, 398, 321

\bibitem[\protect\citeauthoryear{{Haiman}, {Abel} \& {Rees}}{{Haiman}
  et~al.}{2000}]{2000ApJ...534...11H}
{Haiman} Z.,  {Abel} T.,    {Rees} M.~J.,  2000, \apj, 534, 11

\bibitem[\protect\citeauthoryear{{Haiman} \& {Holder}}{{Haiman} \&
  {Holder}}{2003}]{2003ApJ...595....1H}
{Haiman} Z.,  {Holder} G.~P.,  2003, \apj, 595, 1

\bibitem[\protect\citeauthoryear{{Hodges}, {Blumenthal}, {Kofman} \&
  {Primack}}{{Hodges} et~al.}{1990}]{1990NuPhB.335..197H}
{Hodges} H.~M.,  {Blumenthal} G.~R.,  {Kofman} L.~A.,    {Primack} J.~R.,
  1990, Nuclear Physics B, 335, 197

\bibitem[\protect\citeauthoryear{{Iliev}, {Mellema}, {Shapiro} \&
  {Pen}}{{Iliev} et~al.}{2007}]{2007MNRAS.376..534I}
{Iliev} I.~T.,  {Mellema} G.,  {Shapiro} P.~R.,    {Pen} U.-L.,  2007, \mnras,
  376, 534

\bibitem[\protect\citeauthoryear{{Jeong}, {Schmidt} \& {Hirata}}{{Jeong}
  et~al.}{2012}]{2012PhRvD..85b3504J}
{Jeong} D.,  {Schmidt} F.,    {Hirata} C.~M.,  2012, \prd, 85, 023504

\bibitem[\protect\citeauthoryear{{Joudaki}}{{Joudaki}}{2013}]{JoudakiEmail}
{Joudaki} S.,  2013, {private communication}

\bibitem[\protect\citeauthoryear{{Joudaki}, {Dor{\'e}}, {Ferramacho},
  {Kaplinghat} \& {Santos}}{{Joudaki} et~al.}{2011}]{2011PhRvL.107m1304J}
{Joudaki} S.,  {Dor{\'e}} O.,  {Ferramacho} L.,  {Kaplinghat} M.,    {Santos}
  M.~G.,  2011, Physical Review Letters, 107, 131304

\bibitem[\protect\citeauthoryear{{Kamionkowski}, {Verde} \&
  {Jimenez}}{{Kamionkowski} et~al.}{2009}]{2009JCAP...01..010K}
{Kamionkowski} M.,  {Verde} L.,    {Jimenez} R.,  2009, \jcap, 1, 10

\bibitem[\protect\citeauthoryear{{Kofman}, {Blumenthal}, {Hodges} \&
  {Primack}}{{Kofman} et~al.}{1991}]{1991ASPC...15..339K}
{Kofman} L.,  {Blumenthal} G.~R.,  {Hodges} H.,    {Primack} J.~R.,  1991, in
  {Latham} D.~W.,  {da Costa} L.~A.~N.,  eds, Large-scale Structures and
  Peculiar Motions in the Universe Vol.~15 of Astronomical Society of the
  Pacific Conference Series, {Generation of Non-Flat and Non-Gaussian
  Perturbations from Inflation}.
p.~339

\bibitem[\protect\citeauthoryear{{Kohler}, {Gnedin} \& {Hamilton}}{{Kohler}
  et~al.}{2007}]{2007ApJ...657...15K}
{Kohler} K.,  {Gnedin} N.~Y.,    {Hamilton} A.~J.~S.,  2007, \apj, 657, 15

\bibitem[\protect\citeauthoryear{{Komatsu}, {Smith}, {Dunkley}, {Bennett},
  {Gold}, {Hinshaw}, {Jarosik}, {Larson} \& {et al.}}{{Komatsu}
  et~al.}{2011}]{2011ApJS..192...18K}
{Komatsu} E.,  {Smith} K.~M.,  {Dunkley} J.,  {Bennett} C.~L.,  {Gold} B.,
  {Hinshaw} G.,  {Jarosik} N.,  {Larson} D.,    {et al.} 2011, \apjs, 192, 18

\bibitem[\protect\citeauthoryear{{Komatsu} \& {Spergel}}{{Komatsu} \&
  {Spergel}}{2001}]{2001PhRvD..63f3002K}
{Komatsu} E.,  {Spergel} D.~N.,  2001, \prd, 63, 063002

\bibitem[\protect\citeauthoryear{{Kramer}, {Haiman} \& {Oh}}{{Kramer}
  et~al.}{2006}]{2006ApJ...649..570K}
{Kramer} R.~H.,  {Haiman} Z.,    {Oh} S.~P.,  2006, \apj, 649, 570

\bibitem[\protect\citeauthoryear{{Kulkarni} \& {Choudhury}}{{Kulkarni} \&
  {Choudhury}}{2011}]{2011MNRAS.412.2781K}
{Kulkarni} G.,  {Choudhury} T.~R.,  2011, \mnras, 412, 2781

\bibitem[\protect\citeauthoryear{{Lacey} \& {Cole}}{{Lacey} \&
  {Cole}}{1993}]{1993MNRAS.262..627L}
{Lacey} C.,  {Cole} S.,  1993, \mnras, 262, 627

\bibitem[\protect\citeauthoryear{{Lamastra}, {Menci}, {Fiore} \&
  {Santini}}{{Lamastra} et~al.}{2013}]{2013arXiv1302.1363L}
{Lamastra} A.,  {Menci} N.,  {Fiore} F.,    {Santini} P.,  2013, ArXiv e-prints

\bibitem[\protect\citeauthoryear{{Lidz}, {Baxter}, {Adshead} \&
  {Dodelson}}{{Lidz} et~al.}{2013}]{2013arXiv1304.8049L}
{Lidz} A.,  {Baxter} E.~J.,  {Adshead} P.,    {Dodelson} S.,  2013, ArXiv
  e-prints

\bibitem[\protect\citeauthoryear{{Lo Verde}, {Miller}, {Shandera} \&
  {Verde}}{{Lo Verde} et~al.}{2008}]{Lo-Verde:2008rt}
{Lo Verde} M.,  {Miller} A.,  {Shandera} S.,    {Verde} L.,  2008, Journal of
  Cosmology and Astro-Particle Physics, 4, 14

\bibitem[\protect\citeauthoryear{{LoVerde} \& {Smith}}{{LoVerde} \&
  {Smith}}{2011}]{2011JCAP...08..003L}
{LoVerde} M.,  {Smith} K.~M.,  2011, \jcap, 8, 3

\bibitem[\protect\citeauthoryear{{Maggiore} \& {Riotto}}{{Maggiore} \&
  {Riotto}}{2010a}]{2010ApJ...711..907M}
{Maggiore} M.,  {Riotto} A.,  2010a, \apj, 711, 907

\bibitem[\protect\citeauthoryear{{Maggiore} \& {Riotto}}{{Maggiore} \&
  {Riotto}}{2010b}]{2010ApJ...717..515M}
{Maggiore} M.,  {Riotto} A.,  2010b, \apj, 717, 515

\bibitem[\protect\citeauthoryear{{Maggiore} \& {Riotto}}{{Maggiore} \&
  {Riotto}}{2010c}]{2010ApJ...717..526M}
{Maggiore} M.,  {Riotto} A.,  2010c, \apj, 717, 526

\bibitem[\protect\citeauthoryear{{Maldacena}}{{Maldacena}}{2003}]{2003JHEP...0%
5..013M}
{Maldacena} J.,  2003, Journal of High Energy Physics, 5, 13

\bibitem[\protect\citeauthoryear{{Mao}, {D'Aloisio}, {Zhang} \&
  {Shapiro}}{{Mao} et~al.}{2013}]{2013arXiv1305.0313M}
{Mao} Y.,  {D'Aloisio} A.,  {Zhang} J.,    {Shapiro} P.~R.,  2013, ArXiv
  e-prints

\bibitem[\protect\citeauthoryear{{Mao}, {Tegmark}, {McQuinn}, {Zaldarriaga} \&
  {Zahn}}{{Mao} et~al.}{2008}]{2008PhRvD..78b3529M}
{Mao} Y.,  {Tegmark} M.,  {McQuinn} M.,  {Zaldarriaga} M.,    {Zahn} O.,  2008,
  \prd, 78, 023529

\bibitem[\protect\citeauthoryear{{Matarrese} \& {Verde}}{{Matarrese} \&
  {Verde}}{2008}]{2008ApJ...677L..77M}
{Matarrese} S.,  {Verde} L.,  2008, \apjl, 677, L77

\bibitem[\protect\citeauthoryear{{Matarrese}, {Verde} \& {Jimenez}}{{Matarrese}
  et~al.}{2000}]{Matarrese:2000pb}
{Matarrese} S.,  {Verde} L.,    {Jimenez} R.,  2000, \apj, 541, 10

\bibitem[\protect\citeauthoryear{{McQuinn}, {Furlanetto}, {Hernquist}, {Zahn}
  \& {Zaldarriaga}}{{McQuinn} et~al.}{2005}]{2005ApJ...630..643M}
{McQuinn} M.,  {Furlanetto} S.~R.,  {Hernquist} L.,  {Zahn} O.,
  {Zaldarriaga} M.,  2005, \apj, 630, 643

\bibitem[\protect\citeauthoryear{{McQuinn}, {Oh} \&
  {Faucher-Gigu{\`e}re}}{{McQuinn} et~al.}{2011}]{2011ApJ...743...82M}
{McQuinn} M.,  {Oh} S.~P.,    {Faucher-Gigu{\`e}re} C.-A.,  2011, \apj, 743, 82

\bibitem[\protect\citeauthoryear{{Mesinger} \& {Dijkstra}}{{Mesinger} \&
  {Dijkstra}}{2008}]{2008MNRAS.390.1071M}
{Mesinger} A.,  {Dijkstra} M.,  2008, \mnras, 390, 1071

\bibitem[\protect\citeauthoryear{{Mo} \& {White}}{{Mo} \&
  {White}}{1996}]{1996MNRAS.282..347M}
{Mo} H.~J.,  {White} S.~D.~M.,  1996, \mnras, 282, 347

\bibitem[\protect\citeauthoryear{{Musso} \& {Paranjape}}{{Musso} \&
  {Paranjape}}{2012}]{2012MNRAS.420..369M}
{Musso} M.,  {Paranjape} A.,  2012, \mnras, 420, 369

\bibitem[\protect\citeauthoryear{{Musso}, {Paranjape} \& {Sheth}}{{Musso}
  et~al.}{2012}]{2012arXiv1205.3401M}
{Musso} M.,  {Paranjape} A.,    {Sheth} R.~K.,  2012, ArXiv e-prints

\bibitem[\protect\citeauthoryear{{Musso} \& {Sheth}}{{Musso} \&
  {Sheth}}{2012}]{2012MNRAS.423L.102M}
{Musso} M.,  {Sheth} R.~K.,  2012, \mnras, 423, L102

\bibitem[\protect\citeauthoryear{{Namjoo}, {Firouzjahi} \& {Sasaki}}{{Namjoo}
  et~al.}{2013}]{2013EL....10139001N}
{Namjoo} M.~H.,  {Firouzjahi} H.,    {Sasaki} M.,  2013, EPL (Europhysics
  Letters), 101, 39001

\bibitem[\protect\citeauthoryear{{Onken} \& {Miralda-Escud{\'e}}}{{Onken} \&
  {Miralda-Escud{\'e}}}{2004}]{2004ApJ...610....1O}
{Onken} C.~A.,  {Miralda-Escud{\'e}} J.,  2004, \apj, 610, 1

\bibitem[\protect\citeauthoryear{{Paranjape}, {Lam} \& {Sheth}}{{Paranjape}
  et~al.}{2012}]{2012MNRAS.420.1429P}
{Paranjape} A.,  {Lam} T.~Y.,    {Sheth} R.~K.,  2012, \mnras, 420, 1429

\bibitem[\protect\citeauthoryear{{Paranjape} \& {Sheth}}{{Paranjape} \&
  {Sheth}}{2012}]{2012MNRAS.419..132P}
{Paranjape} A.,  {Sheth} R.~K.,  2012, \mnras, 419, 132

\bibitem[\protect\citeauthoryear{{Pawlik}, {Schaye} \& {van
  Scherpenzeel}}{{Pawlik} et~al.}{2009}]{2009MNRAS.394.1812P}
{Pawlik} A.~H.,  {Schaye} J.,    {van Scherpenzeel} E.,  2009, \mnras, 394,
  1812

\bibitem[\protect\citeauthoryear{{Robertson}, {Furlanetto}, {Schneider},
  {Charlot}, {Ellis}, {Stark}, {McLure}, {Dunlop}, {Koekemoer}, {Schenker},
  {Ouchi}, {Ono}, {Curtis-Lake}, {Rogers}, {Bowler} \& {Cirasuolo}}{{Robertson}
  et~al.}{2013}]{2013arXiv1301.1228R}
{Robertson} B.~E.,  {Furlanetto} S.~R.,  {Schneider} E.,  {Charlot} S.,
  {Ellis} R.~S.,  {Stark} D.~P.,  {McLure} R.~J.,  {Dunlop} J.~S.,  {Koekemoer}
  A.,  {Schenker} M.~A.,  {Ouchi} M.,  {Ono} Y.,  {Curtis-Lake} E.,  {Rogers}
  A.~B.,  {Bowler} R.~A.~A.,    {Cirasuolo} M.,  2013, ArXiv e-prints

\bibitem[\protect\citeauthoryear{{Ross}, {Percival}, {Carnero}, {Zhao},
  {Manera}, {Raccanelli}, {Aubourg}, {Bizyaev} \& {et. al.}}{{Ross}
  et~al.}{2013}]{2013MNRAS.428.1116R}
{Ross} A.~J.,  {Percival} W.~J.,  {Carnero} A.,  {Zhao} G.-b.,  {Manera} M.,
  {Raccanelli} A.,  {Aubourg} E.,  {Bizyaev} D.,    {et. al.} 2013, \mnras,
  428, 1116

\bibitem[\protect\citeauthoryear{{Salopek} \& {Bond}}{{Salopek} \&
  {Bond}}{1990}]{1990PhRvD..42.3936S}
{Salopek} D.~S.,  {Bond} J.~R.,  1990, \prd, 42, 3936

\bibitem[\protect\citeauthoryear{{Santos}, {Ferramacho}, {Silva}, {Amblard} \&
  {Cooray}}{{Santos} et~al.}{2010}]{2010MNRAS.406.2421S}
{Santos} M.~G.,  {Ferramacho} L.,  {Silva} M.~B.,  {Amblard} A.,    {Cooray}
  A.,  2010, \mnras, 406, 2421

\bibitem[\protect\citeauthoryear{{Seery} \& {Lidsey}}{{Seery} \&
  {Lidsey}}{2005}]{2005JCAP...06..003S}
{Seery} D.,  {Lidsey} J.~E.,  2005, \jcap, 6, 3

\bibitem[\protect\citeauthoryear{{Shapiro}, {Giroux} \& {Babul}}{{Shapiro}
  et~al.}{1994}]{1994ApJ...427...25S}
{Shapiro} P.~R.,  {Giroux} M.~L.,    {Babul} A.,  1994, \apj, 427, 25

\bibitem[\protect\citeauthoryear{{Shapiro}, {Iliev} \& {Raga}}{{Shapiro}
  et~al.}{2004}]{2004MNRAS.348..753S}
{Shapiro} P.~R.,  {Iliev} I.~T.,    {Raga} A.~C.,  2004, \mnras, 348, 753

\bibitem[\protect\citeauthoryear{{Shapiro}, {Mao}, {Iliev}, {Mellema}, {Datta},
  {Ahn} \& {Koda}}{{Shapiro} et~al.}{2013}]{2012arXiv1211.2036S}
{Shapiro} P.~R.,  {Mao} Y.,  {Iliev} I.~T.,  {Mellema} G.,  {Datta} K.~K.,
  {Ahn} K.,    {Koda} J.,  2013, PRL, 110, 151301

\bibitem[\protect\citeauthoryear{{Shull}, {Harness}, {Trenti} \&
  {Smith}}{{Shull} et~al.}{2012}]{2012ApJ...747..100S}
{Shull} J.~M.,  {Harness} A.,  {Trenti} M.,    {Smith} B.~D.,  2012, \apj, 747,
  100

\bibitem[\protect\citeauthoryear{{Slosar}, {Hirata}, {Seljak}, {Ho} \&
  {Padmanabhan}}{{Slosar} et~al.}{2008}]{2008JCAP...08..031S}
{Slosar} A.,  {Hirata} C.,  {Seljak} U.,  {Ho} S.,    {Padmanabhan} N.,  2008,
  \jcap, 8, 31

\bibitem[\protect\citeauthoryear{{Smith}, {Ferraro} \& {LoVerde}}{{Smith}
  et~al.}{2012}]{2012JCAP...03..032S}
{Smith} K.~M.,  {Ferraro} S.,    {LoVerde} M.,  2012, \jcap, 3, 32

\bibitem[\protect\citeauthoryear{{Sobacchi} \& {Mesinger}}{{Sobacchi} \&
  {Mesinger}}{2013}]{2013arXiv1301.6781S}
{Sobacchi} E.,  {Mesinger} A.,  2013, ArXiv e-prints

\bibitem[\protect\citeauthoryear{{Tashiro} \& {Ho}}{{Tashiro} \&
  {Ho}}{2012}]{2012arXiv1205.0563T}
{Tashiro} H.,  {Ho} S.,  2012, ArXiv e-prints

\bibitem[\protect\citeauthoryear{{Tashiro} \& {Sugiyama}}{{Tashiro} \&
  {Sugiyama}}{2012}]{2012MNRAS.420..441T}
{Tashiro} H.,  {Sugiyama} N.,  2012, \mnras, 420, 441

\bibitem[\protect\citeauthoryear{{Valageas}}{{Valageas}}{2010}]{2010A&amp;A...%
514A..46V}
{Valageas} P.,  2010, \aap, 514, A46

\bibitem[\protect\citeauthoryear{{Verde}, {Jimenez}, {Kamionkowski} \&
  {Matarrese}}{{Verde} et~al.}{2001}]{2001MNRAS.325..412V}
{Verde} L.,  {Jimenez} R.,  {Kamionkowski} M.,    {Matarrese} S.,  2001,
  \mnras, 325, 412

\bibitem[\protect\citeauthoryear{{Verde}, {Wang}, {Heavens} \&
  {Kamionkowski}}{{Verde} et~al.}{2000}]{2000MNRAS.313..141V}
{Verde} L.,  {Wang} L.,  {Heavens} A.~F.,    {Kamionkowski} M.,  2000, \mnras,
  313, 141

\bibitem[\protect\citeauthoryear{{Wang} \& {Kamionkowski}}{{Wang} \&
  {Kamionkowski}}{2000}]{2000PhRvD..61f3504W}
{Wang} L.,  {Kamionkowski} M.,  2000, \prd, 61, 063504

\bibitem[\protect\citeauthoryear{{Wyithe} \& {Cen}}{{Wyithe} \&
  {Cen}}{2007}]{2007ApJ...659..890W}
{Wyithe} J.~S.~B.,  {Cen} R.,  2007, \apj, 659, 890

\bibitem[\protect\citeauthoryear{{Wyithe} \& {Loeb}}{{Wyithe} \&
  {Loeb}}{2003}]{2003ApJ...588L..69W}
{Wyithe} J.~S.~B.,  {Loeb} A.,  2003, \apjl, 588, L69

\bibitem[\protect\citeauthoryear{{Wyithe} \& {Morales}}{{Wyithe} \&
  {Morales}}{2007}]{2007MNRAS.379.1647W}
{Wyithe} J.~S.~B.,  {Morales} M.~F.,  2007, \mnras, 379, 1647

\bibitem[\protect\citeauthoryear{{Xia}, {Baccigalupi}, {Matarrese}, {Verde} \&
  {Viel}}{{Xia} et~al.}{2011}]{2011JCAP...08..033X}
{Xia} J.-Q.,  {Baccigalupi} C.,  {Matarrese} S.,  {Verde} L.,    {Viel} M.,
  2011, \jcap, 8, 33

\bibitem[\protect\citeauthoryear{{Xia}, {Bonaldi}, {Baccigalupi}, {De Zotti},
  {Matarrese}, {Verde} \& {Viel}}{{Xia} et~al.}{2010}]{2010JCAP...08..013X}
{Xia} J.-Q.,  {Bonaldi} A.,  {Baccigalupi} C.,  {De Zotti} G.,  {Matarrese} S.,
   {Verde} L.,    {Viel} M.,  2010, \jcap, 8, 13

\bibitem[\protect\citeauthoryear{{Xia}, {Viel}, {Baccigalupi}, {De Zotti},
  {Matarrese} \& {Verde}}{{Xia} et~al.}{2010}]{2010ApJ...717L..17X}
{Xia} J.-Q.,  {Viel} M.,  {Baccigalupi} C.,  {De Zotti} G.,  {Matarrese} S.,
  {Verde} L.,  2010, \apjl, 717, L17

\bibitem[\protect\citeauthoryear{{Yokoyama} \& {Matsubara}}{{Yokoyama} \&
  {Matsubara}}{2012}]{2012arXiv1210.2495Y}
{Yokoyama} S.,  {Matsubara} T.,  2012, ArXiv e-prints

\bibitem[\protect\citeauthoryear{{Yoo}}{{Yoo}}{2010}]{2010PhRvD..82h3508Y}
{Yoo} J.,  2010, \prd, 82, 083508

\bibitem[\protect\citeauthoryear{{Yoo}, {Fitzpatrick} \& {Zaldarriaga}}{{Yoo}
  et~al.}{2009}]{2009PhRvD..80h3514Y}
{Yoo} J.,  {Fitzpatrick} A.~L.,    {Zaldarriaga} M.,  2009, \prd, 80, 083514

\bibitem[\protect\citeauthoryear{{Zhang}, {Hui} \& {Haiman}}{{Zhang}
  et~al.}{2007}]{2007MNRAS.375..324Z}
{Zhang} J.,  {Hui} L.,    {Haiman} Z.,  2007, \mnras, 375, 324

\end{thebibliography}

\appendix
\section{Derivation of the non-Gaussian ionized density bias in the ESMR}
\label{APP:ESMR}

In this appendix we present our derivation of the non-Gaussian ionized density bias from the ESMR.  The first step is to compute the ionized fraction bias.  We will then use equation (\ref{EQ:bHII2bx}) to convert the ionized fraction bias to the ionized density bias.  We begin with the non-Gaussian ionized fraction contrast, equation (\ref{EQ:deltax}), which contains both scale-independent and -dependent contributions.  In what follows, we define the following functions for notational convenience:

\begin{equation}
 \chi  \equiv \frac{1}{3}  \left\{  \frac{\delta_c-\da_R}{\Smin-S_R} - \frac{1}{\da_c-\da_R} \right\},
\end{equation}

\begin{equation}
  \psi \equiv \frac{1}{S_R} \left[ \da_c - ( \da_c -\da_R) \coth \left( \frac{\da_c^2-\da_c \da_R }{ S_R} \right)  \right].
\end{equation}

\subsection{Scale-independent terms}

The scale-independent contributions can be further divided into the Gaussian term and the non-Gaussian correction. The first term on the right-hand side of equation (\ref{EQ:deltax}) yields the former.  Taylor expanding it to first order about $\da_R = 0$ and taking the limit as $S_R \rightarrow 0$ gives

\begin{equation}
b_x^{\mathrm{G}} = \frac{2} {\da_c} \frac{\pd \ln \avgfcollgm}{\pd \ln \Smin},
\label{EQ:bcollAPP}
\end{equation}  
The non-Gaussian part comes from
\begin{equation}
\da_x^{(i)} = \frac{\langle \da_{\mathrm{min}}^3 \rangle}{\avgfcollgm} \left( \frac{\pd \fcollgm}{\pd \Smin} \chi - \frac{\fcollgm}{\avgfcollgm} \frac{\pd \avgfcollgm}{ \pd \Smin} \chi_0  \right),
\end{equation}
where the ``0" sub-script on $\chi$ again denotes the $\da_R = 0$ and $S_R = 0$ limit.  We Taylor expand $\da_x^{(i)}$ to first order in $\da_R$ about $\da_R=0$.  The constant term is zero, while the linear term is

\begin{align}
\da_x^{(i)} = & \frac{\langle \da_{\mathrm{min}}^3 \rangle}{\avgfcollgm} \Biggl[ \frac{\pd}{\pd \da_R} \left. \left( \chi \frac{\pd \fcollgm}{\pd \Smin} \right)\right|_{\da_R=0} \nonumber \\ & - \frac{\chi_0}{\avgfcollgm}\frac{\pd \avgfcollgm}{\pd \Smin } \left. \frac{\pd \fcollgm}{\pd \da_R} \right|_{\da_R=0} \Biggr] \da_R.
\end{align}
Taking the limit as $S_R \rightarrow 0$ yields

\begin{align} 
 \Delta b_x^{(i)} =     -\frac{\mathcal{S}^{(3)}_{\mathrm{min}} }{6} \Smin  b_x^{\mathrm{G}} \biggl[ \frac{3\da_c \Smin - \da_c^3}{\Smin^2} \nonumber \\  +  b_x^{\mathrm{G}} \left( \frac{\da_c^2}{\Smin}  -1, \right) \biggr],
 \label{EQ:appbcollSI}
 \end{align}
where $\mathcal{S}^{(3)}_{\mathrm{min}} \equiv \langle \damin^3 \rangle / \Smin^2$ denotes the skewness of density fluctuations smoothed on the $\mmin$ scale.  As described in \S\ref{SEC:ESMRGauss}, this is the Lagrangian ionized fraction bias, which is equivalent to the Eulerian ionized fraction bias.  We may therefore apply equation (\ref{EQ:bHII2bx}) to arrive at equation (\ref{EQ:bcfPNG_SI}) for the scale-independent correction to the ionized density bias.

\subsection{The scale-dependent term}

Scale-dependent contributions can come from sub-terms involving $\langle \damin^2 \da_R \rangle$ in the last term of equation (\ref{EQ:deltax}).  There are two such terms,

\begin{equation}
d_1 = \frac{\langle \damin^2 \da_R \rangle}{\Smin} \frac{1}{\avgfcollgm} \frac{\pd  \fcollgm}{ \pd \ln \Smin} \psi ,
\label{EQ:deltaxd}
\end{equation}
and

\begin{equation}
d_2 = - 3 \frac{\langle \damin^2 \da_R \rangle}{\Smin} \frac{1}{\avgfcollgm} \frac{\pd  \fcollgm}{ \pd \ln \Smin} \chi.
\end{equation}  

Equation (\ref{EQ:deltax}) is written in coordinate space whereas the scale-dependence of the bias is manifested in Fourier space.  We employ a convenient method used by \citet{2011PhRvD..84f3512D} for the conversion.  Adapting their strategy to the current task, we take the cross-correlation, $\langle \da_x \da_R \rangle$, between the ionized fraction contrast and the large-scale smoothed density contrast.  We then rearrange the equations to pick off the Fourier space bias parameter.   

\citet{2011PhRvD..84f3512D} considered the statistics of thresholded regions -- regions in the initial density field with peak height above some given value.  They showed that the power spectrum of thresholded regions, which in the Press-Schechter formalism can be interpreted as the collapsed fraction power spectrum, can to first order be expressed with a simple bias relation.  Adopted to our notation, the linear bias relation they found is $\dcoll(\boldsymbol{k}) = \bcoll (k) \damin(\boldsymbol{k})$.  Noting that $\dcoll=\da_x$, and $\bcoll=b_x$ in the ESMR, we can write $\da_x(\boldsymbol{k}) = b_x (k) \damin(\boldsymbol{k})$. We use this relation to write the left-hand side of (\ref{EQ:deltax}) as

\begin{equation}
\langle \da_x \da_R \rangle = \int \frac{\dd^3 \boldsymbol{k}}{(2 \pi)^3} b_x(k) \MM_{\mathrm{min}}(k) \MM_R(k) P_{\phi}(k).
\label{EQ:lhs_sdbias}
\end{equation}  
If we can now rewrite the appropriate terms in the cross-correlation of $\da_R$ with the right-hand side of (\ref{EQ:deltax}) in a similar way, we can simply read off the bias coefficient.    

Let us first consider equation (\ref{EQ:deltaxd}).  We Taylor expand it to first order about $\da_R =0$, neglecting the constant term, since it will vanish upon taking the cross-correlation with $\da_R$.  Taking the limit $S_R \rightarrow 0$, and using the property,

\begin{equation}
\lim_{S_R \rightarrow 0} \left. \psi \right|_{\da_R=0} =0,
\end{equation}
we can write the cross-correlation of $d_1$ with $\da_R$ as

\begin{equation}
\langle d_1 \da_R \rangle = \frac{\langle \damin^2 \da_R \rangle}{\Smin}\frac{\pd \ln \avgfcollgm}{\pd \ln \Smin}  \lim_{S_R\rightarrow0} \left( \left. \frac{\pd \psi}{\pd \da_R }\right|_{\da_R=0} S_R \right)
\end{equation}
Conveniently, 

\begin{equation}
 \lim_{S_R\rightarrow0} \left( \left. \frac{\pd \psi}{\pd \da_R}\right|_{\da_R=0} S_R \right) =1,
\end{equation}
so

\begin{equation}
\langle d_1 \da_R \rangle = \frac{\langle \damin^2 \da_R \rangle}{\Smin} \frac{\da_c b_x^G}{ 2}.
\label{EQ:SD2ndtolast}
\end{equation}
In a similar manner it is straightforward to show that $\langle d_2 \da_R \rangle =0$.

The mixed correlator, $\langle \da_{\mathrm{min}}^2 \da_R \rangle$, can be written in a form similar to equation (\ref{EQ:lhs_sdbias}) [see discussion leading up to equation (9) of \citet{2012arXiv1206.3305D}], 

\begin{align}
\langle \da^2_{\mathrm{min}}  \da_R \rangle  =  \int \frac{\dd^3 \mathbf{k}}{(2 \pi)^3}  \MM_{R}(k,z) P_\phi (k) 4 \Smin ~\mathcal{F}^{(3)}_{\mathrm{min}}(k) .
\label{EQ:mixedcorrelator}
\end{align}
Plugging this into equation (\ref{EQ:SD2ndtolast}), in combination with equation (\ref{EQ:lhs_sdbias}), yields 

\begin{equation}
\Delta b_x^{(d)}(k) = 2 \delta_c b_x^{\mathrm{G}}  \frac{\mathcal{F}^{(3)}_{\mathrm{min}}(k)}{\MM_{\mathrm{min}}(k)},
\label{EQ:appbcollSD}
\end{equation}
where the form factor $\mathcal{F}^{(3)}_{\mathrm{min}}(k)$ is defined in equation (\ref{EQ:F3formfactor}).  We note that this expression is equivalent to equation (43) of \citet{2011PhRvD..84f3512D} for $N=3$.   Their linear bias of thresholded regions is equivalent to our collapsed fraction bias, defined in equation (\ref{EQ:bfcolldef}).  Since the ionized fraction bias is equivalent to the collapsed fraction bias in the ESMR, the former is also equivalent to the linear bias of thresholded regions in  \citet{2011PhRvD..84f3512D}, hence the correspondence between their equation (43) and equation (\ref{EQ:appbcollSD}).  Finally, equations (\ref{EQ:bHII2bx}) and (\ref{EQ:appbcollSD}) yield equation (\ref{EQ:SDcorrection}) for the scale-dependent correction to the ionized density bias.

\section{Details on the source emissivity in the LPTR}
\label{APP:emissivity}

Here we derive expressions for the spatially averaged source emissivity and its first-order perturbation, which appear in equations (\ref{EQ:avgradiativetransfer}) and (\ref{EQ:FOradiativetransfer}) respectively.  For convenience, we define the quantity $n_{\mathrm{H,coll}}(\boldsymbol{r},\mmin,R,\eta) \equiv \nH(\boldsymbol{r},R,\eta) \fcoll(\mmin,R,\da_R,\eta)$, representing the density of hydrogen collapsed into ACHs, smoothed over scale $R$.  Note that $n_{\mathrm{H,coll}}$ appears in equation (\ref{EQ:emissivityB}), and its time derivative appears in equation (\ref{EQ:emissivityA}).       

We write $n_{\mathrm{H,coll}}$ in terms of a spatial average and a first-order perturbation as follows: 

\begin{align}
n_{\mathrm{H,coll}}(\boldsymbol{r},\mmin,R,\eta) = \bnH \biggl[ \avgfcoll(\mmin,\eta) \nonumber \\ + \Delta_{\mathrm{H,coll}}(\mathbf{r},\mmin,R,\eta) \biggr]. 
\label{EQ:nHcoll}
\end{align}
Referring back to equations (\ref{EQ:pertdefs}), (\ref{EQ:emissivityA}), and (\ref{EQ:emissivityB}), the quantity $\bar{\xi}_s$ which appears in equation (\ref{EQ:avgradiativetransfer}) can simply be read off from equation (\ref{EQ:nHcoll}); in source-model A, $\bar{\xi}^A_s = \gamma^A(x_\nu)  \pd \left[ \avgfcoll(\mmin,\eta) \right] / \pd \eta$, while in source-model B, $\bar{\xi}^B_s = \gamma^B(x_\nu)  \avgfcoll(\mmin,\eta) $.  

If we write the smoothed hydrogen number density as $\nH(\mathbf{r},R,\eta) = \bnH(\eta) \left[1+\da_R(\mathbf{r},\eta) \right]$, and the conditional collapsed fraction as $\fcoll(\mmin,R,\da_R,\eta) = \avgfcoll(\mmin,\eta) + \Delta \fcoll(\mmin,R,\da_R,\eta)$, then to leading order $\Delta_{\mathrm{H,coll}} = \da_R(\boldsymbol{r},\eta) \avgfcoll(\mmin,\eta) + \Delta_{\mathrm{fcoll}}(\mmin,R,\da_R,\eta)$, and its Fourier transform is

\begin{align}
\tilde{\Delta}_{\mathrm{H,coll}}(\mathbf{k},\eta) = \avgfcoll(\eta) \tilde{\delta}(\mathbf{k},\eta) + \tilde{\Delta}_{\mathrm{fcoll}}(\mathbf{k},\eta),
\label{EQ:DHcoll}
\end{align} 
where from here on we suppress the $\mmin$ dependence for brevity (it should be understood that the collapsed fraction refers to the fraction of mass in halos above the ACH threshold), and the dependence on $R$ has been dropped under the assumption of small $k$.  The quantity $\tilde{\Delta}_{\mathrm{fcoll}}$ may be written in terms of the collapsed fraction bias, 

\begin{align}
\tilde{\Delta}_{\mathrm{fcoll}}(\mathbf{k},\eta) = \bcoll(\mathbf{k},\eta) \tilde{\da}(\mathbf{k},\eta) \avgfcoll(\eta),
\end{align} 
from which equation (\ref{EQ:DHcoll}) becomes

\begin{align}
\tilde{\Delta}_{\mathrm{H,coll}}(\mathbf{k},\eta) =  \avgfcoll(\eta) \biggl[ 1  + \bcoll(\mathbf{k},\eta) \biggr] \tilde{\delta}(\mathbf{k},\eta). 
\label{EQ:DHcollfinal}
\end{align}
The quantity $\tilde{\Delta}_s$ appearing in equation (\ref{EQ:FOradiativetransfer}) can now be read off, yielding equations (\ref{EQ:DsA}) and (\ref{EQ:DsB}) in source-models A and B respectively.  

The collapsed fraction bias parameters which enter these expressions  follow trivially from the results of appendix \ref{APP:ESMR} since, in the ESMR, the ionized fraction bias is assumed equal to the collapsed fraction bias.  The Gaussian collapsed fraction bias is therefore equation (\ref{EQ:bcollAPP}), while the scale-independent and -dependent terms from PNG are (\ref{EQ:appbcollSI}) and (\ref{EQ:appbcollSD}) respectively, each with the replacement $b_x \rightarrow \bcoll$.

\end{document}